\documentclass{aa}  

\usepackage{graphicx}
\usepackage{txfonts}
\usepackage{lipsum}
\usepackage{subcaption}         
\usepackage{lscape}             
\usepackage{placeins}           

\usepackage{txfonts}
\usepackage[colorlinks=true, citecolor=blue, urlcolor=blue]{hyperref}
\usepackage{placeins}
\usepackage{color, soul}
\usepackage{ulem}
\usepackage{natbib}
\usepackage{lipsum}
\bibpunct{(}{)}{;}{a}{}{,} 
\usepackage{tabularx}  
\usepackage{adjustbox}

\begin{document}

   \title{Population synthesis of Galactic middle-aged pulsar wind nebulae}

   \subtitle{II. Observational signatures of superefficiency}


   \author{D. F. Torres{\inst{1,2,3}}\thanks{dtorres@ice.csic.es}, A. De Sarkar\inst{1}\thanks{desarkar@ice.csic.es},  B. Olmi{\inst{4}},  N. Bucciantini{\inst{4,5,6}}, D. M.-A. Meyer{\inst{1}}
   }

   \institute{Institute of Space Sciences (ICE, CSIC), Campus UAB, Carrer de Can Magrans s/n, 08193 Barcelona, Spain
   \and
   Institut d'Estudis Espacials de Catalunya (IEEC), Gran Capit\`a 2-4, 08034 Barcelona, Spain 
   \and
   Instituci\'o Catalana de Recerca i Estudis Avan\c cats (ICREA), 08010 Barcelona, Spain 
             \and INAF - Osservatorio Astrofisico di Arcetri, Largo E. Fermi 5, I-50125 Firenze, Italy
            \and Università degli Studi di Firenze, Via Sansone 1, 50019, Sesto F.no (Firenze), Italy 
            \and INFN - Sezione di Firenze, Via G. Sansone 1, I-50019 Sesto F.no (Firenze), Italy
}

   \date{Received XXXX XX, 20XX}
 
  \abstract{
  Pulsar wind nebulae (PWNe) interacting with the host supernova remnants (SNRs) can enter the reverberation phase in which reverse-shock–driven compression amplifies the magnetic field and rapidly reprocesses particles, sometimes producing “superefficiency,” where the radiative output in a given frequency band exceeds the pulsar’s instantaneous spin-down power. We investigate the prevalence of this phenomenon in the Galactic population by modeling PWNe with the hybrid \texttt{TIDE+L} framework, which self-consistently follows dynamical evolution, particle spectra, and emission from radio to PeV energies. We track superefficiency across frequency bands and evolutionary stages, analyzing both individual objects and ensemble properties, including compression-resolved samples and population spectral energy distributions. Superefficiency is most common in the far-infrared, but emerges across frequencies and evolutionary phases. It is enhanced in systems where accumulated low-energy electrons radiate in magnetically amplified nebulae.
We predict substantially more superefficient sources than a purely thin-shell model
would, with differences ranging from factors of a few in FIR and GeV bands to more than an order of magnitude in several optical/UV/X-ray bands.
  }

   \titlerunning{Population synthesis of middle-aged PWNe - II}
   \authorrunning{Torres et al.}
   \maketitle
   \nolinenumbers

\section{Introduction}

Pulsar wind nebulae (PWNe) are bubbles of relativistic plasma produced by the interaction of pulsar outflow with its surrounding medium. 
The dynamical and radiative evolution of a PWN is strongly influenced by the evolutionary stage of the parent supernova remnant (SNR). 
While young PWNe expand freely into the cold, self-similar ejecta, the majority of Galactic PWNe are older, with ages over $\sim10^{4}$ yr. 
In this stage, the reverse shock of the SNR, moving towards the explosion centre, collides with the PWN. 
This initiates the so-called reverberation phase, during which the nebula undergoes significant compression, re-expansion, and in some cases multiple oscillatory cycles, see e.g., \cite{reynolds84, gelfand09, martin16, bandiera23III}. 
Reverberation alters the energy balance of the nebula: adiabatic compression increases the internal pressure and magnetic field strength, while radiative losses become more severe. 

One remarkable possible outcome of reverberation is 
superefficiency, which
labels the period when the integrated radiative luminosity of a PWN in a given frequency band at a given epoch exceeds its contemporaneous pulsar spin-down power \citep{torres18,torres19}.
This phenomenon arises because, during reverse-shock-driven compression, energy is transferred from the surrounding medium to the nebula, adiabatically heating the particles and amplifying the magnetic field, which together enhance its emission beyond the pulsar’s instantaneous spin-down power (see  e.g. \cite{Vorster13,torres17}).
The consequence is that the instantaneous radiative efficiency can temporarily exceed unity, in stark contrast to the usual expectation that $L_{\rm band} \ll \dot{E}_{\rm SD}$ at all times in all frequency bands.
Superefficiency therefore challenges the commonly adopted picture that the spin-down luminosity sets a strict upper bound on the observable non-thermal luminosity of PWNe.  

From a theoretical standpoint, superefficiency redefines how we interpret the energy budget of PWNe and demands that models account for external energy transfer from the SNR reverse shock, along with the resulting adiabatic heating and magnetic field amplification during compression. 
Observationally, it complicates the classification of high-energy sources: a superefficient PWN may appear far more luminous than its pulsar would suggest, potentially leading to misidentifications in catalogs or to the inference of anomalously efficient acceleration processes. 
Superefficiency could also explain PWNe exhibiting unusually high luminosity ratios between X-ray and $\gamma$-ray bands.
It could also help explain why some PWNe might be unusually bright for their pulsar's present-day $\dot{E}_{\rm SD}$ , whereas other pulsars with comparable $\dot{E}_{\rm SD}$ show no detectable nebular counterpart.

However, one-zone purely thin-shell models impose simplified prescriptions for this phase, limiting their reliability in describing the detailed dynamics of reverberation \citep[see, e.g.,][]{pacini73, reynolds84, becker07, zhang08, qiao09, gelfand09, fang10, tanaka10, tanaka11, bucciantini11, martin12, Vorster13, torres13, torres14, temim15, martin16, kolb17,vanrensburg18, zhu18, torres18, torres19,fiori20, desarkar22, martin22}. 
Consequently, the prevalence and observational characteristics of superefficient PWNe across the Galaxy remain largely unconstrained.  

To address this gap, we employ the hybrid \texttt{TIDE+L} framework \citep{bandiera20, bandiera21, bandiera23II, bandiera23III}, which combines the computational efficiency of the thin-shell approximation during the free-expansion stage with the fidelity of a Lagrangian treatment of the SNR during reverberation. 
This approach self-consistently follows the dynamical and spectral evolution of the PWN, while remaining fast enough for large-scale simulations. 
In the companion study \citep{desarkar26}, we applied \texttt{TIDE+L} to evolve a synthetic population of Galactic middle-aged PWNe to quantify their detectability with current and upcoming $\gamma$-ray observatories. 
In this second paper, we investigate the manifestation of superefficiency across the same synthetic Galactic population of PWNe.

Assuming a realistic SN explosion rate, we first estimate the number of superefficient PWNe expected at the present Galactic epoch across multiple frequency bands, distinguishing systems undergoing active compression from those in post-compression stages. 
Once the superefficient PWNe are identified, we study them in detail. 
We repeat the same analysis for the same source population using \texttt{TIDE}, a purely thin-shell model without the Lagrangian reverberation modules implemented in \texttt{TIDE+L}, in order to isolate the effect of a more detailed reverberation treatment.
We follow the evolution of the superefficient fraction with age and frequency band. 
To identify the physical mechanisms behind the population trends, we examine representative individual systems with different compression factors, as well as model sources that are superefficient in FIR, X-rays, and GeV gamma rays. 
We also construct grouped SEDs and electron spectra for selected superefficient subsets, including compression-resolved samples, to relate the spectral signatures of superefficiency to particle accumulation, cooling, and  magnetic field amplification.

By systematically exploring the conditions under which superefficiency arises within a population synthesis framework, this work provides the first statistical baseline for interpreting candidate superefficient PWNe and guiding targeted multi-wavelength (MWL) follow-up observations.

\section{The population and its evolution}\label{pop}

The generation of the initial population and the computation of its subsequent evolution follow the same prescription adopted in \cite{desarkar26}. 
We briefly summarize the assumptions here and refer the reader to \cite{desarkar26} for a detailed description.

We generate a synthetic Galactic PWN population consistent with pulsar birth properties and core-collapse SN progenitors, focusing on middle-aged systems ($\leq 10^{5}$ yr). 
From a total of 2400 simulated sources, we randomly select 1600 objects per realization, corresponding to the expected number of PWNe formed over the past $10^{5}$ yr for the best-fit Galactic core-collapse SN rate of $\sim$1.6 per century \citep{rozwa21}, neglecting older, highly diluted systems. 
To capture stochastic sampling uncertainties and the impact of the assumed SN rate, we perform 1000 random realizations of populations containing 1600 sources, from which we derive the mean expected number of superefficient PWNe and the associated population variance, unless stated otherwise.

\begin{figure*}
\centering
\includegraphics[width=0.3\linewidth]{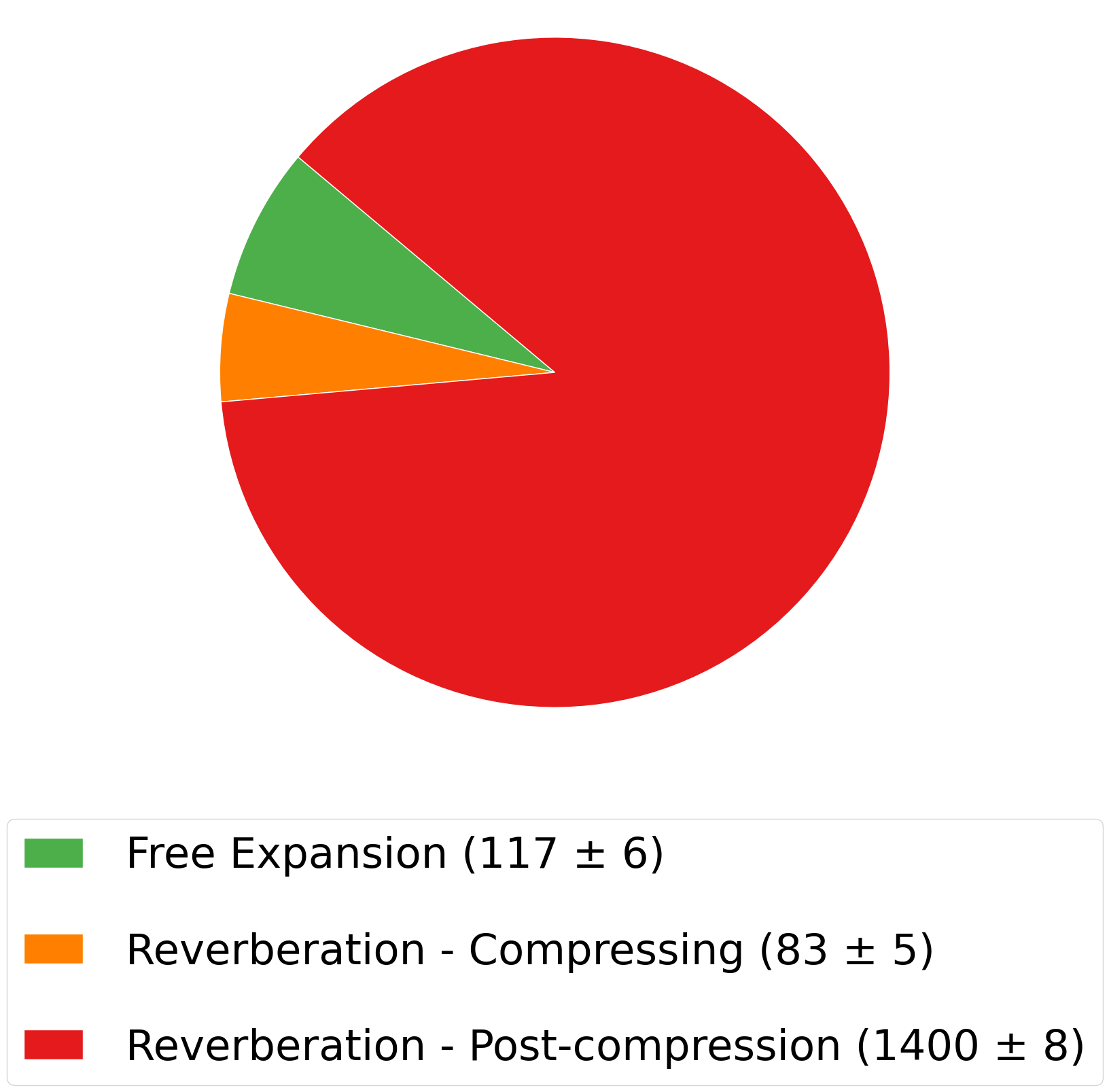}
\includegraphics[width=0.6\linewidth]{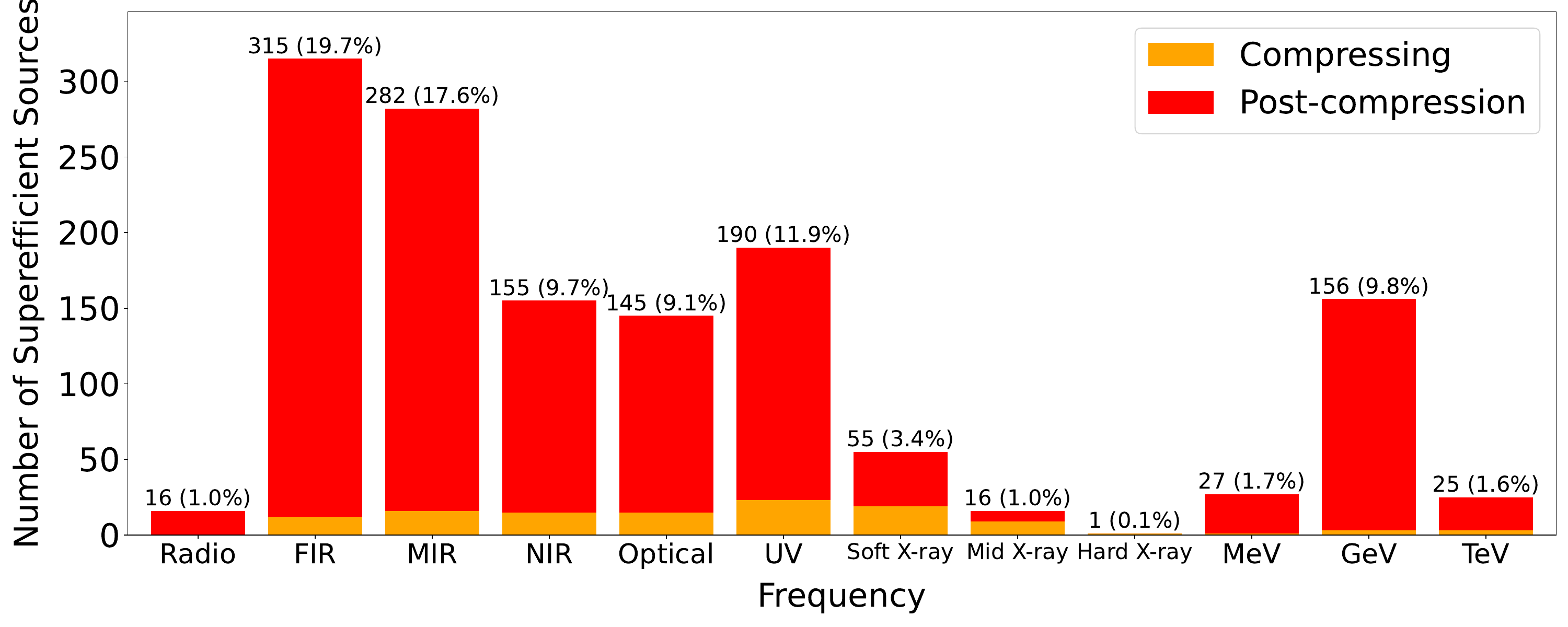}
\includegraphics[width=0.3\linewidth]{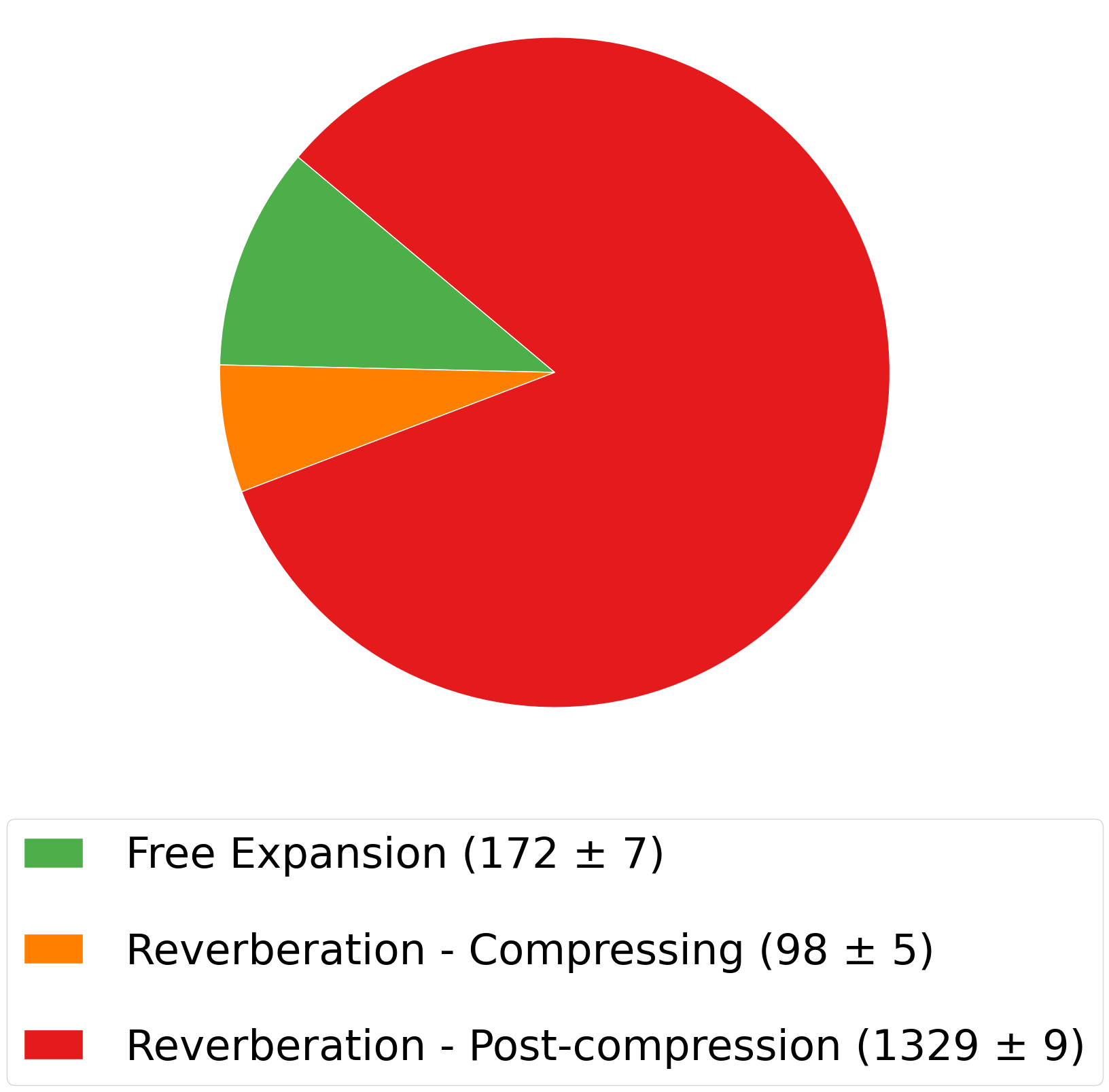}
\includegraphics[width=0.6\linewidth]{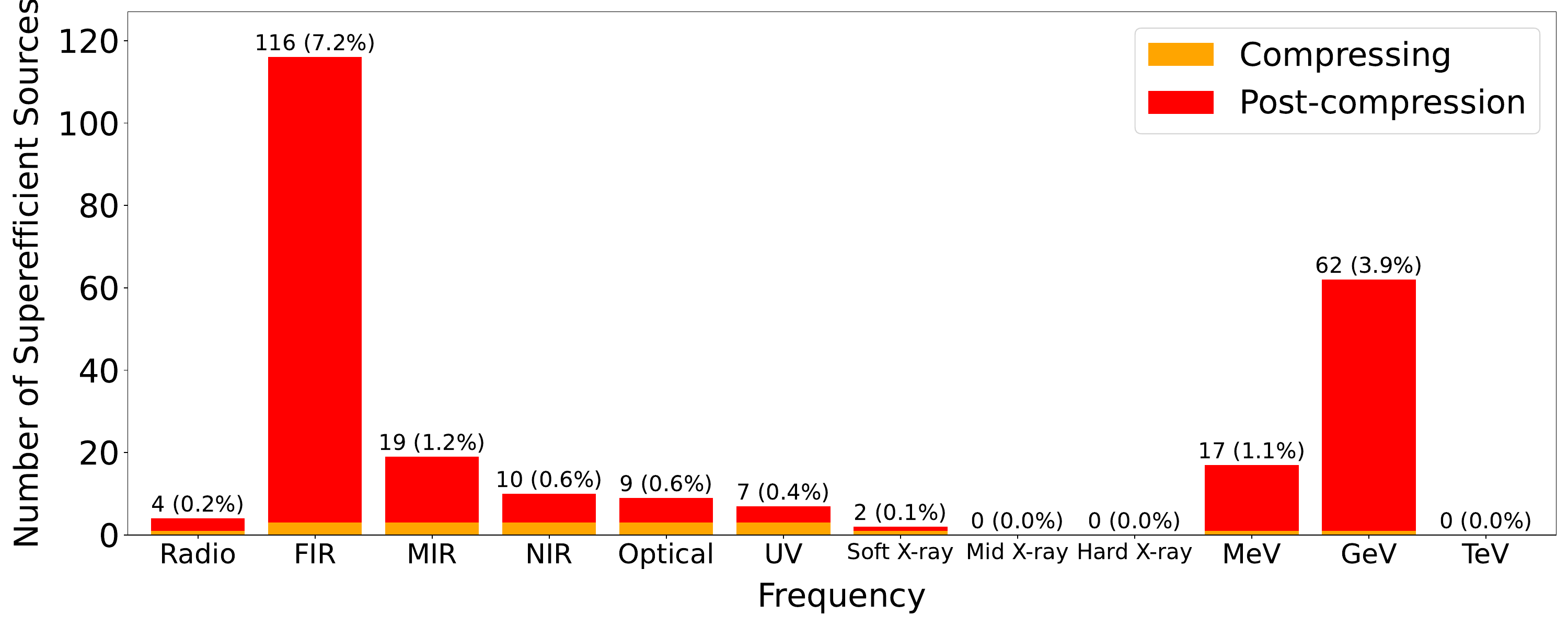}
\caption{The pie charts show the distribution of PWNe across evolutionary stages-free expansion (green), reverberation-compressing (orange), and reverberation-post-compression (red). The accompanying bar plots display the frequency-dependent distribution of superefficient sources, separated by compression state using the same color scheme. The top and bottom panels correspond to the population simulated with \texttt{TIDE+L} and with \texttt{TIDE}, respectively. All results refer to the current epoch. For the pie charts, the reported counts represent mean values obtained from 1000 realizations of randomly drawn 1600 PWNe at the final simulation time, with the corresponding $1\sigma$ uncertainties indicated in the legend. The bar plots show the corresponding mean number and fractional contribution of superefficient PWNe in each frequency range obtained via the same sampling procedure.}
\label{fig: pie_super}
\end{figure*}

\begin{table*}
\caption{Predicted estimation of the number of superefficient PWNe in different frequency bands for two modeling assumptions.}
\label{tab:sup_counts}
\centering
\scriptsize
\begin{tabular*}{0.5\textwidth}{@{\extracolsep{\fill}}lrrr}
\hline\hline
\multicolumn{4}{c}{\textbf{\texttt{TIDE+L}}} \\
\hline
Band & Total & Compressing & Post-compression \\
\hline
Radio        & 16 $\pm$ 2   & 0 $\pm$ 0  & 16 $\pm$ 2 \\
FIR          & 315 $\pm$ 9 & 12 $\pm$ 2 & 303 $\pm$ 9 \\
MIR          & 282 $\pm$ 9 & 16 $\pm$ 2 & 266 $\pm$ 9 \\
NIR          & 155 $\pm$ 7   & 15 $\pm$ 2 & 140 $\pm$ 7 \\
Optical      & 145 $\pm$ 7   & 15 $\pm$ 2 & 130 $\pm$ 6 \\
UV           & 190 $\pm$ 7  & 23 $\pm$ 3 & 167 $\pm$ 7 \\
Soft X-ray   & 55 $\pm$ 4   & 19 $\pm$ 2 & 36 $\pm$ 3 \\
Mid X-ray    & 16 $\pm$ 2    & 9 $\pm$ 2  & 7 $\pm$ 2 \\
Hard X-ray   & 1 $\pm$ 1   & 1 $\pm$ 1  & 0 $\pm$ 0 \\
MeV          & 27 $\pm$ 3    & 1 $\pm$ 1  & 26 $\pm$ 3 \\
GeV          & 156 $\pm$ 7   & 3 $\pm$ 1  & 153 $\pm$ 7 \\
TeV          & 25 $\pm$ 3    & 3 $\pm$ 1  & 22 $\pm$ 3 \\
\hline\hline
\multicolumn{4}{c}{\textbf{\texttt{TIDE}}} \\
\hline
Band & Total & Compressing & Post-compression \\
\hline
Radio        & 4 $\pm$ 1  & 1 $\pm$ 1 & 3 $\pm$ 1 \\
FIR          & 116 $\pm$ 6 & 3 $\pm$ 1 & 113 $\pm$ 6 \\
MIR          & 19 $\pm$ 3 & 3 $\pm$ 1 & 16 $\pm$ 2 \\
NIR          & 10 $\pm$ 2  & 3 $\pm$ 1 & 7 $\pm$ 2 \\
Optical      & 9 $\pm$ 2  & 3 $\pm$ 1 & 6 $\pm$ 1 \\
UV           & 7 $\pm$ 2  & 3 $\pm$ 1 & 4 $\pm$ 1 \\
Soft X-ray   & 2 $\pm$ 1 & 1 $\pm$ 1 & 1 $\pm$ 1 \\
Mid X-ray    & 0 $\pm$ 0  & 0 $\pm$ 0 & 0 $\pm$ 0 \\
Hard X-ray   & 0 $\pm$ 0  & 0 $\pm$ 0 & 0 $\pm$ 0 \\
MeV          & 17 $\pm$ 2  & 1 $\pm$ 1 & 16 $\pm$ 2 \\
GeV          & 62 $\pm$ 4 & 1 $\pm$ 1 & 61 $\pm$ 4 \\
TeV          & 0 $\pm$ 0  & 0 $\pm$ 0 & 0 $\pm$ 0 \\
\hline
\end{tabular*}
\tablefoot{The table reports the mean number of superefficient sources across 1000 realizations, together with the corresponding 1$\sigma$ uncertainties.}
\end{table*}

\begin{table*}
\centering
\scriptsize
\caption{Pairwise number of superefficient PWNe between frequency bands at the current age.}
\label{tab:supereff_matrix}
\setlength{\tabcolsep}{4pt}
\renewcommand{\arraystretch}{1.0}

\begin{tabular}{lcccccccccccc}
\hline
 & Radio & FIR & MIR & NIR & Opt. & UV & Soft  & Mid  & Hard  & MeV & GeV & TeV \\
  &  & &  &  &  & & X-ray & X-ray & X-ray \\
\hline
Radio	& 17 & 16 & 14	& 8	& 6	& 6	& 0	& 0	& 0	& 8	& 7	& 0\\
FIR	    & 16 & 314 & 233 & 138	& 120 & 118 & 26 & 7 & 0 & 22 & 102 & 16\\
MIR	    & 14	& 233 & 285 & 161	& 145 & 148 & 39 & 11 & 0 & 13 & 82 & 18\\
NIR	    & 8	& 138 & 161 & 161 & 142 & 134 & 37 & 11 & 0 & 6 & 63 & 16\\
Optical	& 6	& 120 & 145 & 142 & 148 & 138 & 40 & 12 & 0 & 4 & 54 & 17\\
UV	    & 6	& 118 & 148 & 134 & 138 & 197 & 58 & 18 & 0 & 4 & 53 & 21\\
Soft X-ray	& 0	& 26	& 39 & 37 & 40 & 58 & 58 & 18 & 0 & 1 & 15 & 15\\
Mid X-ray	& 0	& 7	& 11 & 11 & 12 & 18 & 18 & 18 & 0 & 0 & 3 & 9\\
Hard X-ray	& 0	& 0	& 0	& 0	& 0	& 0	& 0	& 0	& 0	& 0	& 0	& 0\\
MeV	        & 8	& 22	& 13	& 6	& 4	& 4	& 1	& 0	& 0	& 28	& 23	& 1\\
GeV	        & 7	& 102	& 82	& 63	& 54	& 53	& 15	& 3	& 0	& 23	& 157	& 16\\
TeV	        & 0	& 16	& 18	& 16	& 17	& 21	& 15	& 9	& 0	& 1	& 16	& 28\\
\hline
\end{tabular}
\tablefoot{The table shows the matrix $\mathcal{N}_{ij}(t)$ at the current age, where $\mathcal{N}_{ij}$
denotes the number of sources superefficient in both bands $i$ and $j$, independent of evolutionary state. Note that the content of this table corresponds to a single random realization of 1600 PWNe.}
\end{table*}

\begin{figure*}
    \includegraphics[width=\linewidth]{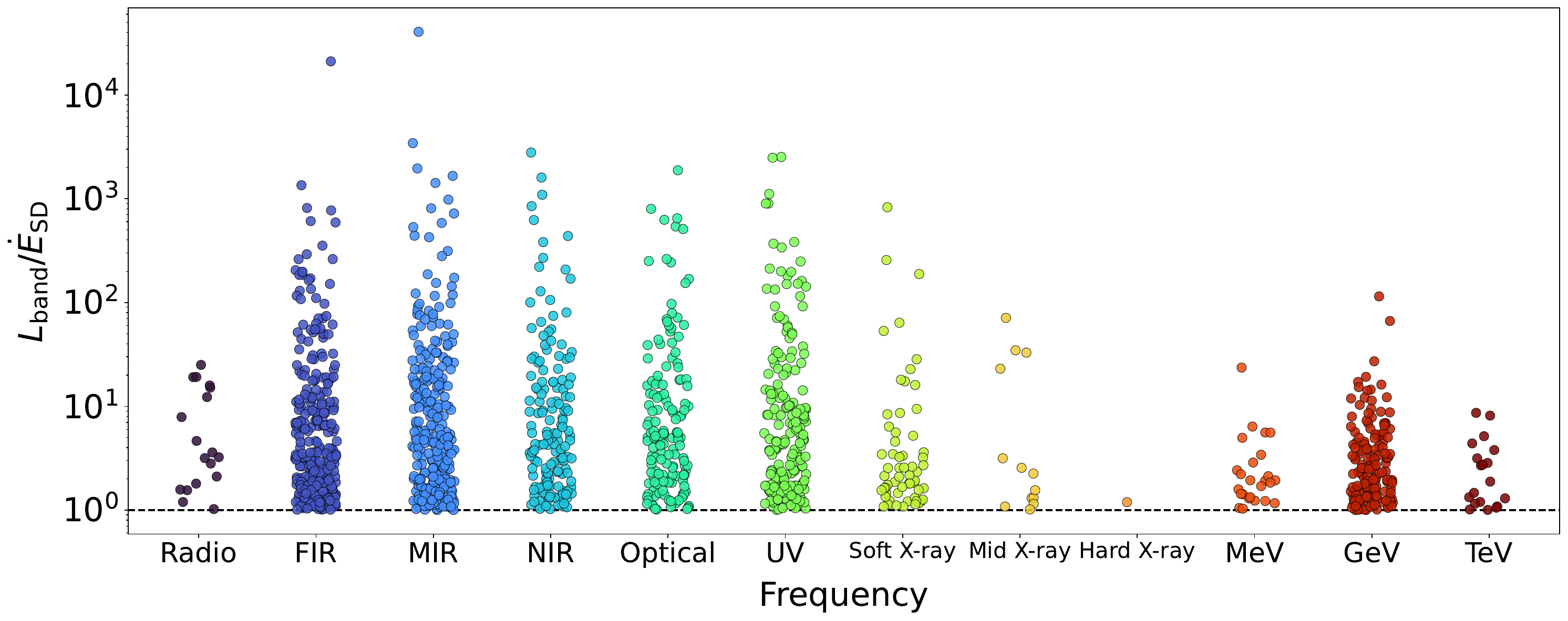}
    \caption{Distribution of the superefficiency ratio, $L_{\rm band}/\dot{E}_{\rm SD}$, for the superefficient PWNe in a random realization of 1600 sources at the current age. Each point corresponds to an individual source that satisfies $L_{\rm band}/\dot{E}_{\rm SD}>1$ in the corresponding frequency. The different bands are shown along the horizontal axis, while the vertical axis gives the ratio between the integrated luminosity in the corresponding frequency band and the pulsar spin-down power at the current age. The dashed horizontal line marks $L_{\rm band}/\dot{E}_{\rm SD}=1$, i.e. the threshold above which a source is classified as superefficient.}
\label{fig: Lband/Edot}
\end{figure*}

\begin{figure*}
    \includegraphics[width=0.33\linewidth]{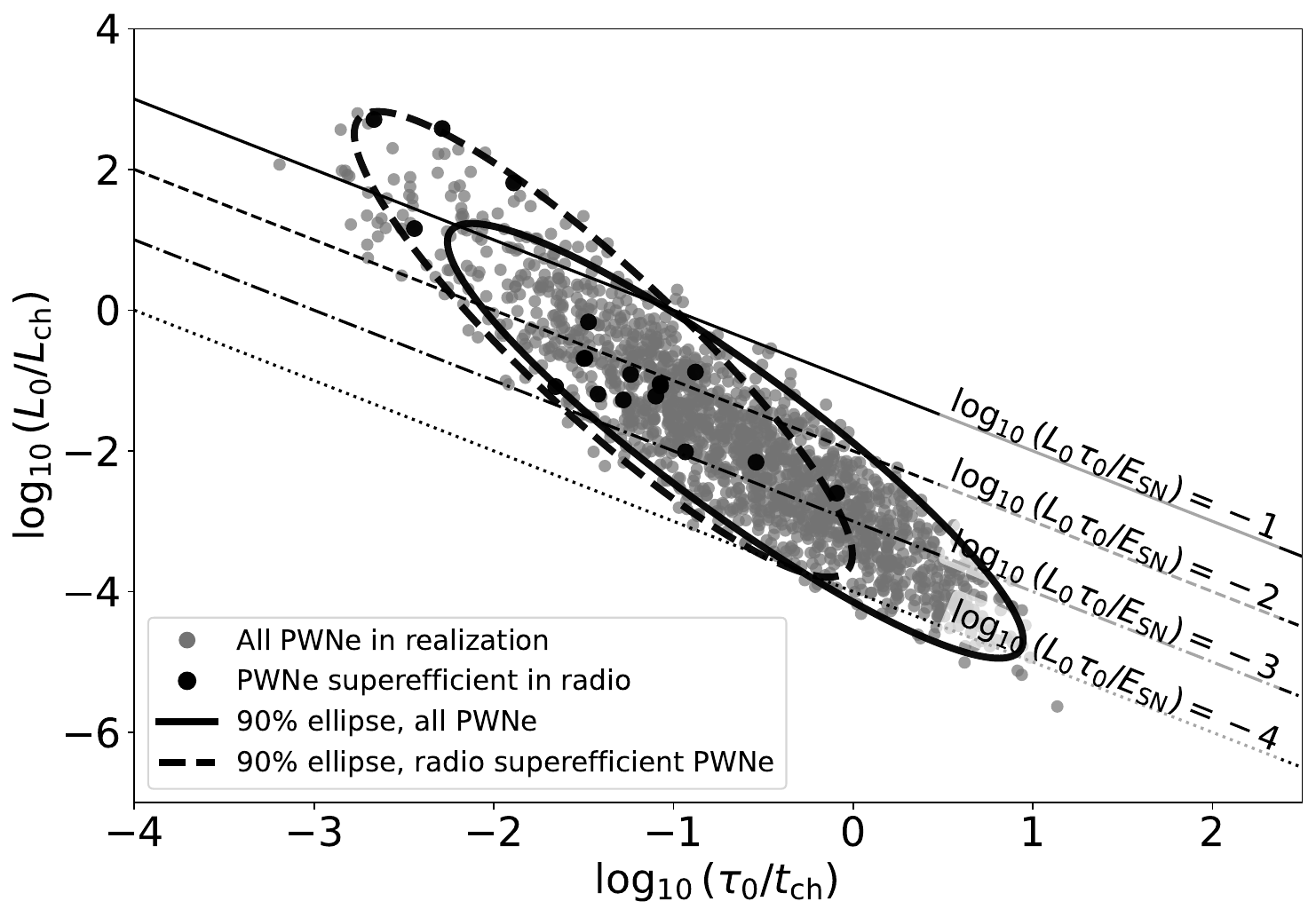}
     \includegraphics[width=0.33\linewidth]{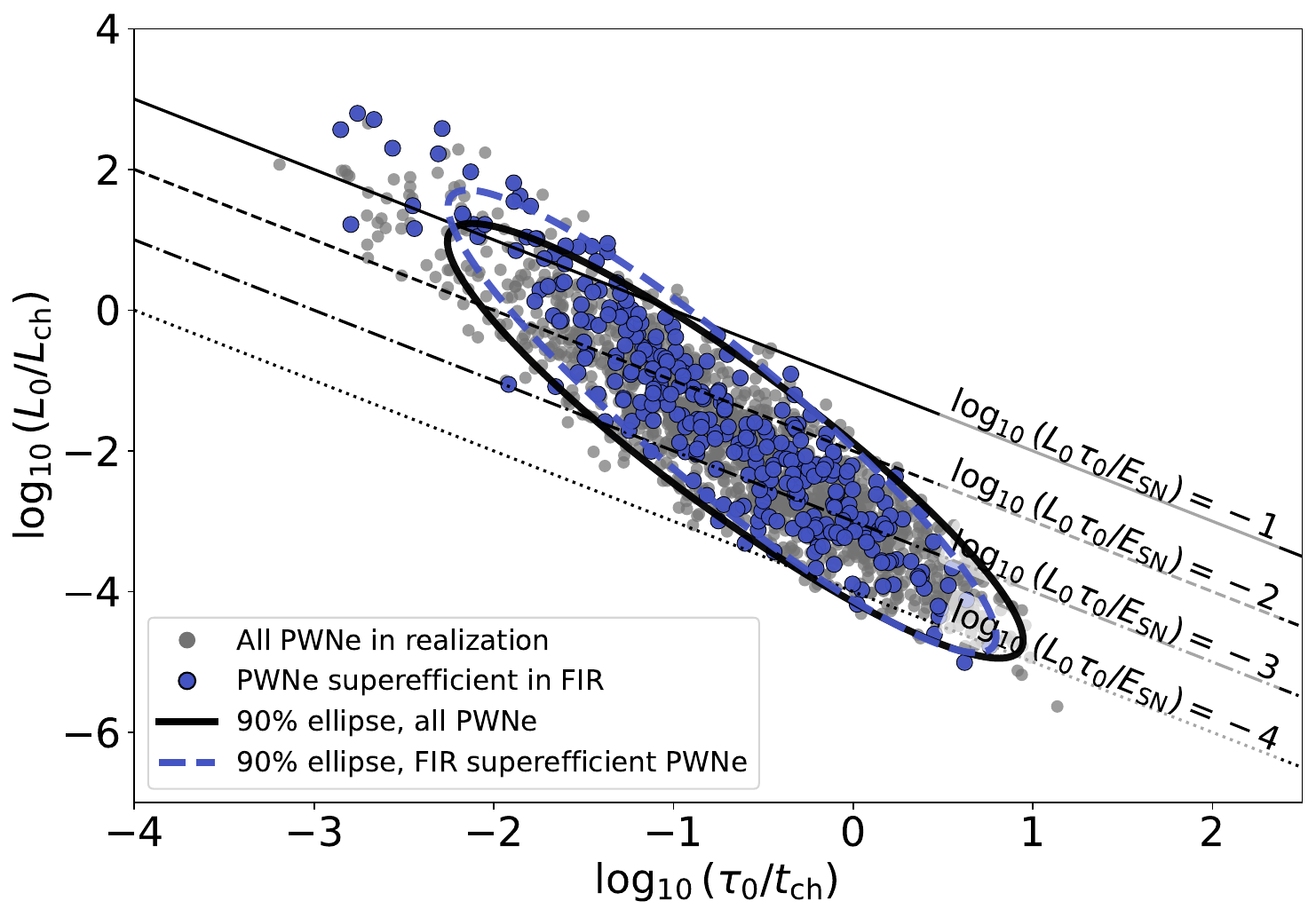}
    \includegraphics[width=0.33\linewidth]{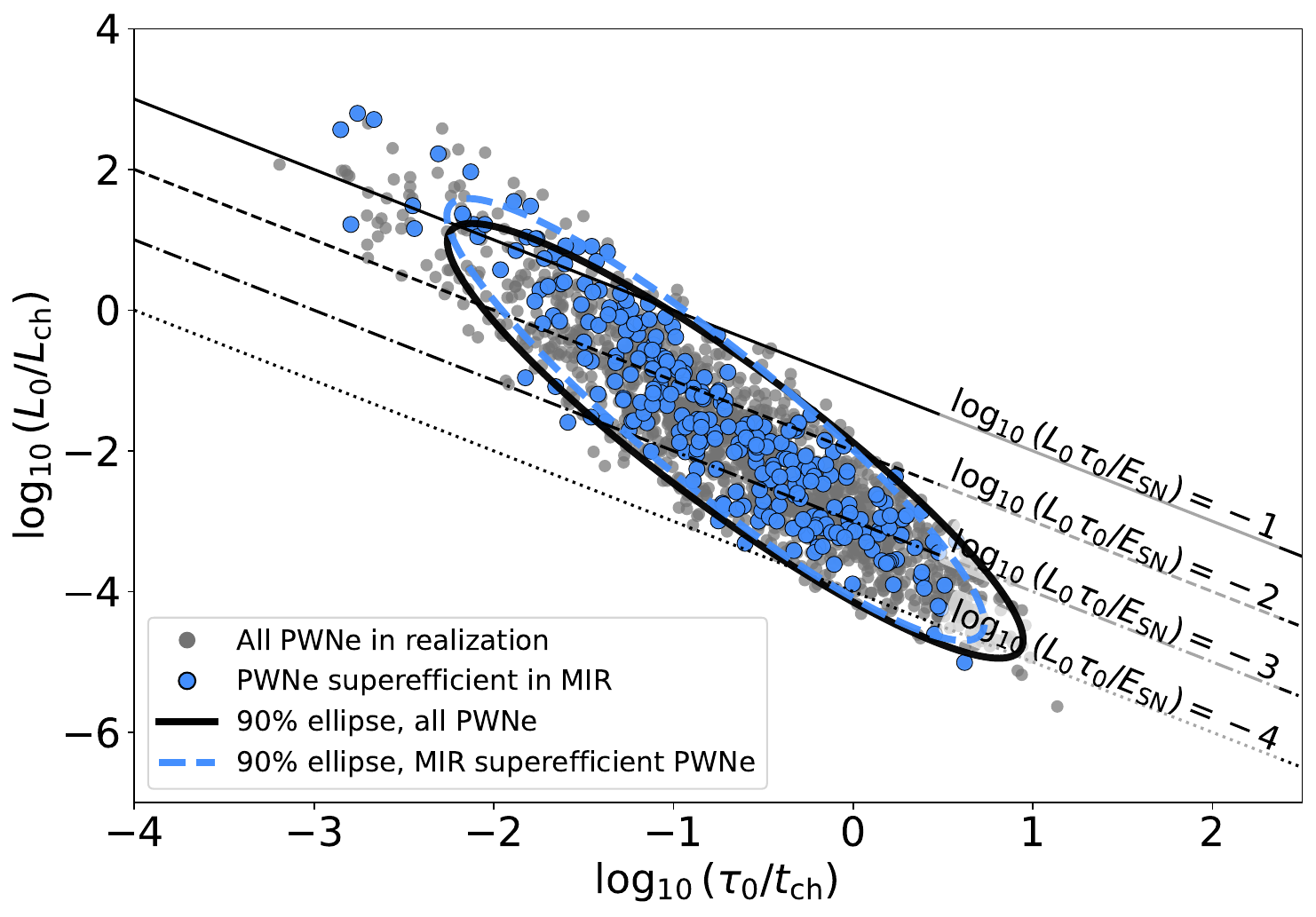}
    \includegraphics[width=0.33\linewidth]{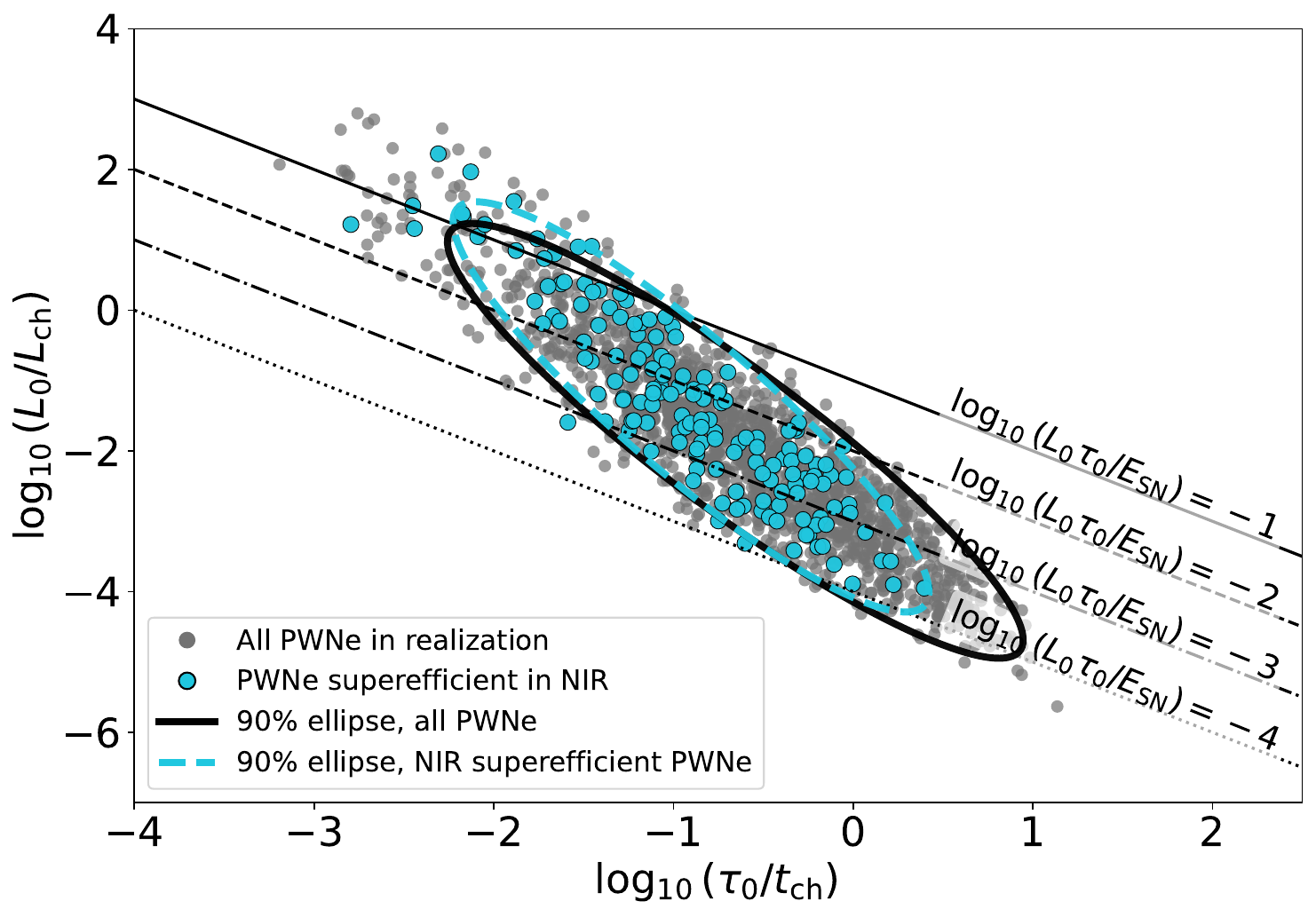}
    \includegraphics[width=0.33\linewidth]{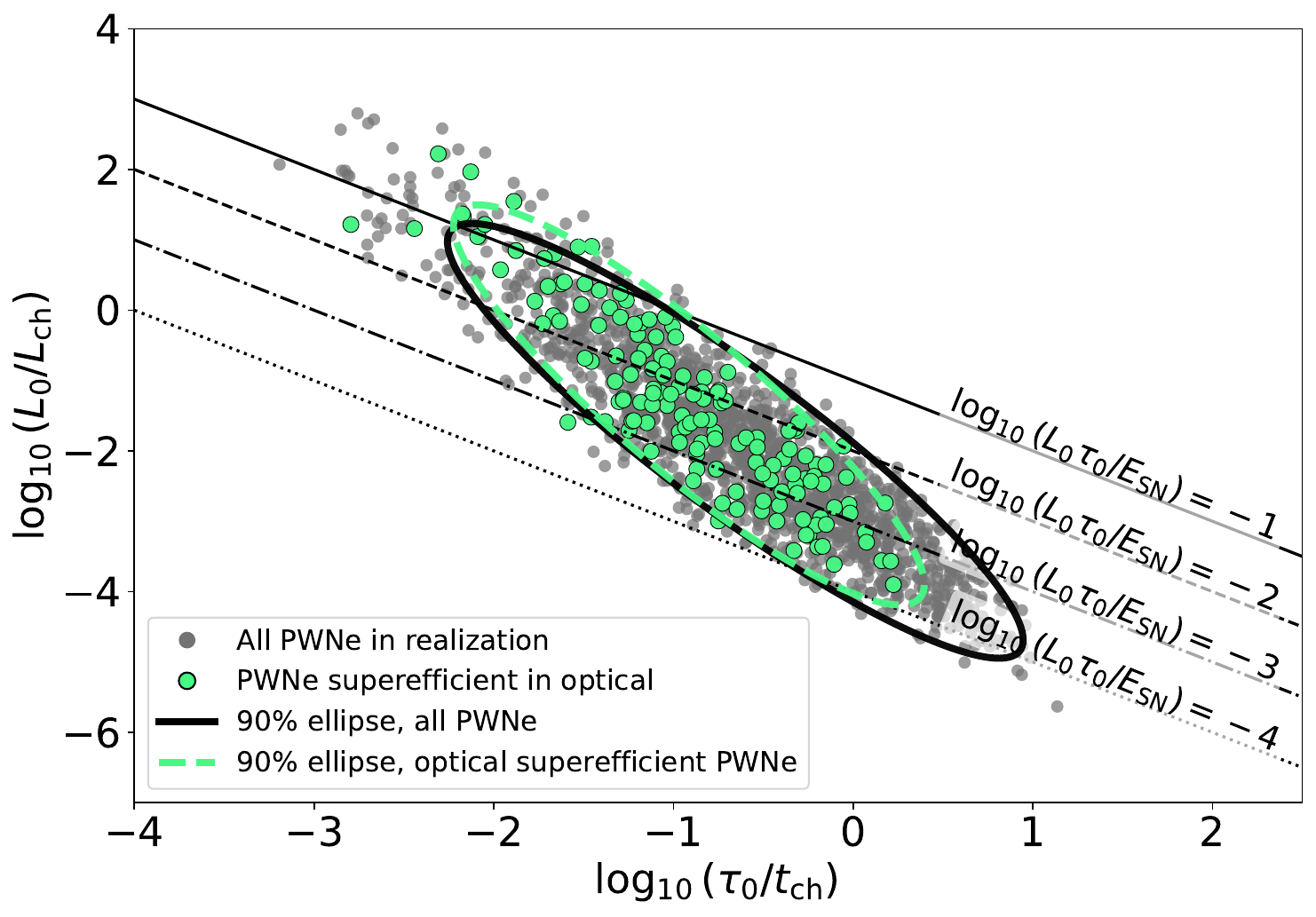}
    \includegraphics[width=0.33\linewidth]{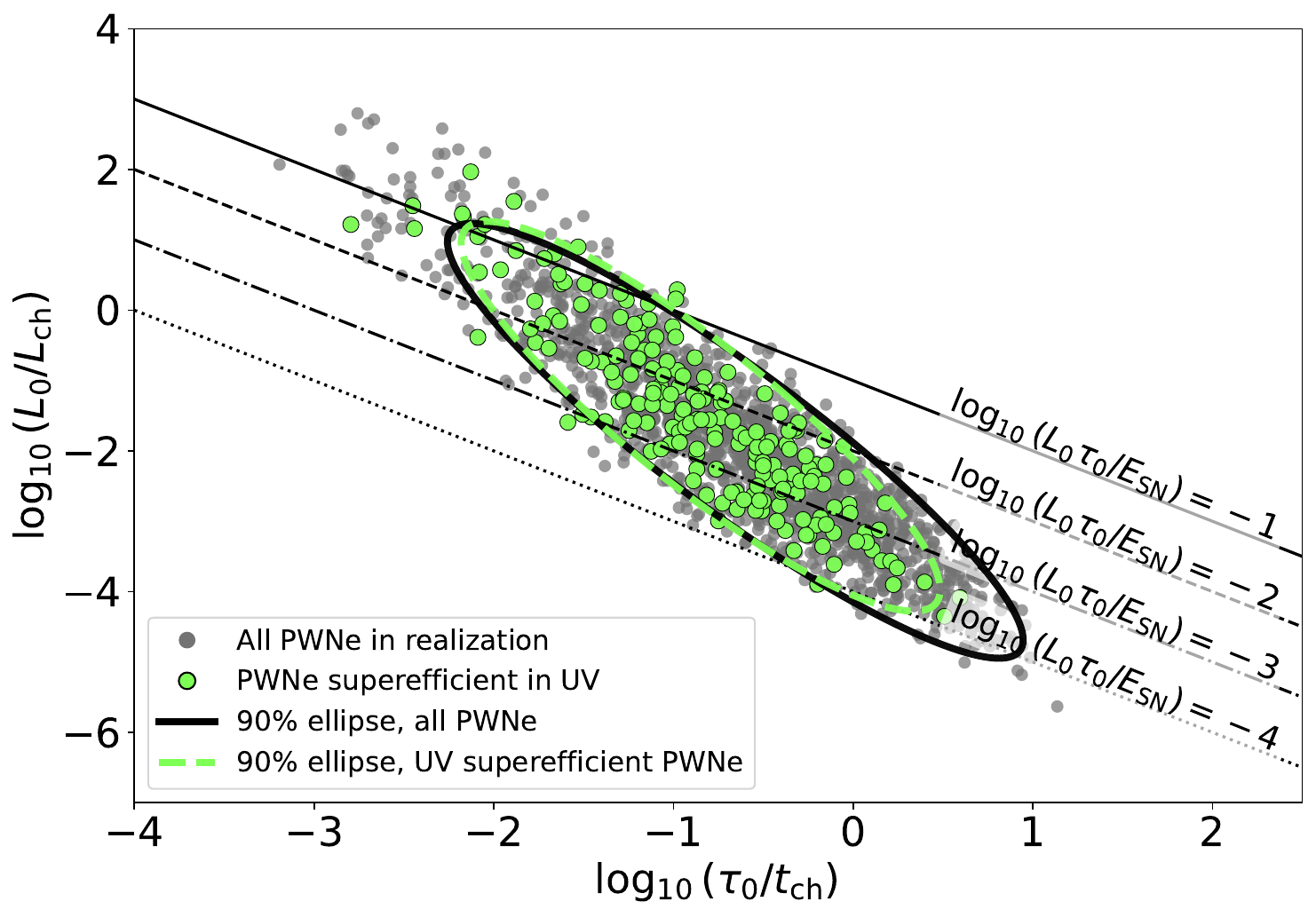}
    \includegraphics[width=0.33\linewidth]{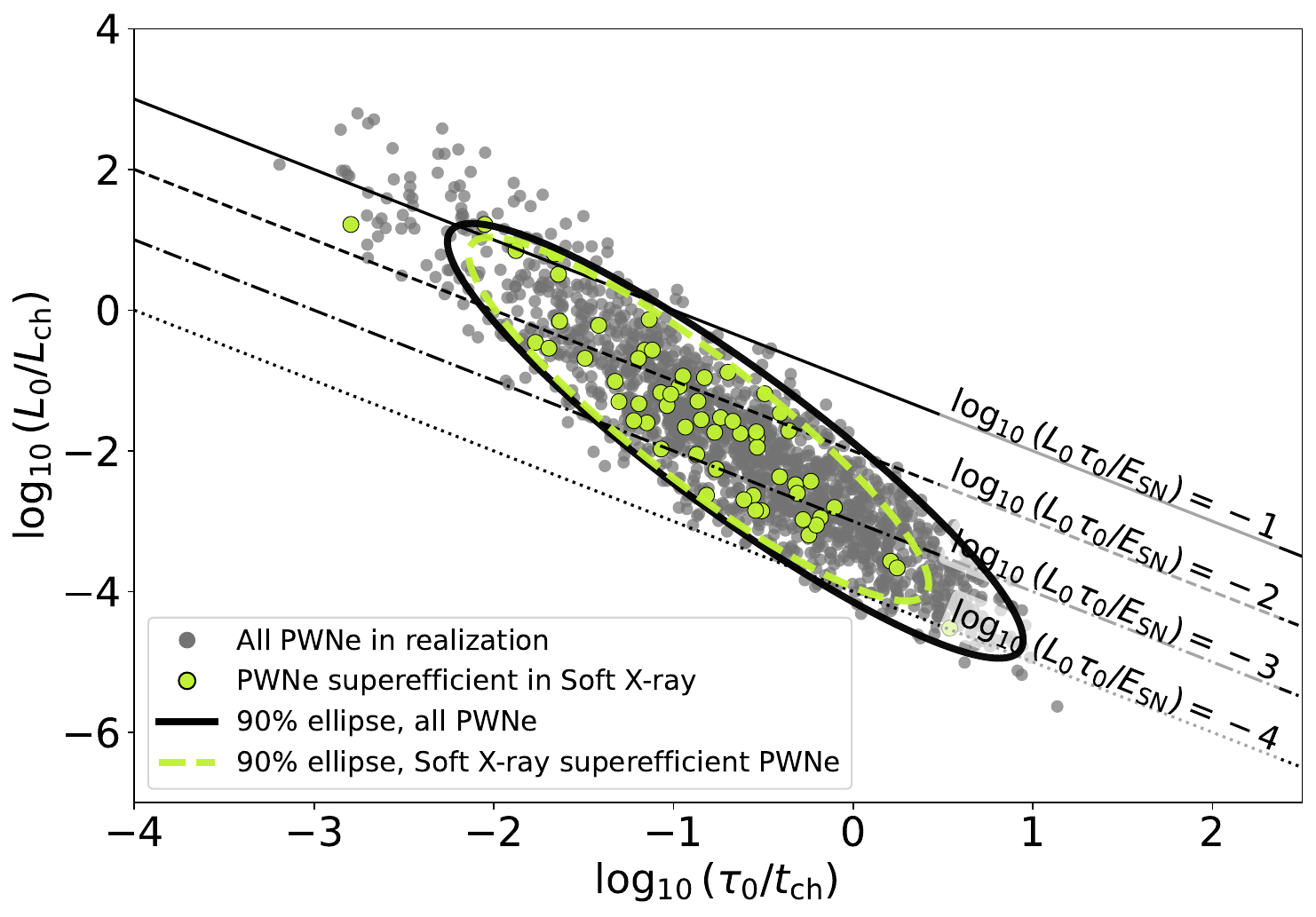}
    \includegraphics[width=0.33\linewidth]{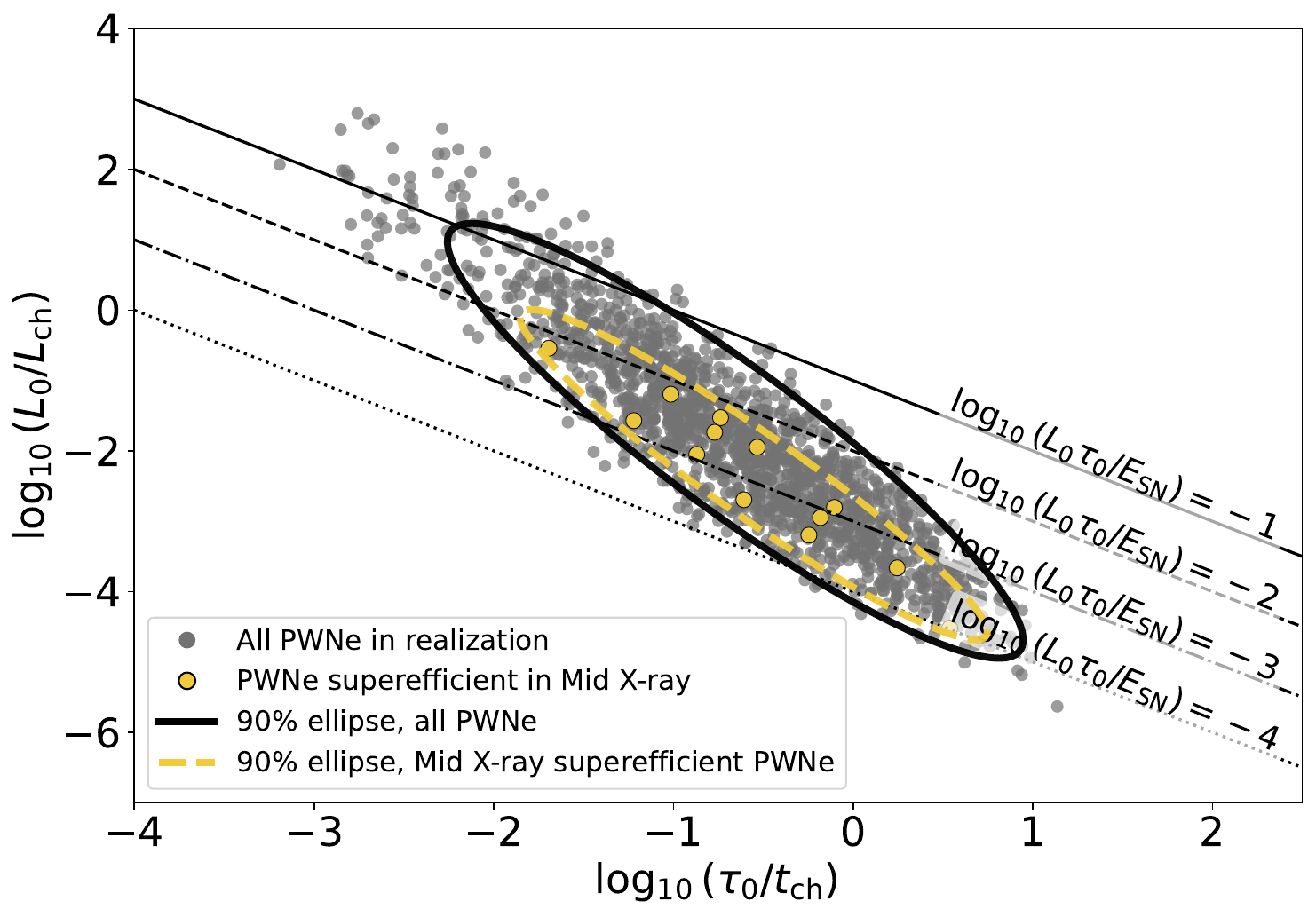}
    \includegraphics[width=0.33\linewidth]{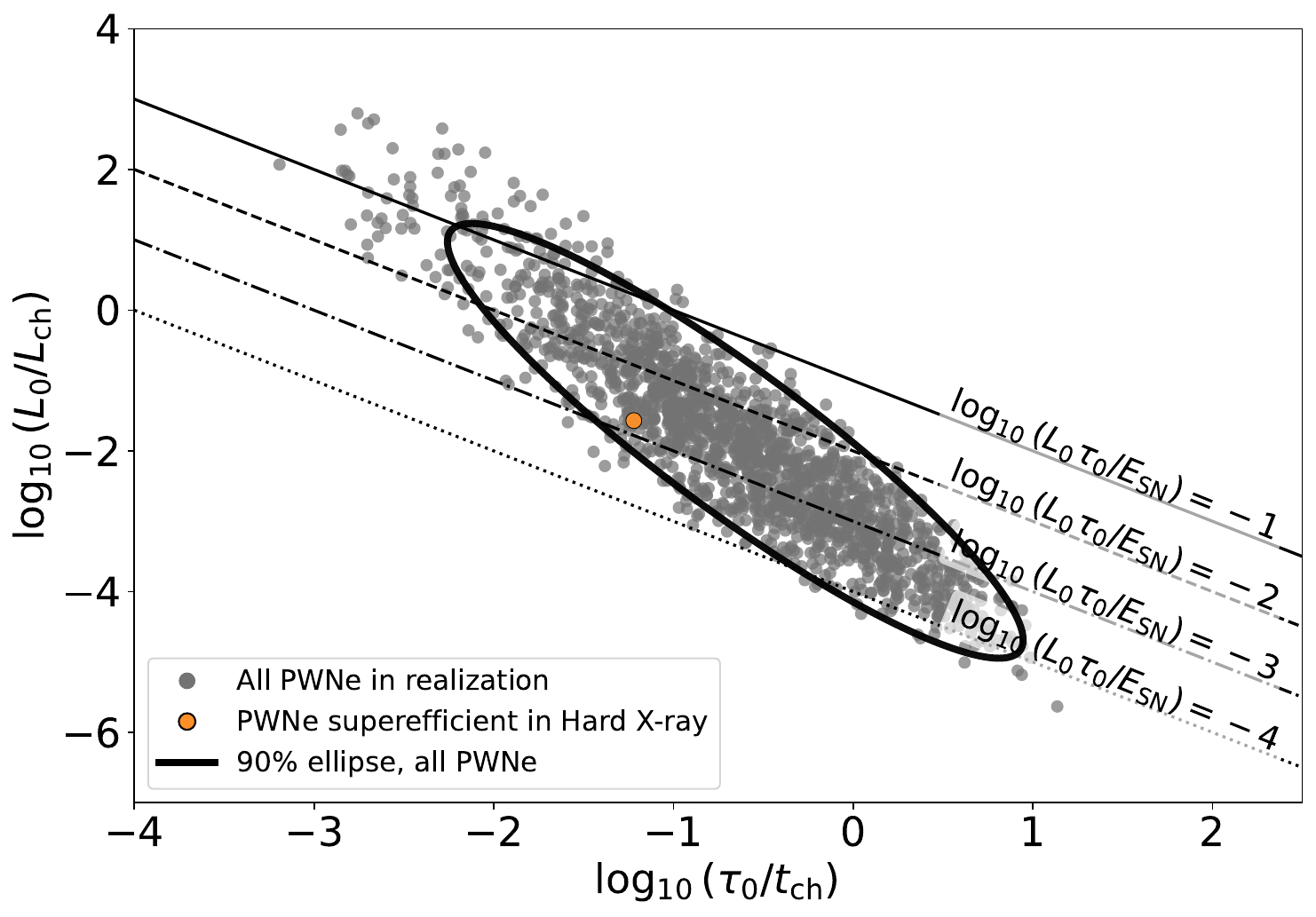}
    \includegraphics[width=0.33\linewidth]{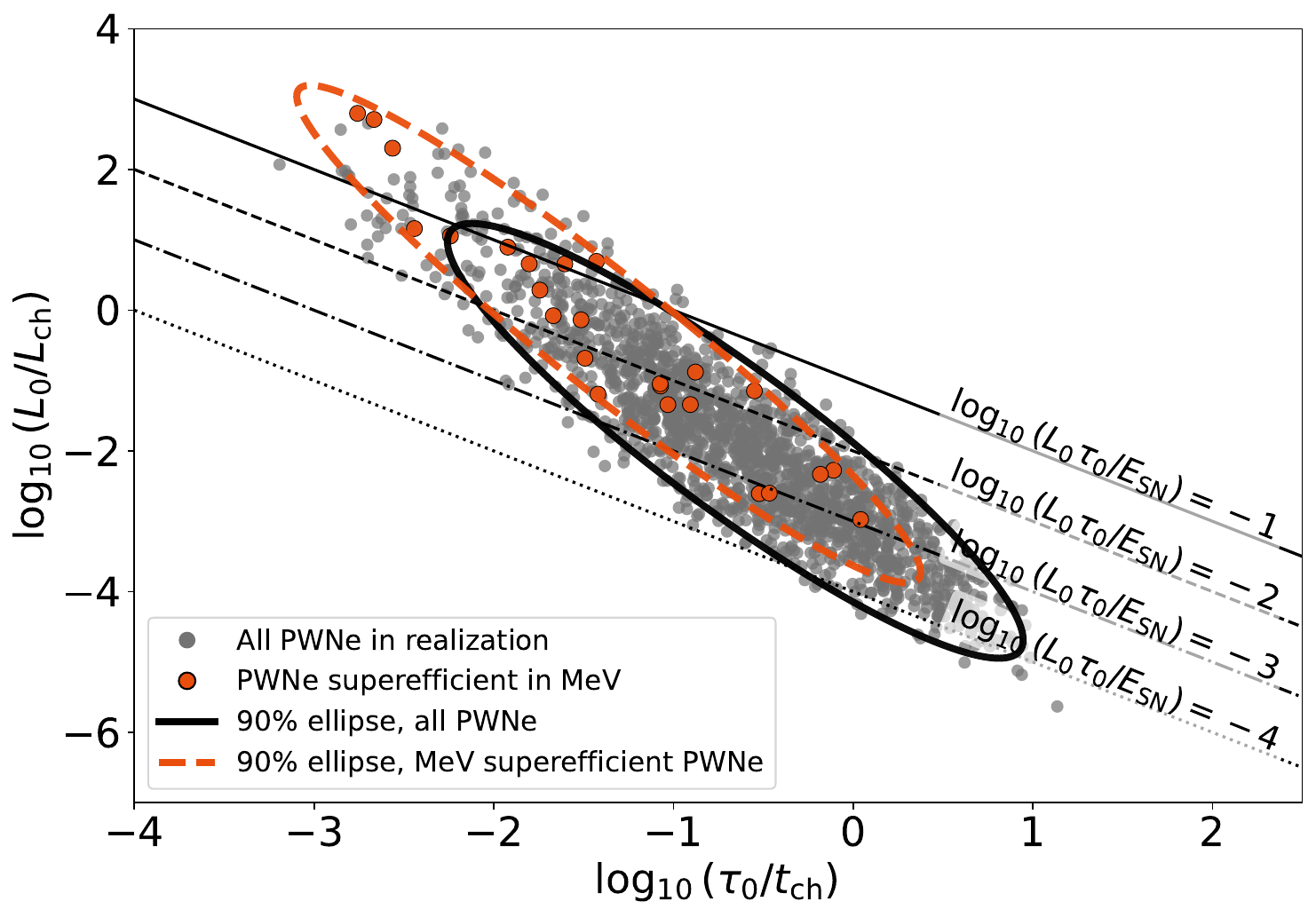}
    \includegraphics[width=0.33\linewidth]{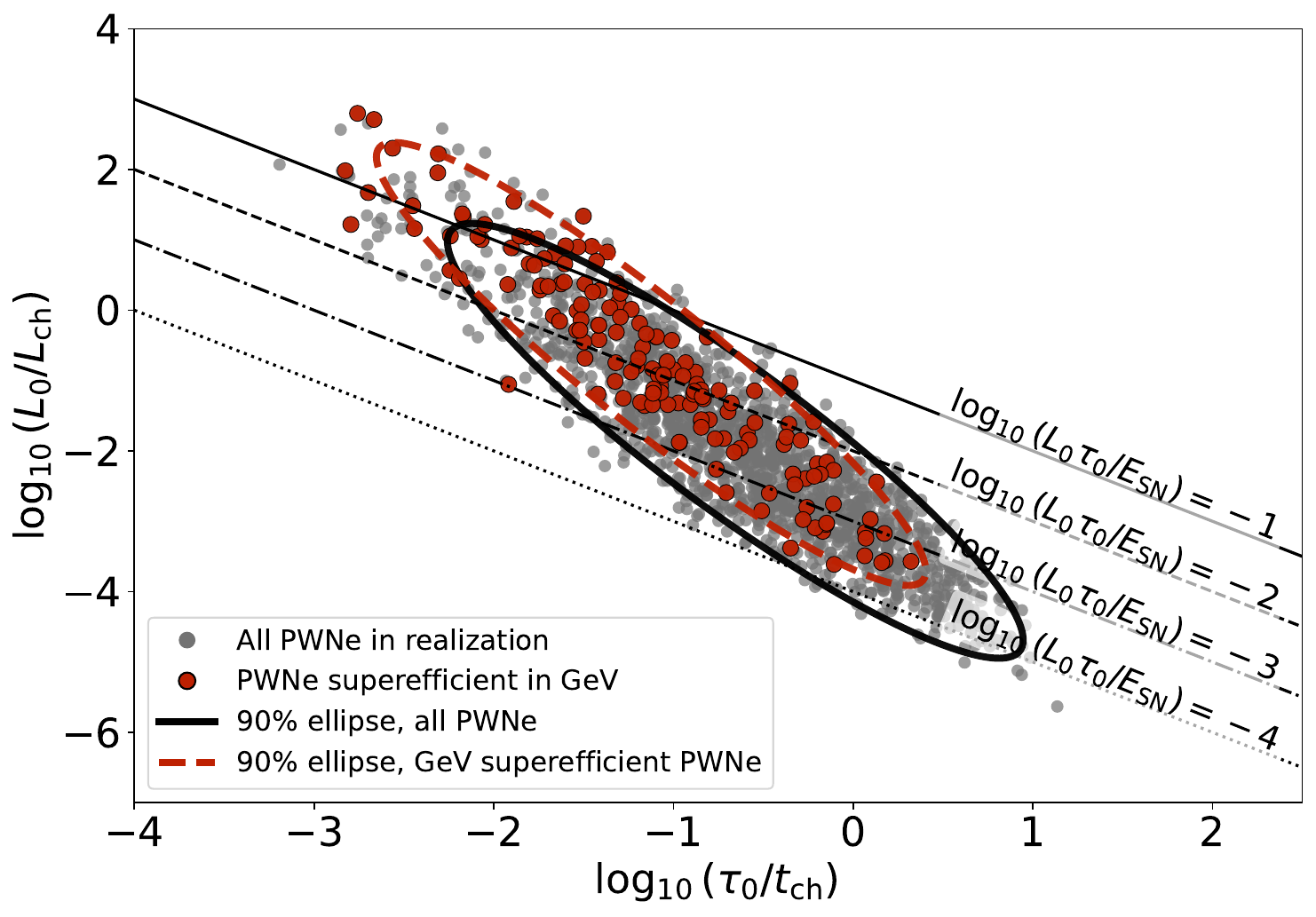}
    \includegraphics[width=0.33\linewidth]{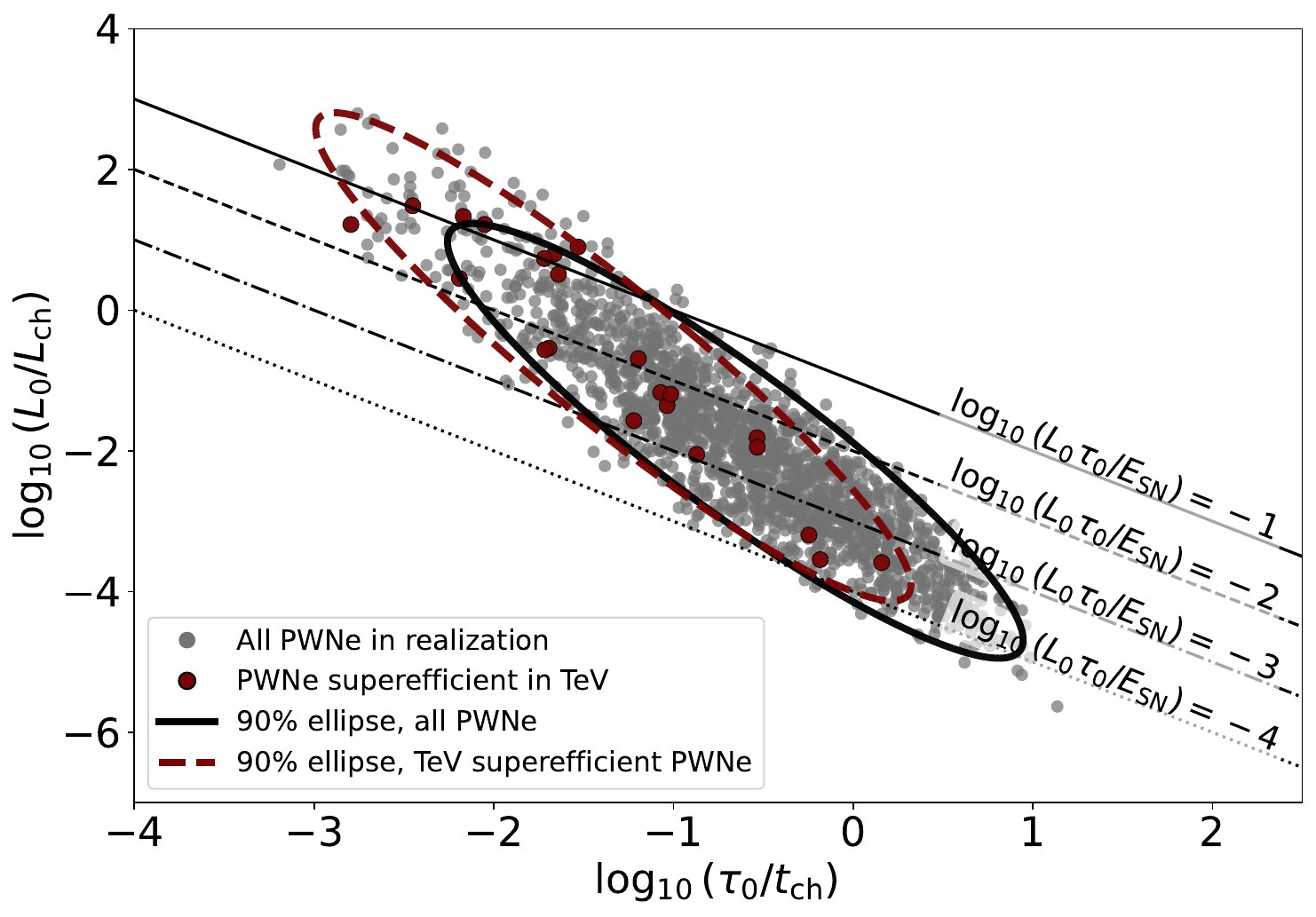}
    \caption{Distribution of sources in the $\log_{10}(L_0/L_{\rm ch})$-$\log_{10}(\tau_0/t_{\rm ch})$ plane for a random realization of 1600 PWNe at the current age. Grey points show all sources in the realization, while the colored points mark the subset that is superefficient in the corresponding frequency. The diagonal lines indicate constant values of $\log_{10}(L_0\tau_0/E_{\rm SN})$. 90$\%$ enclosing ellipses are also shown in the figure; in black solid line for all the sources in the realization and dashed line for the superefficient sources, color coded as in Fig. \ref{fig: Lband/Edot}. Note that no ellipse is shown if there are less than three superefficient sources for a frequency. 
}
\label{fig: L0_tau0}
\end{figure*}

\begin{figure*}
    \includegraphics[width=0.5\linewidth]{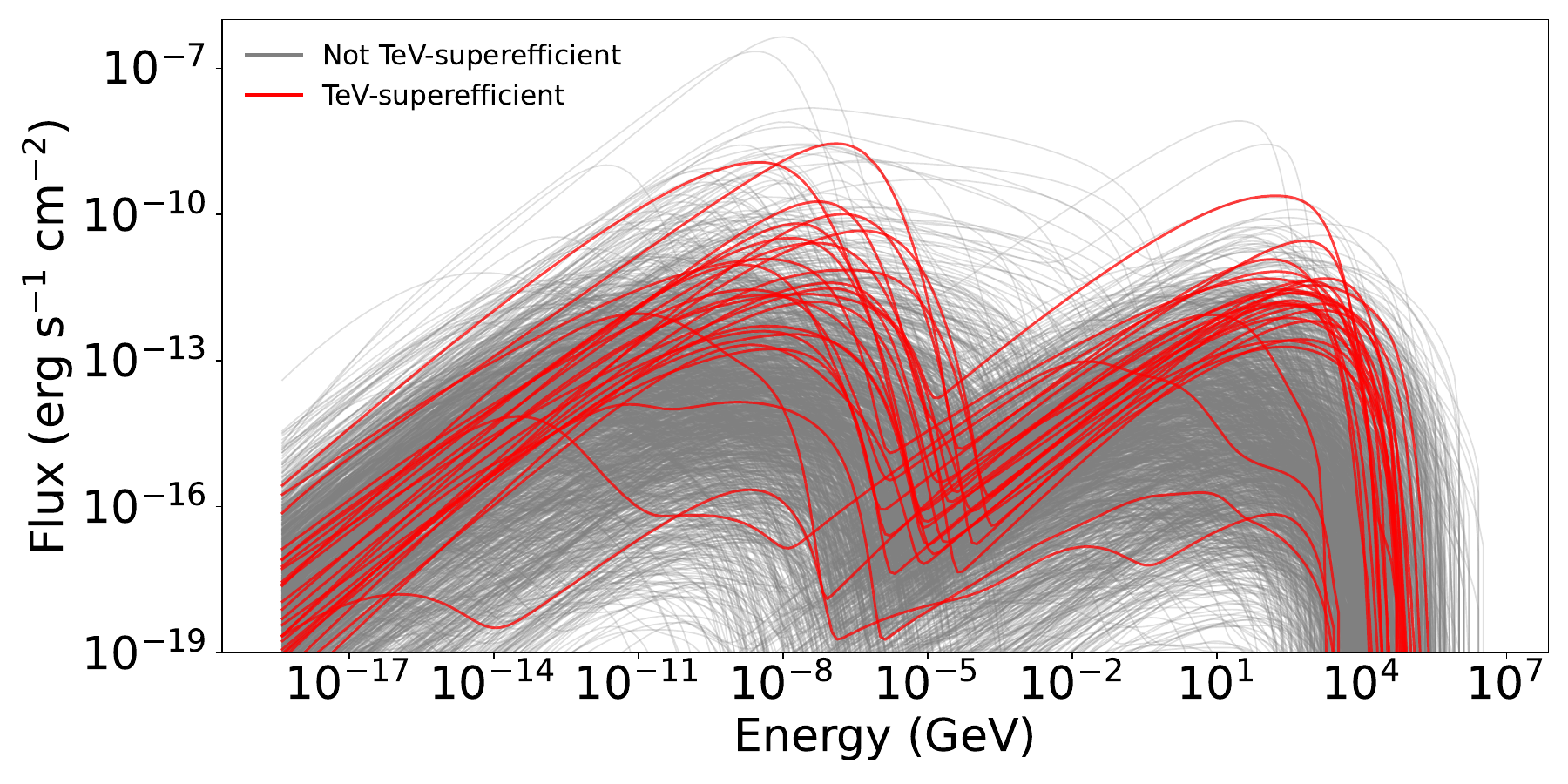}
    \includegraphics[width=0.5\linewidth]{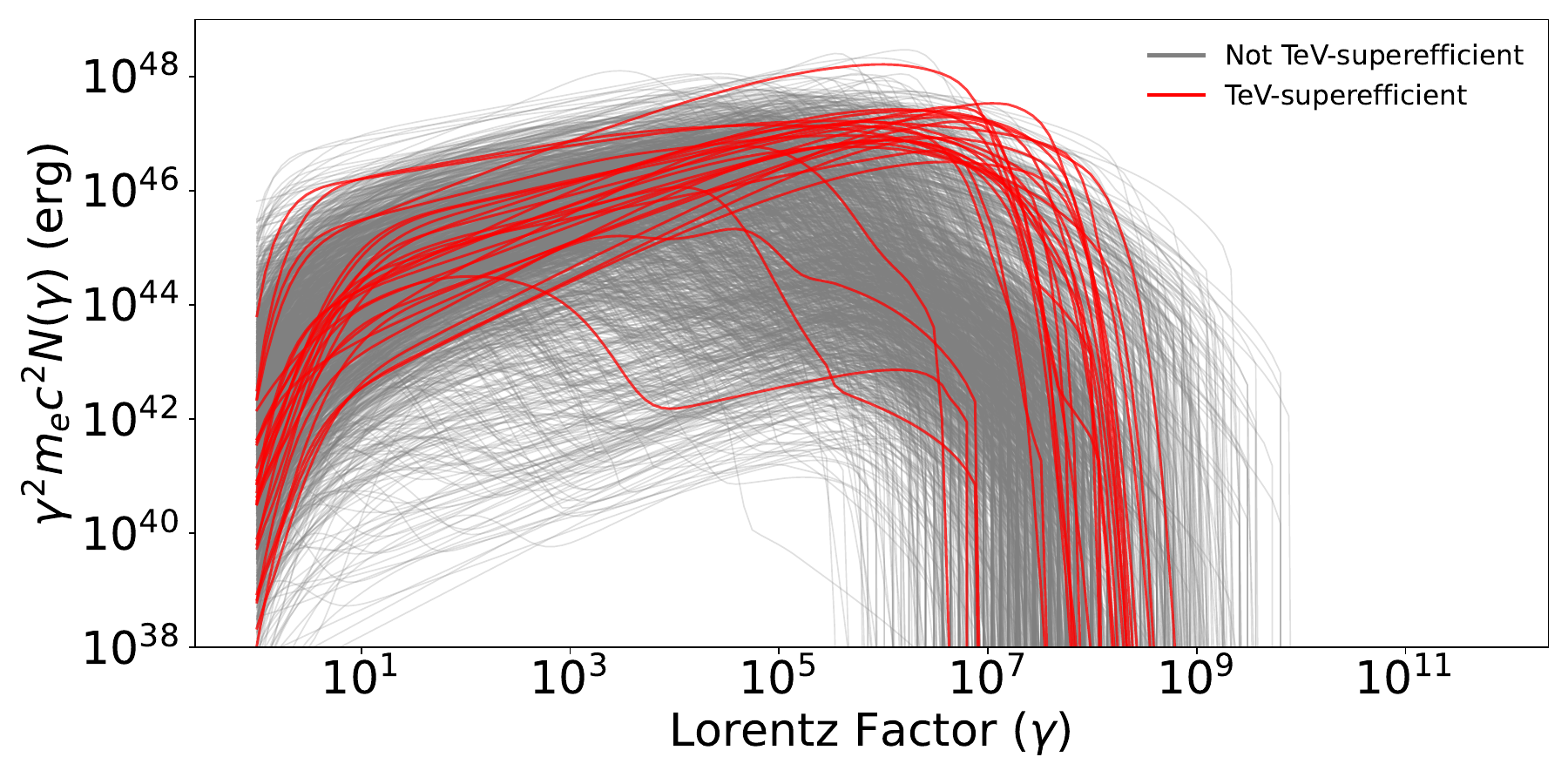}
    \caption{The left panel shows the SEDs, and the right panel shows the corresponding electron spectra of TeV-superefficient and non-TeV-superefficient PWNe in a random realization of 1600 sources at the current age. Grey curves show sources that are not TeV-superefficient, while red curves mark sources satisfying $L_{\rm TeV}/\dot{E}_{\rm SD}>1$ in the 1-10 TeV range.}
\label{fig: tev_superefficient}
\end{figure*}

The spatial distribution of PWNe follows that of Galactic core-collapse SNRs and pulsars \citep{fgk06, cristofari17, fiori22}, ensuring a realistic spread in Galactic longitude and latitude. 
Ages were sampled from the SN birth rate, excluding systems younger than 100 yr. 
The SN explosion energy was drawn from a truncated normal distribution bounded between $10^{50}$ and $10^{52}$ erg, and with a mean of 
$\langle \log_{10}(E_{\rm SNR}/{\rm erg}) \rangle = 51$ 
and a standard deviation of $0.54$ \citep{batzofin24,batzofin25, martinez22}, while the ambient density $n_{\rm ISM}$ was uniformly distributed between $0.01$ and $10$ cm$^{-3}$. 

Progenitor zero-age main-sequence (ZAMS) masses were sampled from a Salpeter Initial Mass Function (IMF) ($p(M)\propto M^{-2.35}$) between 8-50 M$_\odot$ using inverse-transform sampling \citep{devroye86}. 
Ejecta masses $M_{\rm ej}$ were subsequently interpolated from the GENEVA stellar evolution models \citep{ekstrom12, batzofin24}, ensuring physical consistency between progenitor and remnant properties. 
This approach replaces the simplified Gaussian ejecta mass distributions used in earlier works \citep[e.g.][]{ctagps24, fiori22}, and is more realistic given that ZAMS masses are better constrained.

Each PWN is powered by a pulsar whose initial spin period and magnetic field are drawn from observationally motivated distributions \citep{bandiera23II}: 
$P_{0}$ follows a Gaussian distribution with mean 100 ms and $\sigma=80$ ms (truncated at 10 ms), while $B_{0}$ follows a lognormal distribution with $\langle \log_{10}(B_{0}/{\rm G}) \rangle = 12.3$ and $\sigma = 0.25$. 
A fixed braking index $n=2.33$, representative of the Crab pulsar \citep{lyne15, horvath19}, is adopted. 
These parameters set the initial spin-down luminosity and characteristic timescale through standard magnetic dipole spin-down relations \citep{gaensler06}.

Other nebular parameters follow distributions from previous studies \citep{fiori22, desarkar26}: 
the particle break energy (log-normal distribution, with $\log_{10}(\gamma_{\rm b})$ having mean 5.74 and standard deviation 0.37), low- and high-energy spectral indices (uniform distribution in $1.0<\alpha_{1}<1.7$ and $2.0<\alpha_{2}<2.7$, respectively), magnetic fraction (log-normal distribution with $\log_{10}(\eta)=-1.80$ and standard deviation $0.35$), and containment factor (uniform distribution in $0.01<R_{\rm L}/R_{\rm TS}<0.9$). 
Local interstellar radiation fields (ISRFs) are interpolated from the tabulated near- and far-infrared (NIR and FIR, respectively) energy densities \citep{porter06}, multiplied by random factors between 2 and 10 to match the higher photon densities inferred from PWN spectral modeling \citep{torres14}.
The temperatures of the FIR and NIR fields were fixed at $T_{\rm NIR} = 3000$ K and $T_{\rm FIR} = 70$ K, respectively.

The dynamical and radiative evolution of each system is computed using the hybrid \texttt{TIDE+L} framework \citep{bandiera20, bandiera21, bandiera23II, bandiera23III}. 
This combines the one-zone \texttt{TIDE} code \citep{martin12, torres14, martin16, martin22} with Lagrangian modules that accurately model reverberation and feedback during PWN-SNR interaction. 
Unlike purely thin-shell models, which use analytical prescriptions for surrounding SNR structure, \texttt{TIDE+L} self-consistently evolves the hydrodynamic (HD) structure of the surrounding SNR, and captures shell thickening, delayed compression, and multiple shock reflections, providing a more realistic treatment of the dynamical and radiative coupling as verified with 1D HD simulations.
The particle population evolves under synchrotron, inverse-Compton (IC), bremsstrahlung, synchrotron self-Compton (SSC), and adiabatic losses, with injected power partitioned between magnetic and particle energy according to the magnetic fraction. 

Following the approach adopted in \cite{desarkar26}, we classify the evolutionary state of each PWN-SNR system into three distinct stages based on the evolution of the PWN radius relative to the SN reverse shock. 
%
%
We distinguish three stages. 
First, in the free-expansion phase, the PWN expands freely within the SNR and has not yet been overtaken by the reverse shock. 
Second, in the reverberation-compressing phase, the reverse shock has reached the nebula and the PWN radius is decreasing, but has not yet reached its first minimum. 
Third, the reverberation-post-compression phase begins after this first minimum.
%
%

\section{Results and discussion}\label{results}

\subsection{Estimation of superefficient PWNe at the current epoch}

To robustly quantify the relative contributions of each evolutionary stage and to characterize the distribution of superefficient sources across frequency bands, we apply a sampling procedure to the synthetic population, following the methodology of \cite{desarkar26}. 
From the full set of 2400 simulated PWNe, we draw 1000 independent realizations of fixed-size subsamples of 1600 sources, according to the assumed Galactic SN rate, sampling without replacement in each realization. 
For every realization, we record the number of systems in each evolutionary phase as well as the subset of sources satisfying the superefficiency criterion. 
The reported counts correspond to mean values over all realizations, with the associated $1\sigma$ scatter reflecting purely stochastic sampling variance.\footnote{
Formally, for each realization $k$, we denote by $N_k$ the number of systems in a given evolutionary phase or satisfying the superefficiency criterion. The mean count is then given by
\begin{equation}
\langle N \rangle = \frac{1}{N_{\mathrm{real}}} \sum_{k=1}^{N_{\mathrm{real}}} N_k ,
\end{equation}
where $N_{\mathrm{real}}$ is the total number of realizations. The corresponding $1\sigma$ uncertainty is quantified using the sample standard deviation,
\begin{equation}
\sigma_N = \sqrt{\frac{1}{N_{\mathrm{real}}-1} \sum_{k=1}^{N_{\mathrm{real}}} \left( N_k - \langle N \rangle \right)^2 } ,
\end{equation}
which captures the realization-to-realization scatter arising solely from stochastic sampling.
}

Superefficiency is evaluated across 12 distinct frequency bands spanning the electromagnetic spectrum, selected to probe the radiative output of PWNe from radio to very-high-energy $\gamma$-rays.
These bands are radio (1.4 GHz), far-infrared (FIR; $3\times10^{11}$-$1.2\times10^{13}$ Hz), mid-infrared (MIR; $1.2\times10^{13}$-$2\times10^{14}$ Hz), near-infrared (NIR; $1.0\times10^{14}$-$3.5\times10^{14}$ Hz), optical ($3.5\times10^{14}$-$7.5\times10^{14}$ Hz), ultraviolet (UV; $7.5\times10^{14}$-$3\times10^{16}$ Hz), soft X-rays (0.1-1 keV), mid X-rays (1-10 keV), hard X-rays (10-100 keV),  MeV (0.3-30 MeV), GeV $\gamma$-rays (1-10 GeV), and TeV $\gamma$-rays (1-10 TeV). 
Together, these intervals sample synchrotron emission from radio through X-rays and inverse-Compton emission at $\gamma$-ray energies, enabling a unified assessment of superefficiency across the full non-thermal spectral energy distribution (SED).

Because superefficiency is physically induced by the reverberation phase, as discussed earlier, superefficient sources are further subdivided according to their dynamical state. 
In particular, we distinguish between PWNe that are still undergoing active compression and those that have already passed their first minimum radius and entered the post-compression phase. 
The resulting distributions of superefficient sources as a function of frequency band and compression state are summarized in the bar plots shown in Fig. \ref{fig: pie_super}, along with pie charts showing the number of PWNe in each evolutionary stages at the current age.
Table \ref{tab:sup_counts} summarizes our estimates of the number of superefficient PWNe across the full set of frequency bands for two different modeling assumptions. 
We first focus on the reference case of the full Galactic population modeled with \texttt{TIDE+L}, and then discuss how the results change when reverberation is treated using the usual purely thin-shell approach via the \texttt{TIDE} model.
The sampling procedure discussed above was considered in all cases.
The corresponding pie charts and superefficiency distribution bar plots are shown in Fig. \ref{fig: pie_super} for the same population assumptions.

We find that the largest number of superefficient PWNe at the present epoch occurs in the FIR band. 
This agrees with previous suggestions by \cite{bandiera23III}, reflecting the abundance of long-lived, low-energy electrons in the predominantly middle-aged ($\gtrsim10^{4}$ yr) Galactic PWN population. 
Unlike the high-energy particles responsible for X-ray emission, the synchrotron cooling timescales of FIR-emitting electrons are comparatively long. 
These particles can therefore accumulate over extended periods, making the FIR synchrotron luminosity particularly sensitive to even moderate changes in the nebular magnetic field. 
In addition, during reverberation, radiative and adiabatic losses can shift particles initially injected at higher energies toward the lower-energy range relevant for FIR emission, further increasing the number of particles contributing to this band. 
This combination of long particle lifetimes, accumulated relic populations, and magnetic field enhancement makes FIR superefficiency the most common outcome.

In the X-ray band, superefficiency is more closely connected to the reverberation history of the nebula. 
X-ray emission is produced by higher-energy electrons, whose synchrotron cooling times are short, especially when the magnetic field is amplified during compression. 
During active compression, the increase in magnetic field strength can temporarily enhance the synchrotron luminosity, allowing some systems to exceed their contemporaneous spin-down power. 
However, if the compression is too strong, the same amplified magnetic field also produces severe synchrotron losses, rapidly depleting the X-ray-emitting particles. 
As a result, X-ray superefficiency does not simply trace stronger compression. 
Rather, it depends on the competition between compression-driven luminosity enhancement and the survival of high-energy particles. 
For systems in which synchrotron losses are less catastrophic, a relic population of energetic particles can survive the main compression episode, allowing X-ray superefficiency to persist into the post-compression phase.
As a result, we find almost comparable numbers of X-ray-superefficient sources in the compressing and post-compression phases.

At GeV $\gamma$-ray energies, the situation is different because the emission is dominated by IC upscattering rather than synchrotron radiation. 
The GeV luminosity is therefore less directly tied to the instantaneous magnetic field strength and depends strongly on the accumulated electron population. 
When synchrotron losses are moderate, particles injected over the past evolution of the pulsar can survive to late times and continue producing IC emission even after the present-day spin-down power has substantially declined. 
Since superefficiency is defined relative to the contemporaneous $\dot{E}_{\rm SD}$, a persistent relic IC luminosity combined with a low current $\dot{E}_{\rm SD}$ can naturally lead to a large number of GeV-superefficient systems. 
Thus, GeV superefficiency is not necessarily a signature of an instantaneous compression-driven brightening, but can instead reflect the long-term survival of particles accumulated throughout the PWN evolution, particularly in post-compression systems.

\begin{figure*}
\centering
\includegraphics[width=0.3\linewidth]{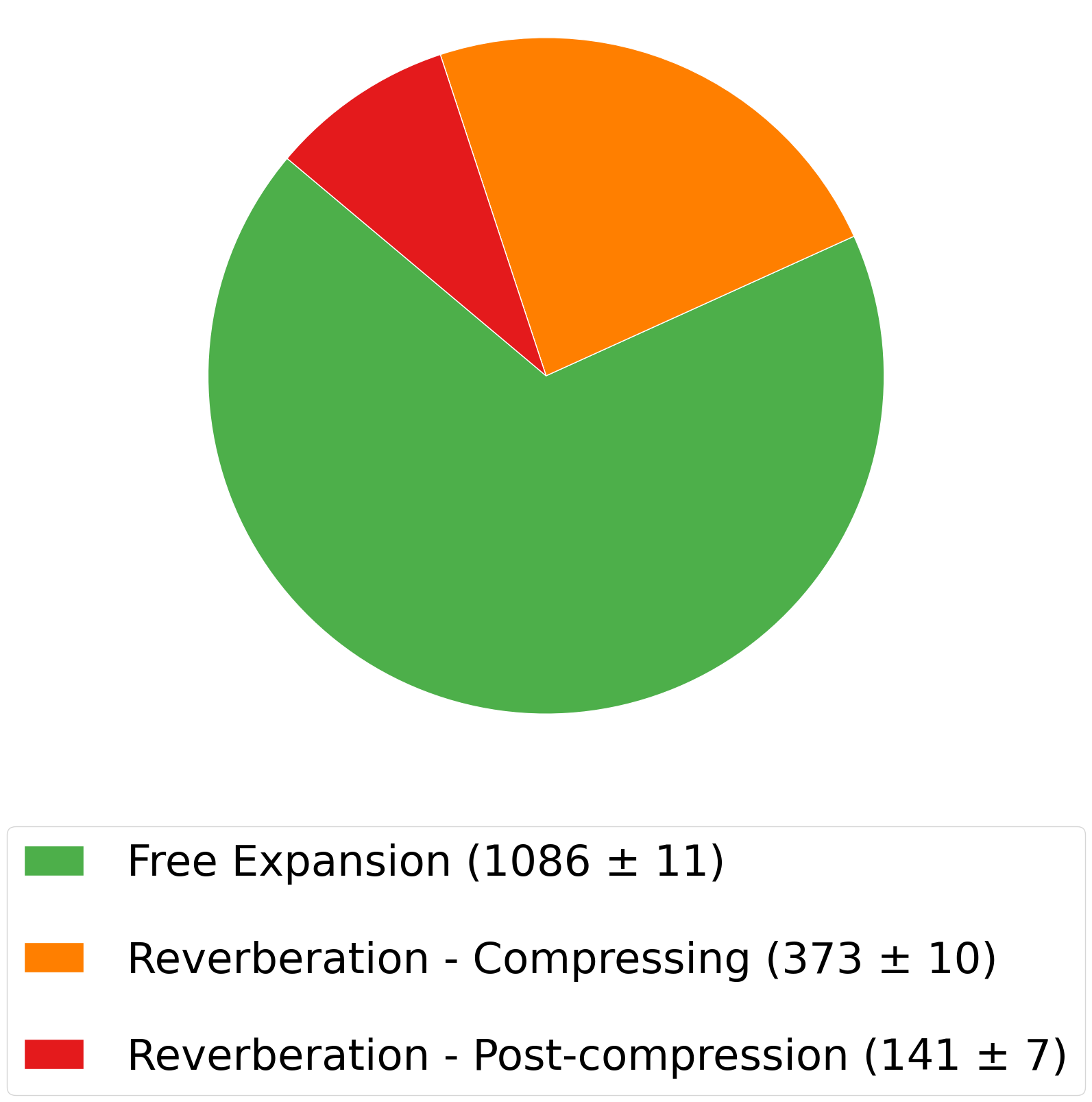}
\includegraphics[width=0.6\linewidth]{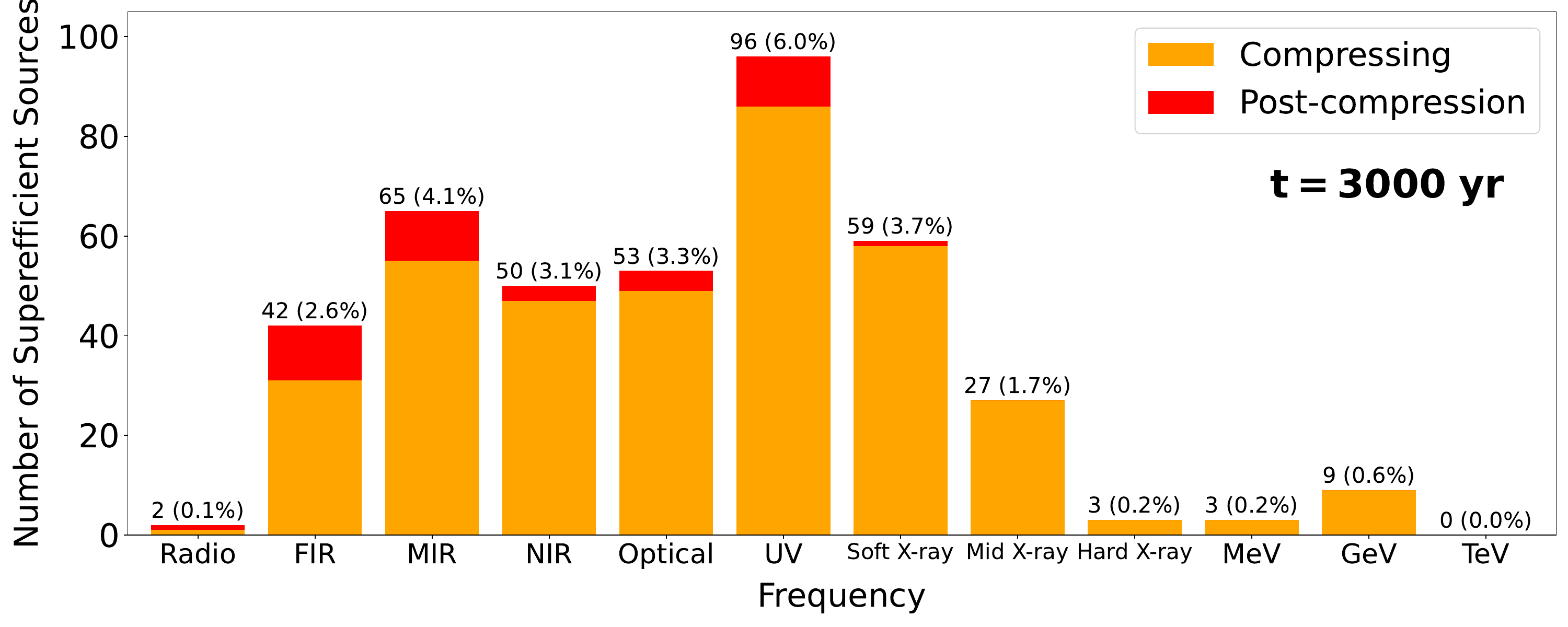}\\
\includegraphics[width=0.3\linewidth]{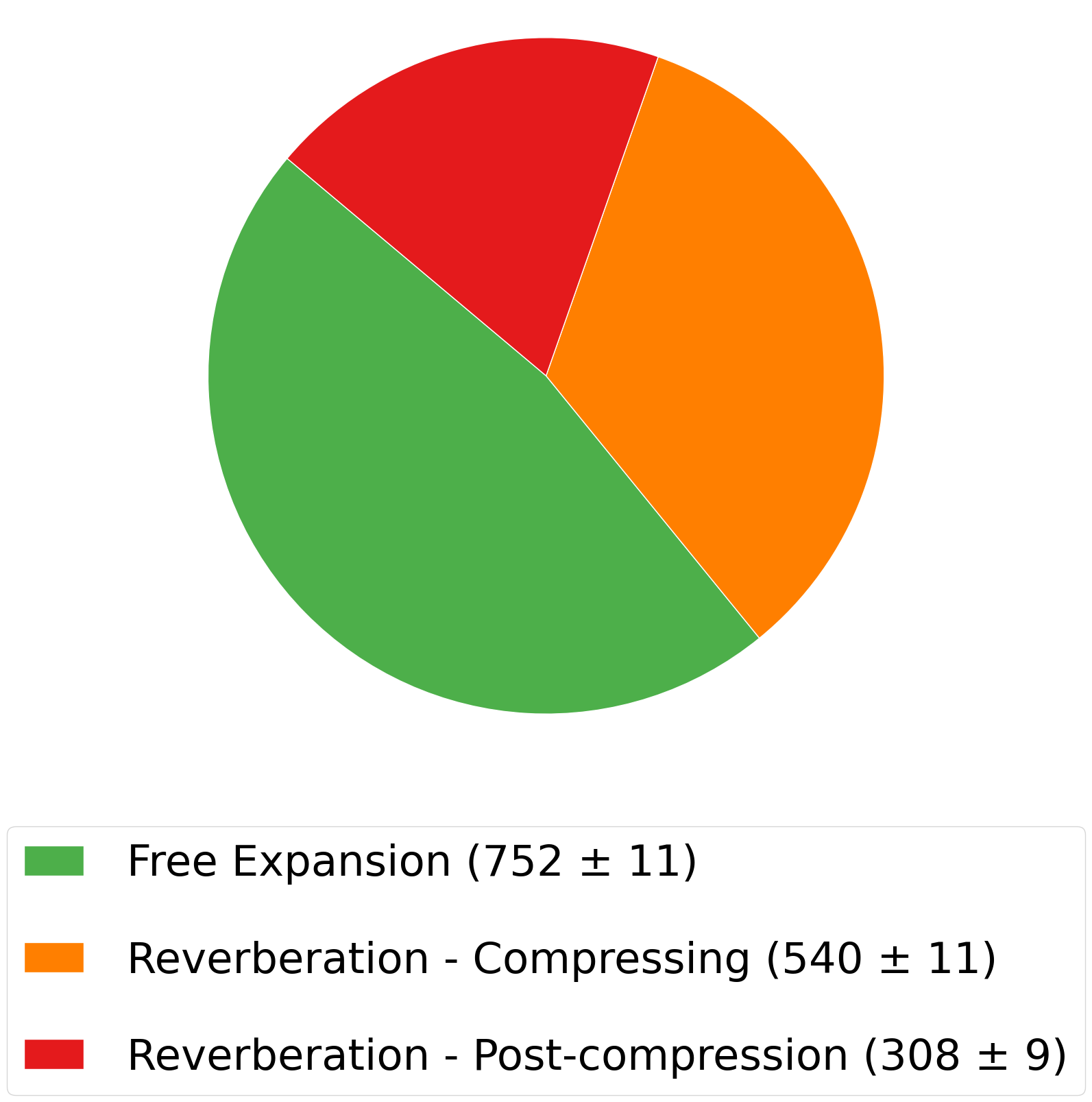}
\includegraphics[width=0.6\linewidth]{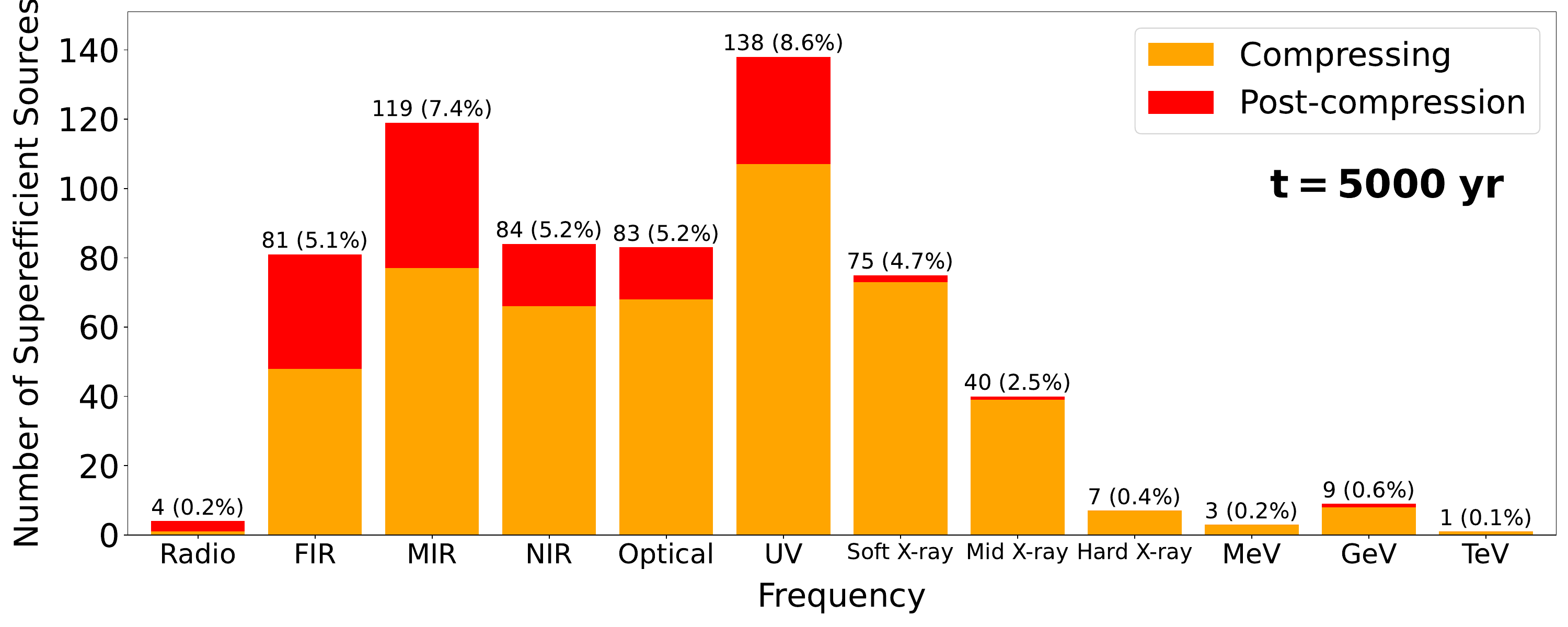}\\
\includegraphics[width=0.3\linewidth]{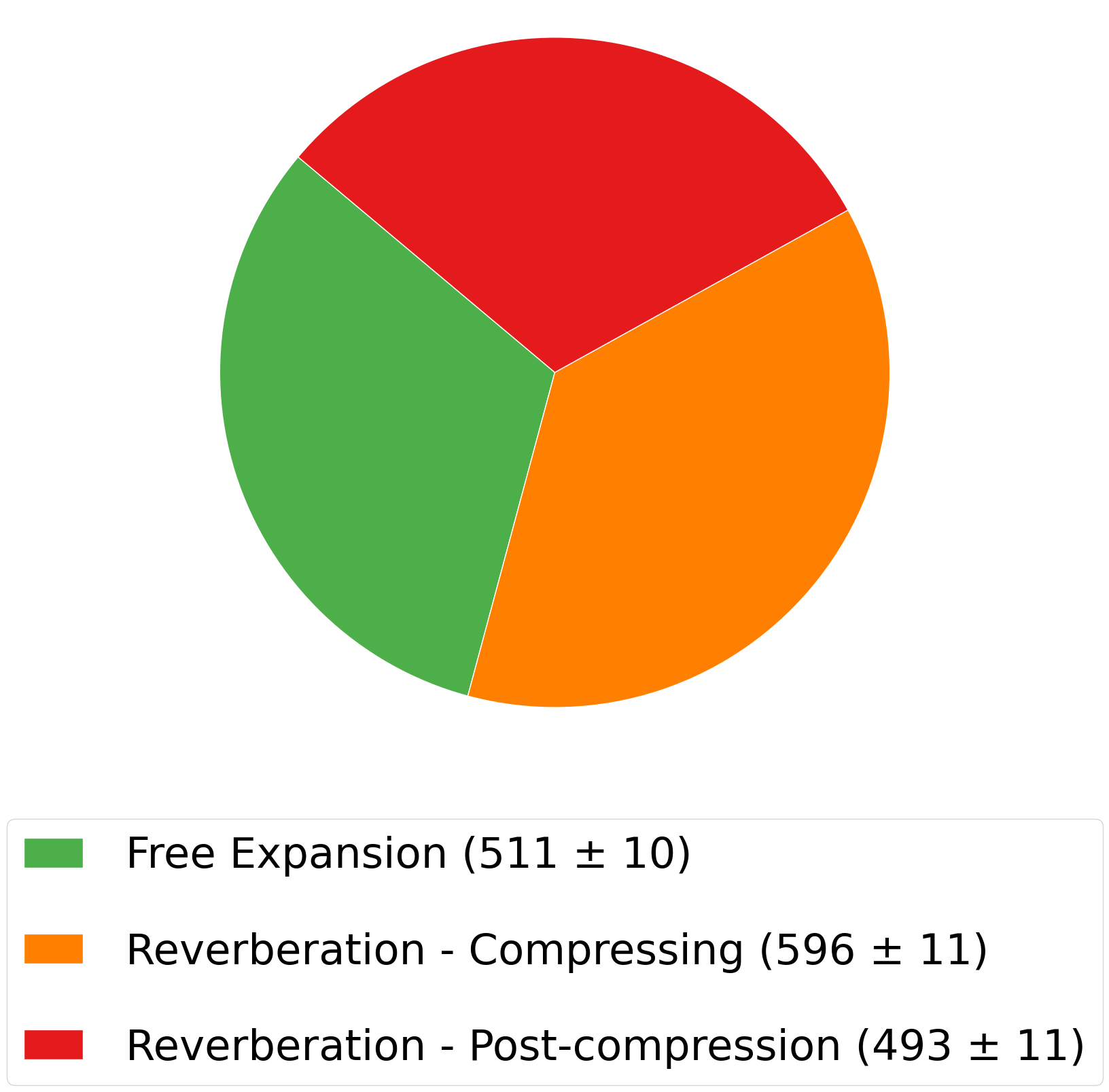}
\includegraphics[width=0.6\linewidth]{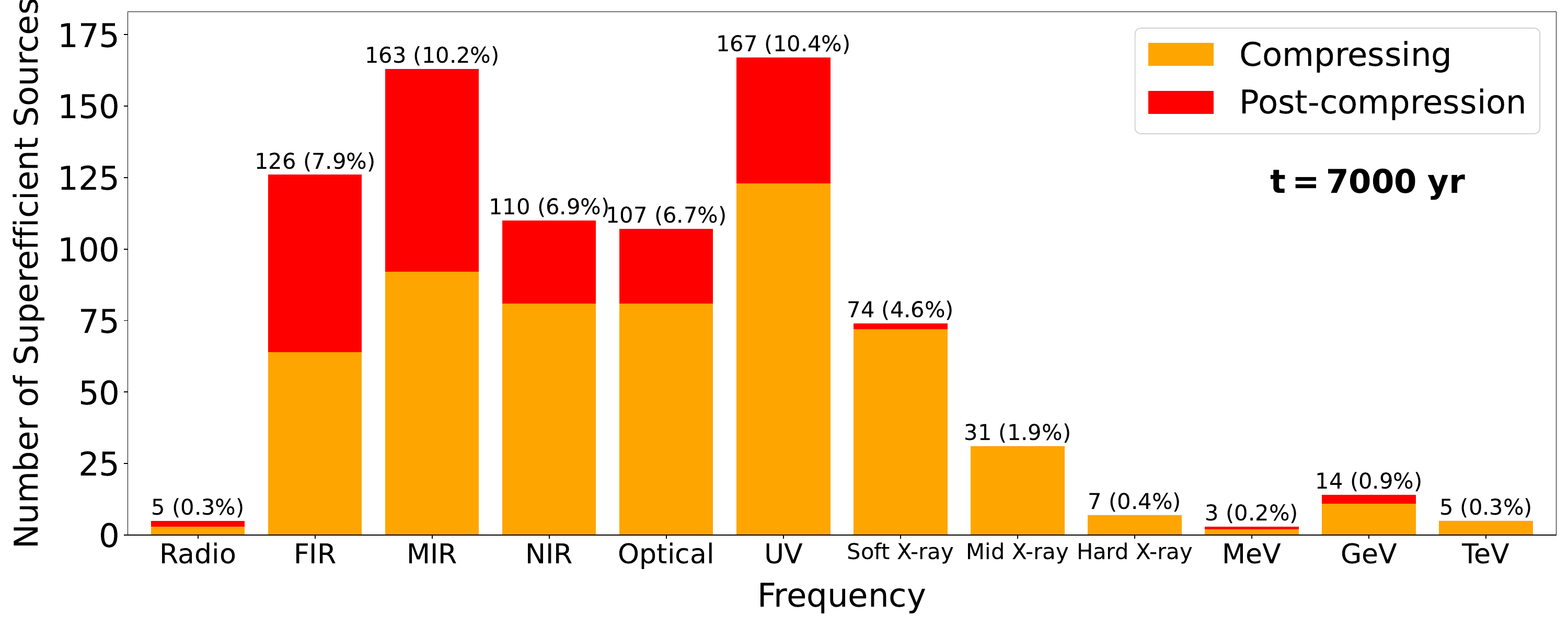} \\
\includegraphics[width=0.3\linewidth]{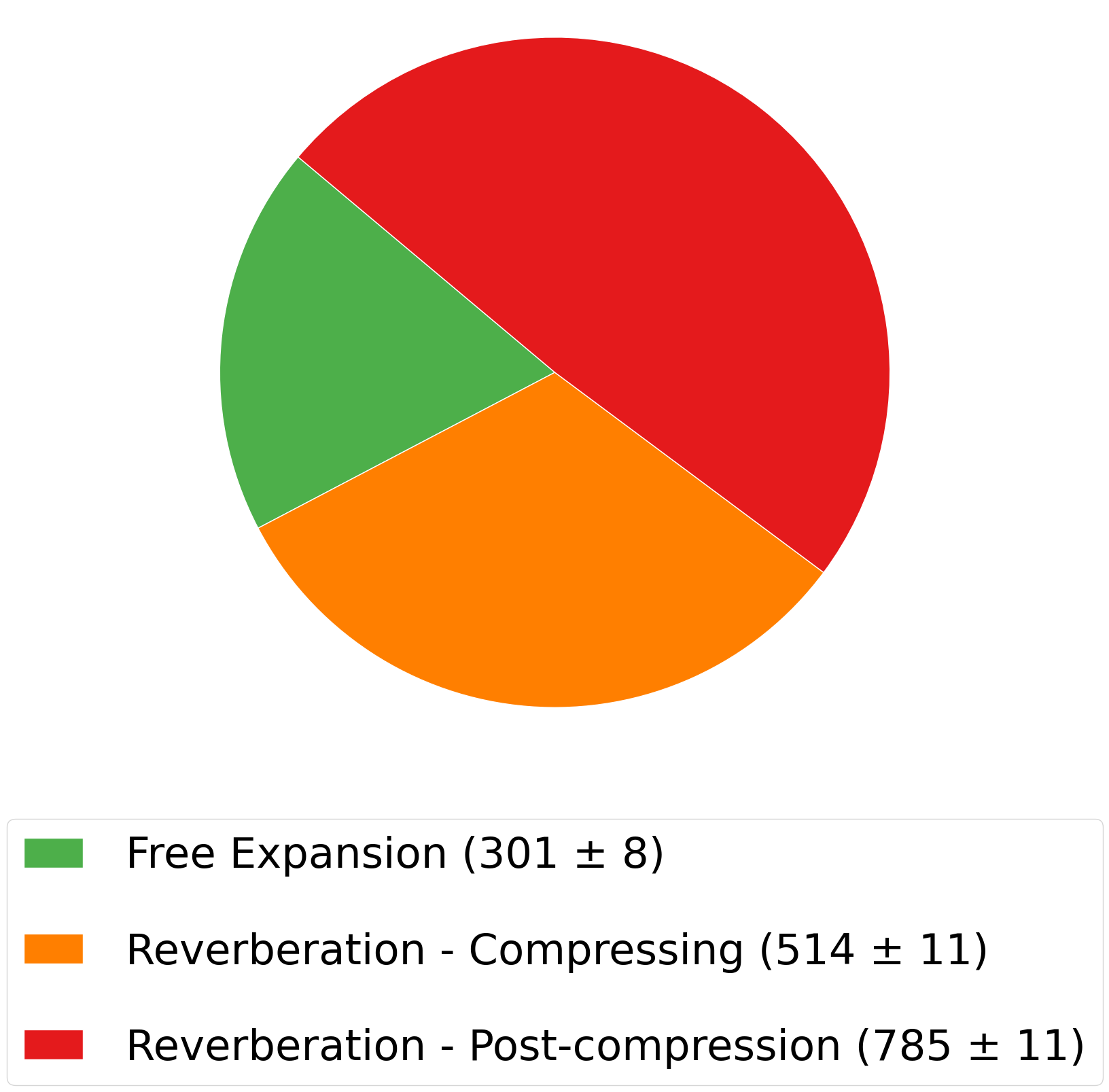}
\includegraphics[width=0.6\linewidth]{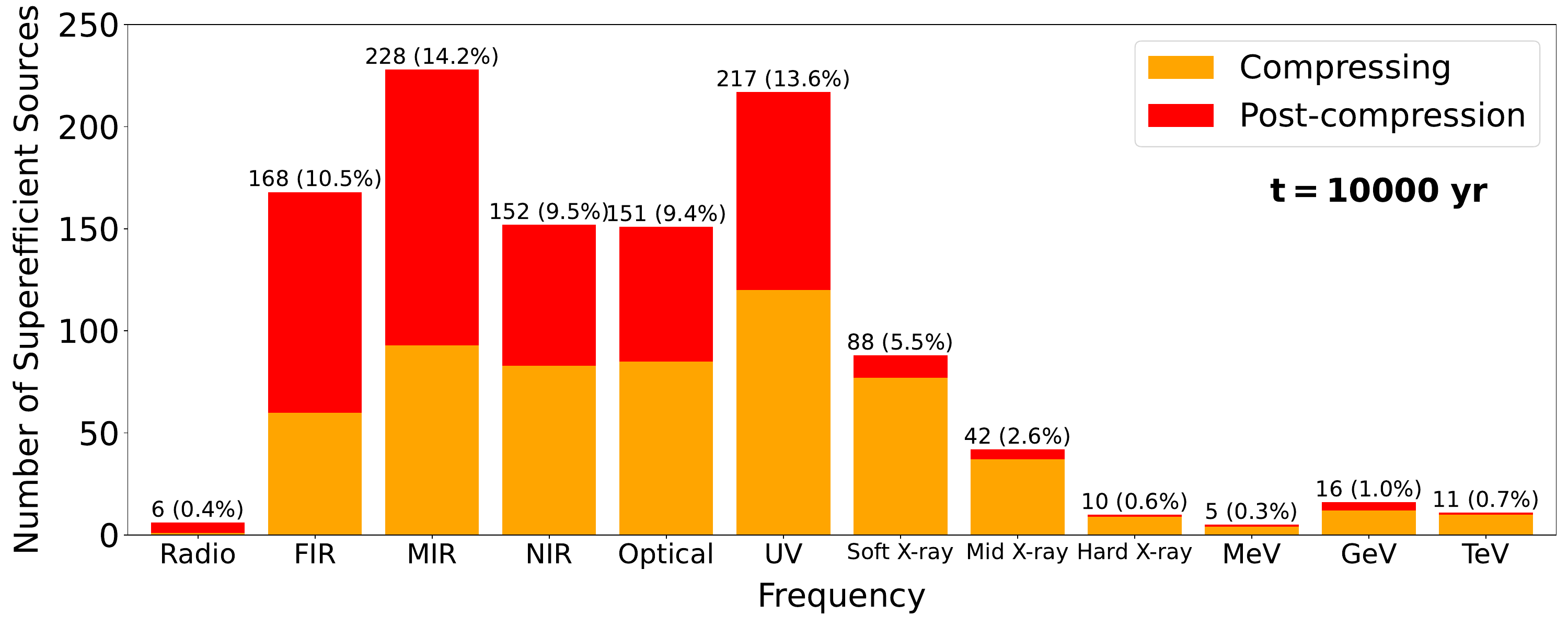}\\

\caption{The pie charts in the left panels follow the same format as Fig. \ref{fig: pie_super}, but at 3000, 5000, 7000, and 10000 years (from top to bottom panels). The corresponding mean number and fractional contribution of superefficient sources with frequency ranges at these same ages, along with the compression states, are given as bar plots in the right panel, similar to Fig. \ref{fig: pie_super}.
}
\label{fig: pie_super_time_series}
\end{figure*}


\begin{table*}
\centering
\scriptsize
\caption{Pairwise number of superefficient PWNe between frequency bands at different ages.}
\label{tab:supereff_matrix_age}
\setlength{\tabcolsep}{4pt}
\renewcommand{\arraystretch}{1.0}

\begin{tabular}{lcccccccccccc}
\hline\hline

\multicolumn{13}{c}{$t = 3000 \mathrm{yr}$} \\
\hline
 & Radio & FIR & MIR & NIR & Opt. & UV & Soft  & Mid  & Hard  & MeV & GeV & TeV \\
  &  & &  &  &  & & X-ray & X-ray & X-ray \\
\hline
Radio	& 1	& 1	& 1	& 0	& 0	& 0	& 0	& 0	& 0	& 0	& 0	& 0\\
FIR	    & 1	& 41 & 33 & 25	& 24 & 23 & 8 & 2 & 0 & 3 & 9 & 0\\
MIR	    & 1	& 33 & 64 & 47	& 48 & 48 & 29 & 12 & 1 & 3 & 9 & 0\\
NIR	    & 0	& 25 & 47 & 48 & 47 & 45 & 27 & 10 & 1 & 1 & 9 & 0\\
Optical	& 0	& 24 & 48 & 47 & 50 & 48 & 30 & 13 & 2 & 1 & 9 & 0\\
UV	    & 0	& 23 & 48 & 45 & 48 & 98 & 63 & 30 & 4 & 0 & 8 & 0\\
Soft X-ray	& 0	& 8	& 29 & 27 & 30 & 63 & 63 & 30 & 4 & 0 & 0 & 0\\
Mid X-ray	& 0	& 2	& 12 & 10 & 13 & 30 & 30 & 30 & 4 & 0 & 0 & 0\\
Hard X-ray	& 0	& 0	& 1	& 1	& 2	& 4	& 4	& 4	& 4	& 0	& 0	& 0\\
MeV	        & 0	& 3	& 3	& 1	& 1	& 0	& 0	& 0	& 0	& 3	& 1	& 0\\
GeV	        & 0	& 9	& 9	& 9	& 9	& 8	& 0	& 0	& 0	& 1	& 9	& 0\\
TeV	        & 0	& 0	& 0	& 0	& 0	& 0	& 0	& 0	& 0	& 0	& 0	& 0\\
\hline

\multicolumn{13}{c}{$t = 5000 \mathrm{yr}$} \\
\hline
 & Radio & FIR & MIR & NIR & Opt. & UV & Soft  & Mid  & Hard  & MeV & GeV & TeV \\
  &  & &  &  &  & & X-ray & X-ray & X-ray \\
\hline
Radio	& 6	& 6	& 2	& 1	& 1	& 2	& 1	& 2	& 2	& 2	& 1	& 0\\
FIR	    & 6	& 66 & 54 & 41 & 38 & 38 & 18 & 11 & 3 & 3 & 7 & 0\\
MIR	    & 2	& 54 & 109 & 76 & 72 & 75 & 40 & 21 & 5 & 3 & 7 & 0\\
NIR	    & 1	& 41 & 76 & 76 & 71 & 69 & 39 & 20 & 4 & 2 & 7 & 0\\
Optical	& 1	& 38 & 72 & 71 & 74 & 72 & 41 & 21 & 5 & 2 & 7 & 0\\
UV	    & 2 & 38 & 75 & 69 & 72 & 138 & 73 & 36 & 9	& 2	& 7	& 0\\
Soft X-ray	& 1	& 18 & 40 & 39 & 41 & 73 & 75 & 37 & 8	& 1	& 3	& 0\\
Mid X-ray	& 2	& 11 & 21 & 20 & 21 & 36 & 37 & 39 & 9 & 1 & 2 & 0\\
Hard X-ray	& 2	& 3	& 5	& 4	& 5	& 9	& 8	& 9	& 10 & 2 & 1 & 0\\
MeV	        & 2	& 3	& 3	& 2	& 2	& 2	& 1	& 1	& 2	& 3	& 2	& 0\\
GeV	        & 1	& 7	& 7	& 7	& 7	& 7	& 3	& 2	& 1	& 2	& 7	& 0\\
TeV	        & 0	& 0	& 0	& 0	& 0	& 0	& 0	& 0	& 0	& 0	& 0	& 0\\
\hline

\multicolumn{13}{c}{$t = 7000 \mathrm{yr}$} \\
\hline
 & Radio & FIR & MIR & NIR & Opt. & UV & Soft  & Mid  & Hard  & MeV & GeV & TeV \\
  &  & &  &  &  & & X-ray & X-ray & X-ray \\
\hline
Radio	& 6	& 4	& 3	& 1	& 1	& 1	& 0	& 0	& 1	& 2	& 1	& 0\\
FIR	    & 4	& 114 & 91 & 63	& 58 & 53 & 22 & 7 & 1 & 3 & 12	& 2\\
MIR	    & 3	& 91 & 151 & 102 & 99 & 97 & 42	& 16 & 3 & 3 & 12 & 5\\
NIR	    & 1	& 63 & 102 & 103 & 98 & 92 & 40	& 16 & 2 & 2 & 12 & 5\\
Optical	& 1	& 58 & 99 & 98 & 101 & 95 & 42 & 16	& 2	& 2	& 12 & 5\\
UV	     & 1 & 53 & 97 & 92	& 95 & 154 & 73	& 28 & 6 & 1 & 11 & 5\\
Soft X-ray	& 0	& 22 & 42 & 40 & 42	& 73 & 73 & 28 & 6 & 0 & 3 & 5\\
Mid X-ray	& 0	& 7	& 16 & 16 & 16 & 28	& 28 & 28 & 6 & 0 & 2 & 5\\
Hard X-ray	& 1	& 1	& 3	& 2	& 2	& 6	& 6	& 6	& 7	& 1	& 0	& 0\\
MeV	        & 2	& 3	& 3	& 2	& 2	& 1	& 0	& 0	& 1	& 3	& 2	& 0\\
GeV	        & 1	& 12 & 12 & 12 & 12	& 11 & 3 & 2 & 0 & 2 & 12 & 2\\
TeV	        & 0	& 2	& 5	& 5	& 5	& 5	& 5	& 5	& 0	& 0	& 2	& 5\\
\hline

\multicolumn{13}{c}{$t = 10000 \mathrm{yr}$} \\
\hline
 & Radio & FIR & MIR & NIR & Opt. & UV & Soft  & Mid  & Hard  & MeV & GeV & TeV \\
  &  & &  &  &  & & X-ray & X-ray & X-ray \\
\hline
Radio	& 7	& 5	& 4	& 3	& 2	& 2	& 1	& 1	& 0	& 2	& 2	& 1\\
FIR	    & 5	& 172 & 142	& 91 & 83 & 80 & 33	& 18 & 2 & 7 & 16 & 10\\
MIR	    & 4	& 142 & 239	& 158 & 149	& 149 & 67 & 31 & 6	& 7	& 16 & 14\\
NIR	    & 3	& 91 & 158 & 158 & 147 & 138 & 65 & 29 & 6 & 7 & 16	& 14\\
Optical	& 2	& 83 & 149 & 147 & 154 & 145 & 66 & 30 & 6 & 7 & 16	& 14\\
UV	     & 2 & 80 & 149	& 138 & 145	& 220 & 92 & 43 & 10 & 6 & 14 & 15\\
Soft X-ray	& 1	& 33 & 67 & 65 & 66	& 92 & 93 & 44 & 10	& 3	& 8	& 15\\
Mid X-ray	& 1	& 18 & 31 & 29 & 30	& 43 & 44 & 44 & 10	& 2 & 4	& 15\\
Hard X-ray	& 0	& 2	& 6	& 6	& 6	& 10 & 10 & 10 & 10	& 0	& 0	& 5\\
MeV	        & 2	& 7	& 7	& 7	& 7	& 6	& 3	& 2	& 0	& 7	& 6	& 2\\
GeV	        & 2	& 16 & 16 & 16 & 16	& 14 & 8 & 4 & 0 & 6 & 16 & 4\\
TeV	        & 1	& 10 & 14 & 14 & 14	& 15 & 15 & 15 & 5 & 2 & 4	& 15\\
\hline
\end{tabular}
\tablefoot{Each sub-block corresponds to the matrix $\mathcal{N}_{ij}(t)$ at age $t$, where $\mathcal{N}_{ij}$
denotes the number of sources superefficient in both bands $i$ and $j$, independent of evolutionary state. Note that the content of this table corresponds to a single random realization of 1600 PWNe, and it is given as an example of the behavior.}
\end{table*}

\begin{figure*}
    \centering
    \includegraphics[width=\columnwidth]{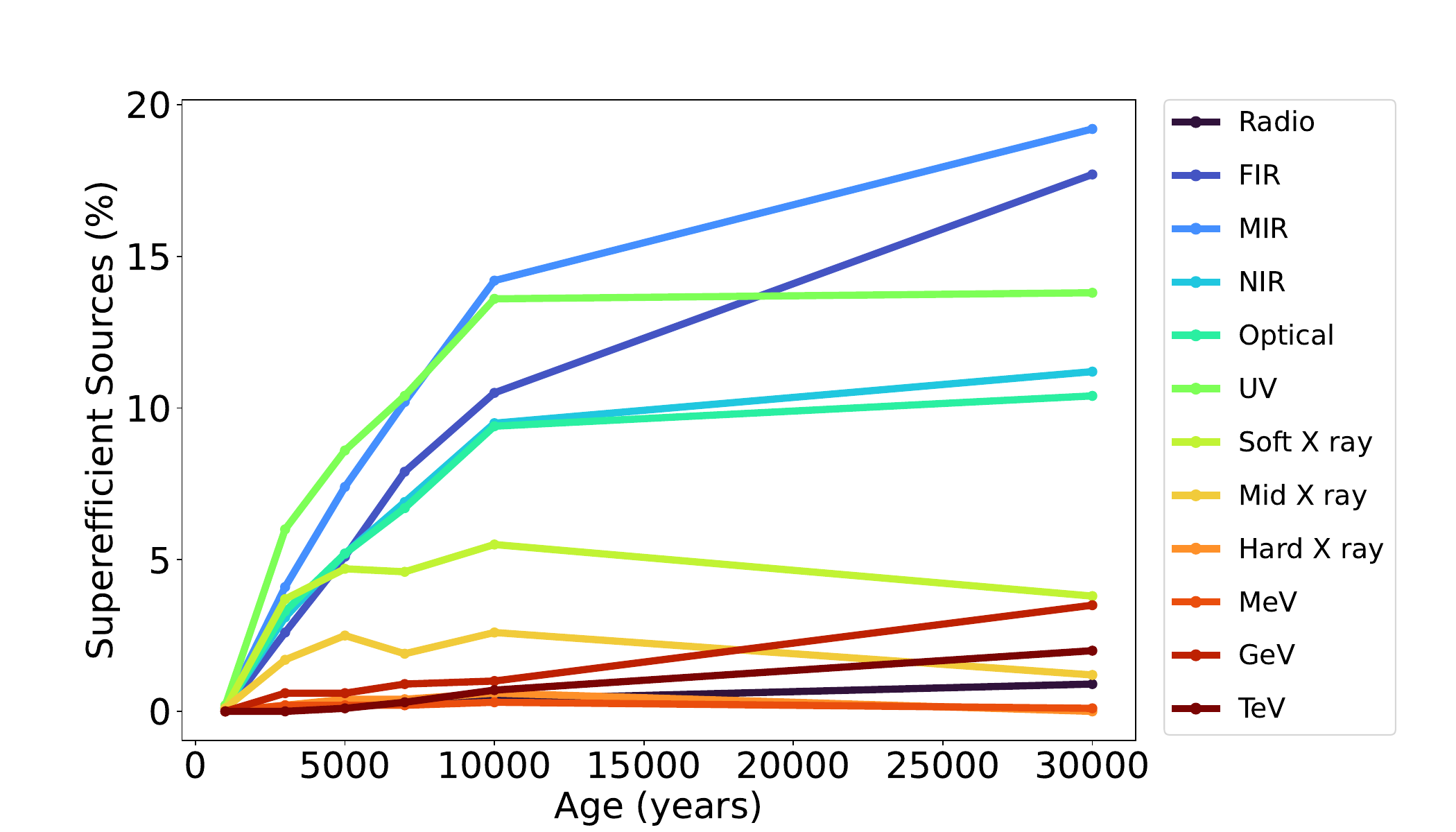}
    \includegraphics[width=\columnwidth]{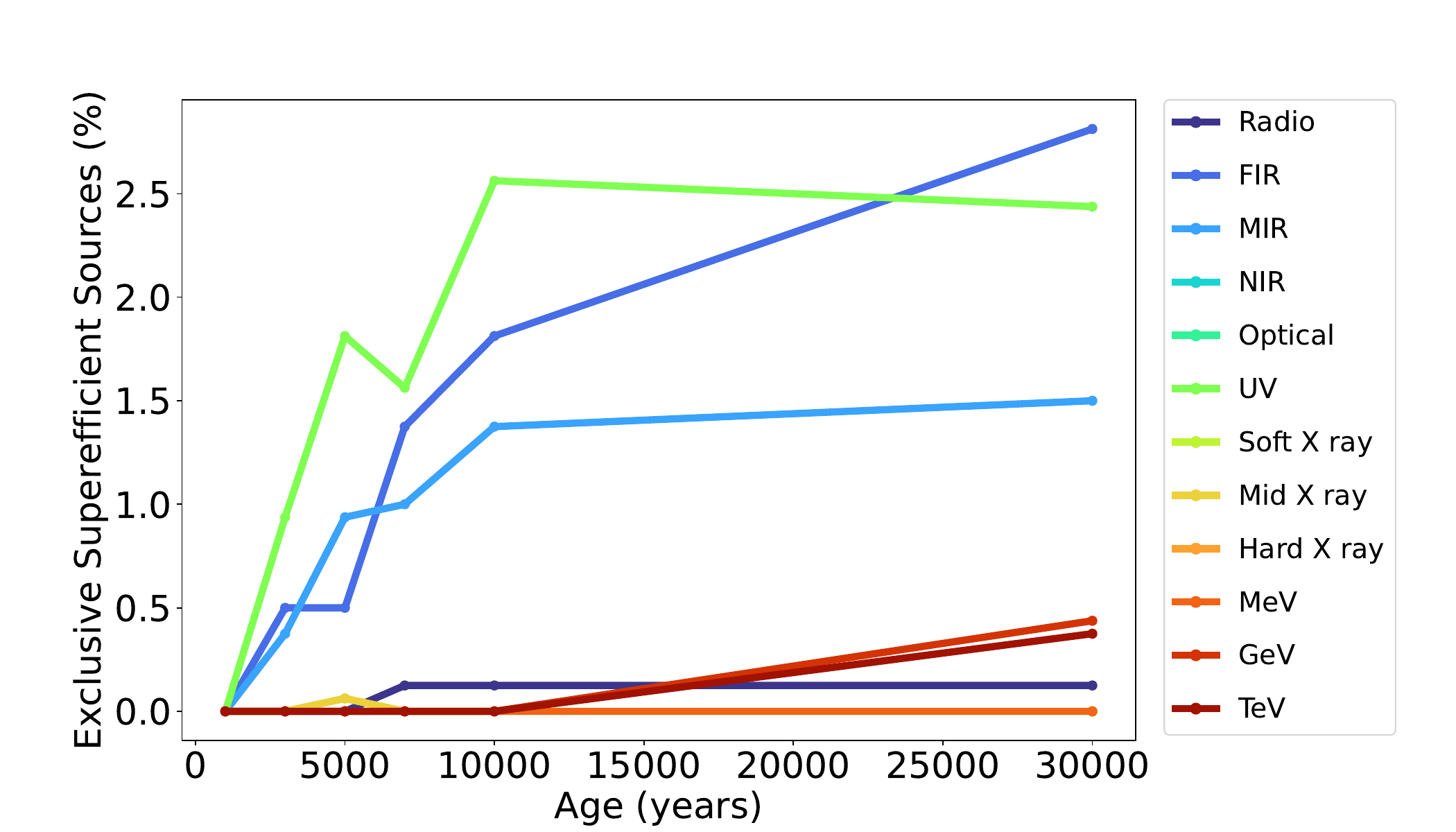}
    \caption{The left panel shows the time evolution of the mean percentage of superefficient sources across different energy bands, as obtained from Fig. \ref{fig: pie_super_time_series}. The right panel, on the other hand, shows sources that are exclusively superefficient in any given frequency band at any time for a single random realization of 1600 PWNe.
}
    \label{fig: sup_percentage}
\end{figure*}


\begin{figure*}
    \includegraphics[width=0.33\linewidth]{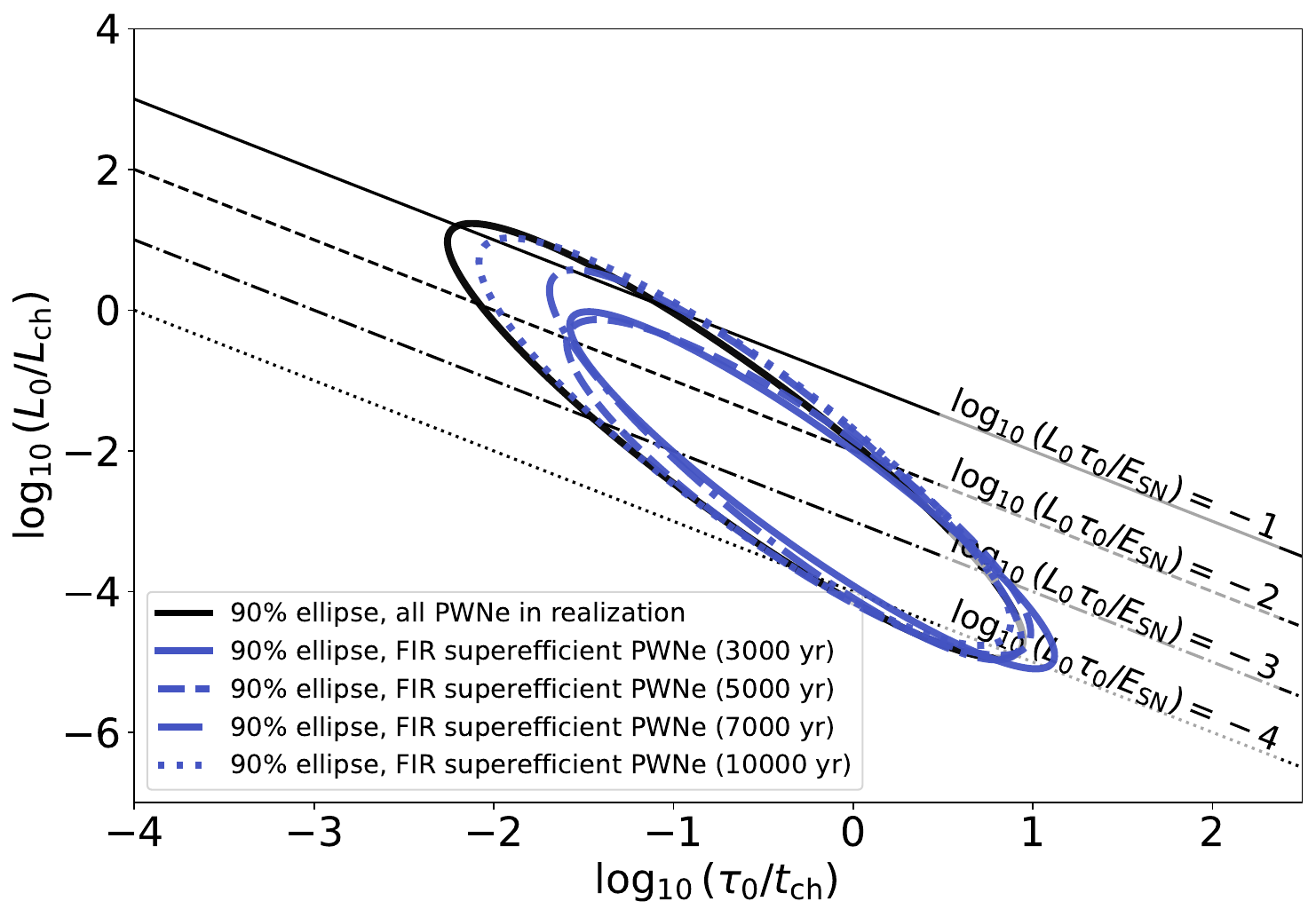}
    \includegraphics[width=0.33\linewidth]{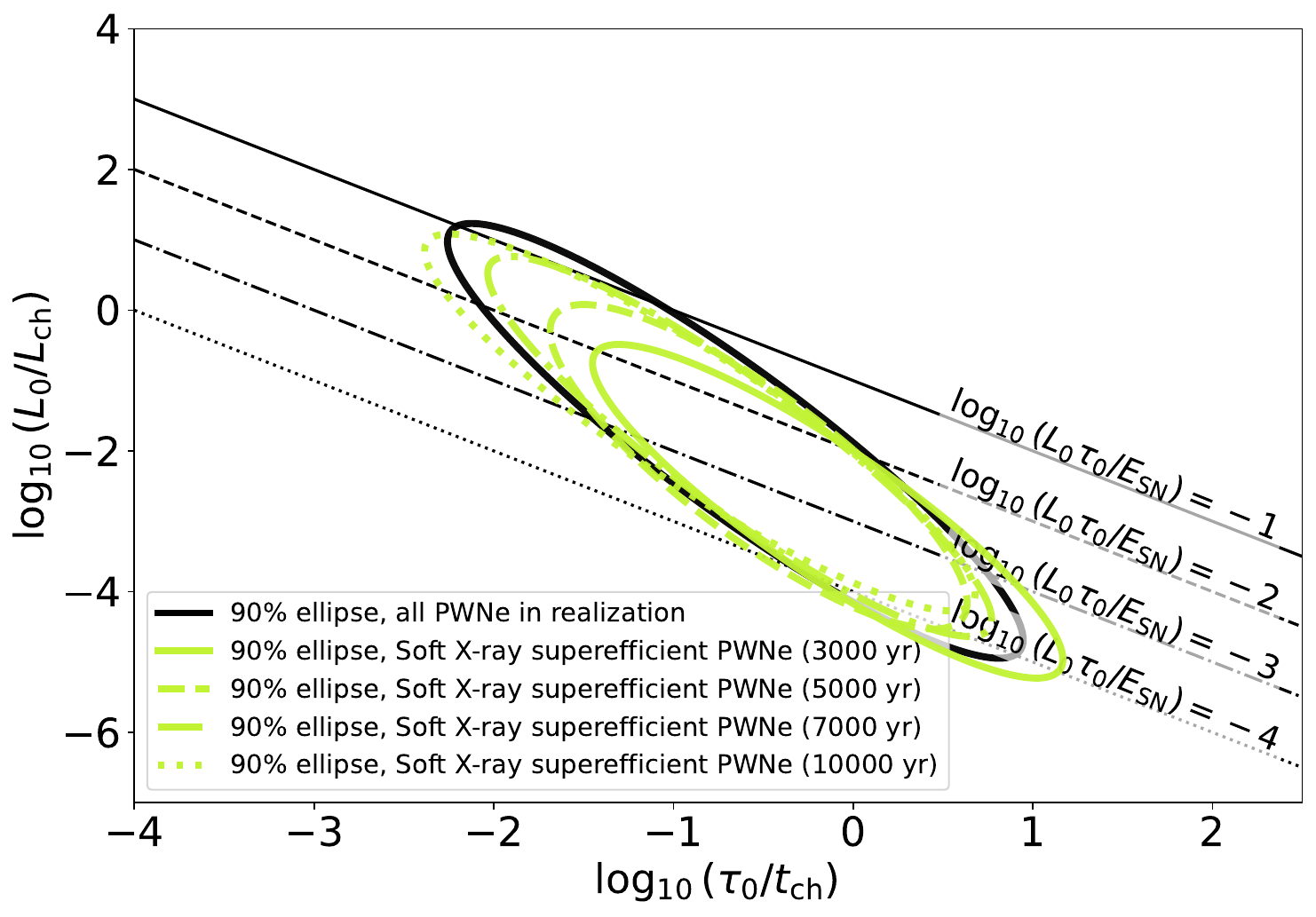}
    \includegraphics[width=0.33\linewidth]{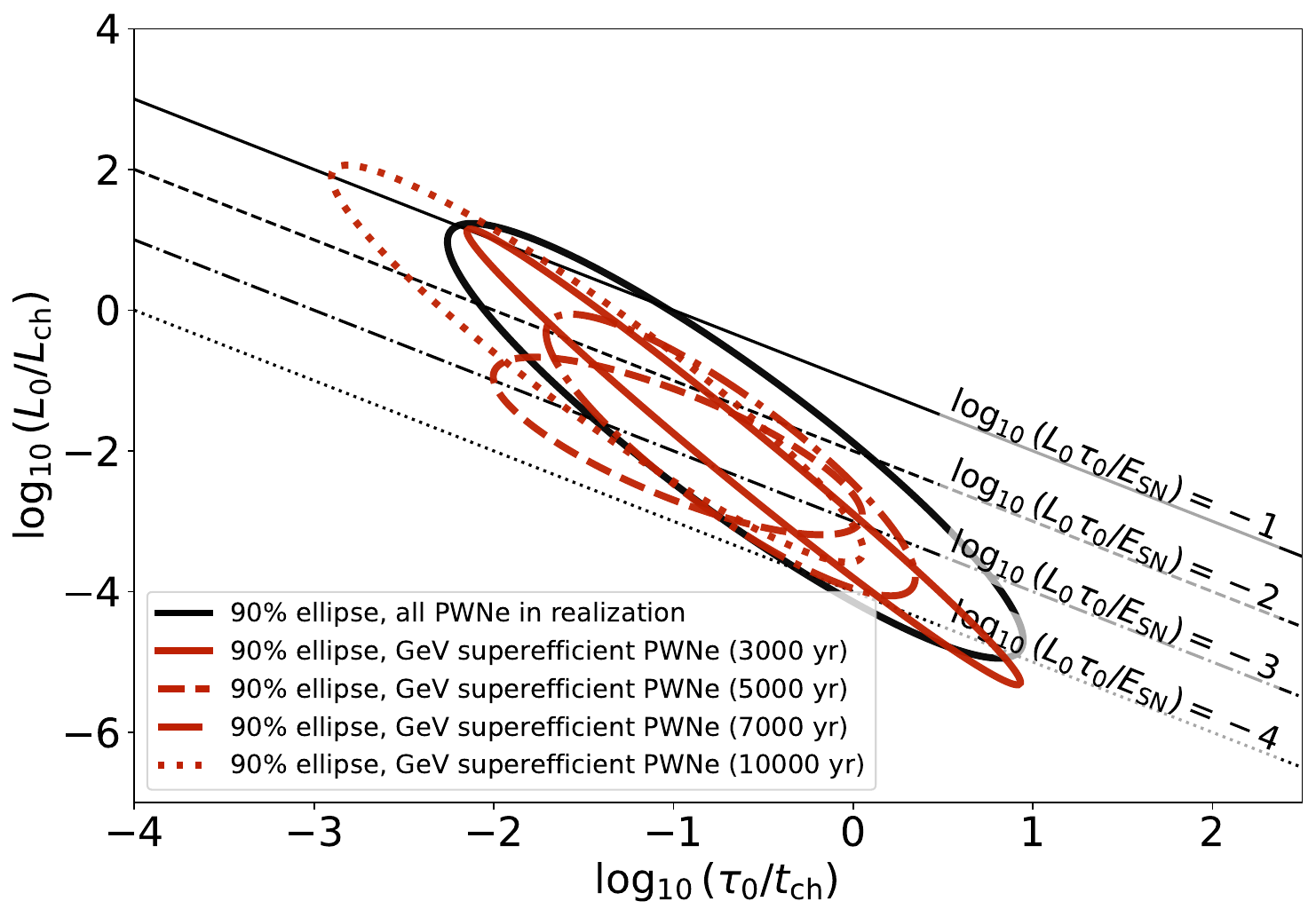}
    \caption{The evolution of 90$\%$ ellipses of superefficient PWNe in FIR (left panel), soft X-ray (middle panel), and GeV (right panel) bands on the $\log_{10}(L_0/L_{\rm ch})$-$\log_{10}(\tau_0/t_{\rm ch})$ plane 
    at different evolutionary ages, for a single random realization of 1600 sources. }
\label{fig: L0_tau0_age}
\end{figure*}

In contrast, the results obtained for the full population 
modeled with a less detailed treatment of reverberation
(in our case using \texttt{TIDE}) differ both quantitatively and qualitatively. 
Quantitatively, as shown in Table \ref{tab:sup_counts} and Fig. \ref{fig: pie_super}, the total number of superefficient PWNe is significantly reduced in all frequency bands, including in the FIR band, and there are essentially no superefficient systems predicted at X-ray energies. 
The lack of a detailed treatment of reverberation in purely thin-shell models misrepresents PWN compression, particle reprocessing, and magnetic field amplification,
thus resulting in a reduced predicted number of superefficient PWNe (for more details see \cite{bandiera23III}).
With a less detailed description, superefficient sources are found predominantly in the FIR band. 
By contrast, a more consistent treatment (\texttt{TIDE+L}) places substantial superefficiency from the MIR to the UV range, in addition to the FIR. 
%
%
In the simplified thin-shell approach, the compression during reverberation can become excessively strong, producing unrealistically large magnetic fields and, consequently, severe synchrotron losses. 
These losses preferentially deplete the comparatively higher-energy particles responsible for MIR-UV emission, leaving mainly the lower-energy particle population that emits in the FIR. 
In \texttt{TIDE+L}, the more realistic treatment of reverberation leads to a milder and more gradual compression. 
As a result, MIR-UV-emitting particles can survive more effectively, increasing the fraction of superefficient sources in these bands.

We further present the superefficient PWNe distribution across different frequency bands in Table \ref{tab:supereff_matrix}, which shows a matrix of superefficient sources in various frequency bands considered.
Each element of the matrix highlights the number of sources that are simultaneously superefficient at the corresponding pair of bands at the current age.

To highlight the degree of superefficiency, i.e., how far the luminosities in the corresponding frequency bands exceed the contemporaneous spin-down luminosity across different frequencies, we show the distribution of the ratio $L_{\rm band}/\dot{E}_{\rm SD}$ for different frequencies in Fig. \ref{fig: Lband/Edot} at the current age.
As expected, the highest ratios ($> 10^3$) were obtained at the lower energies (FIR to UV), where the superefficiency is more prevalent due to the accumulation of lower energy particles radiating in an increased magnetic field.

Additionally, in Fig. \ref{fig: L0_tau0}, we show the position of the superefficient sources at the current age in various frequency bands on the characteristic $\log_{10}(L_0/L_{\rm ch})$-$\log_{10}(\tau_0/t_{\rm ch})$ plane.
Superefficient sources appear across the plane, showing that it is not an uncommon event, nor that it is associated only to a particular combination of possible phase-space parameters.

Interestingly, $25 \pm 3$ sources are found to be superefficient in the considered TeV range (1-10 TeV) in our synthetic population. 
%
Fig. \ref{fig: tev_superefficient} shows the SEDs and electron spectra of the TeV-superefficient sources for one random realization of 1600 sources in red, on the backdrop of rest of the population in grey.
%
Most of the TeV-superefficient sources are very luminous in the TeV range.
%
To quantify the detectability of the TeV-superefficient sources,
we adopted the same detectability criteria as \cite{desarkar26}, namely sky coverage and flux thresholds for the Cherenkov Telescope Array Observatory (CTAO), High Energy Stereoscopic System (H.E.S.S.), and Large High Altitude Air Shower Observatory (LHAASO), 
 In the representative realization considered here, 26 out of 1600 PWNe are TeV-superefficient. 
2/26 are in the compression stage and 24/26 are in the post-compression stage. 
Among the TeV-superefficient subset, 20 sources are detectable by at least one of the considered TeV instruments; 6 sources would remain undetected. 
CTAO could detect 17 TeV-superefficient sources, followed by H.E.S.S. with 9 and LHAASO-WCDA with 8.
No TeV-superefficient source in this realization is detectable by LHAASO-KM2A. 
These instrument counts are not mutually exclusive, as all H.E.S.S.-detected sources are also CTAO-detected, while LHAASO-WCDA contributes 3 unique detections in addition to sources overlapping with CTAO and/or H.E.S.S. 
%
%
%
Note that we do not find any hint of superefficiency in ultra-high-energy $\gamma$-ray range (50 - 500 TeV), hence, we do not show any bar corresponding to that in Fig. \ref{fig: pie_super}.

In summary, FIR is the most promising band for superefficiency and surveys in this band may be the best suited for identifying superefficient PWNe. 
Observational exploitation will require accounting for confusion, angular resolution, and pulsar/PWN association criteria.
X-ray detections, instead, are more sensitive to the reverberation history: they can be enhanced during active compression, but their persistence depends on whether the high-energy particles survive the associated synchrotron losses. 
GeV observations are especially useful for probing evolved and post-compression systems, where relic particles can continue to produce IC emission even after the present-day spin-down power has declined.

\begin{figure*}[t!]
\centering
\raisebox{2.5mm}{\includegraphics[width=0.33\textwidth]{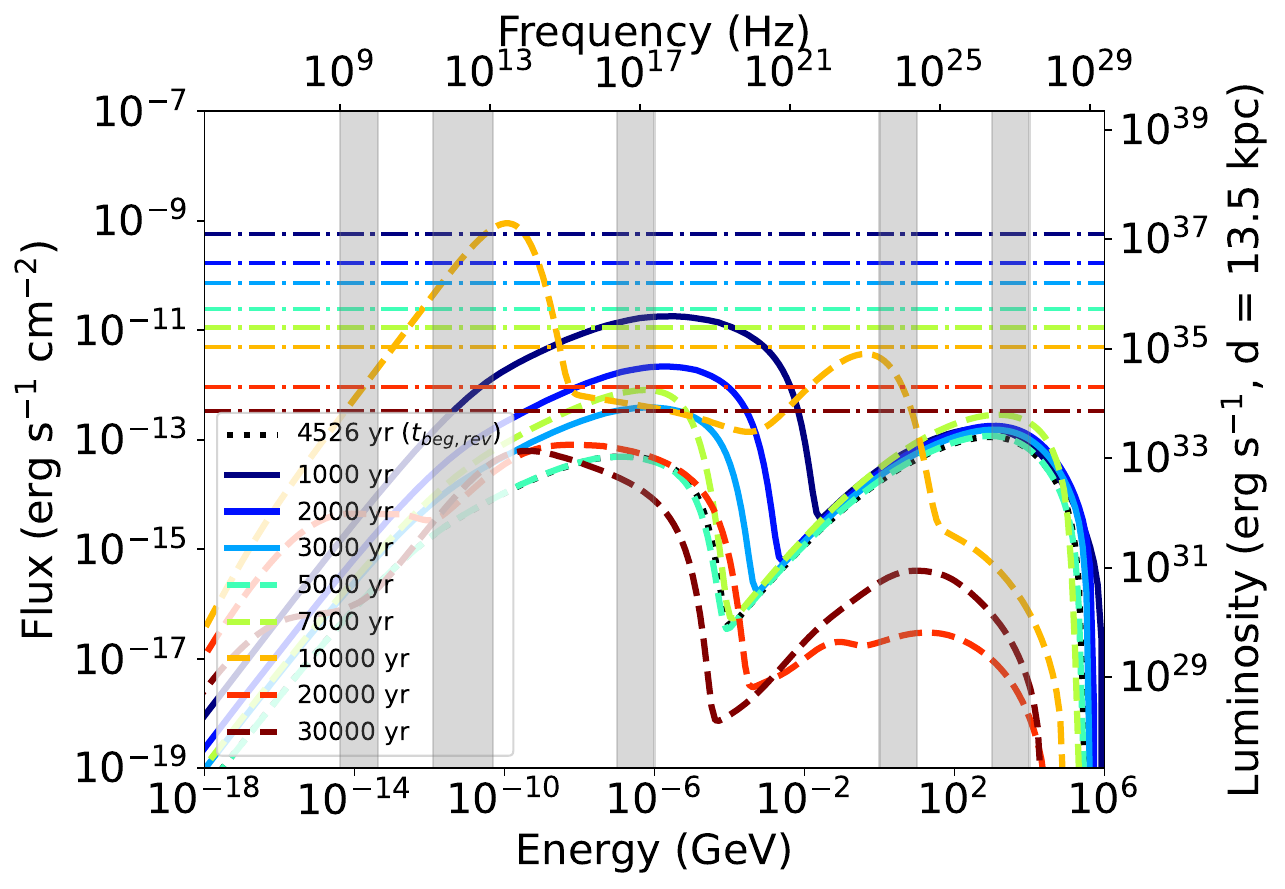}}
\includegraphics[width=0.33\textwidth]{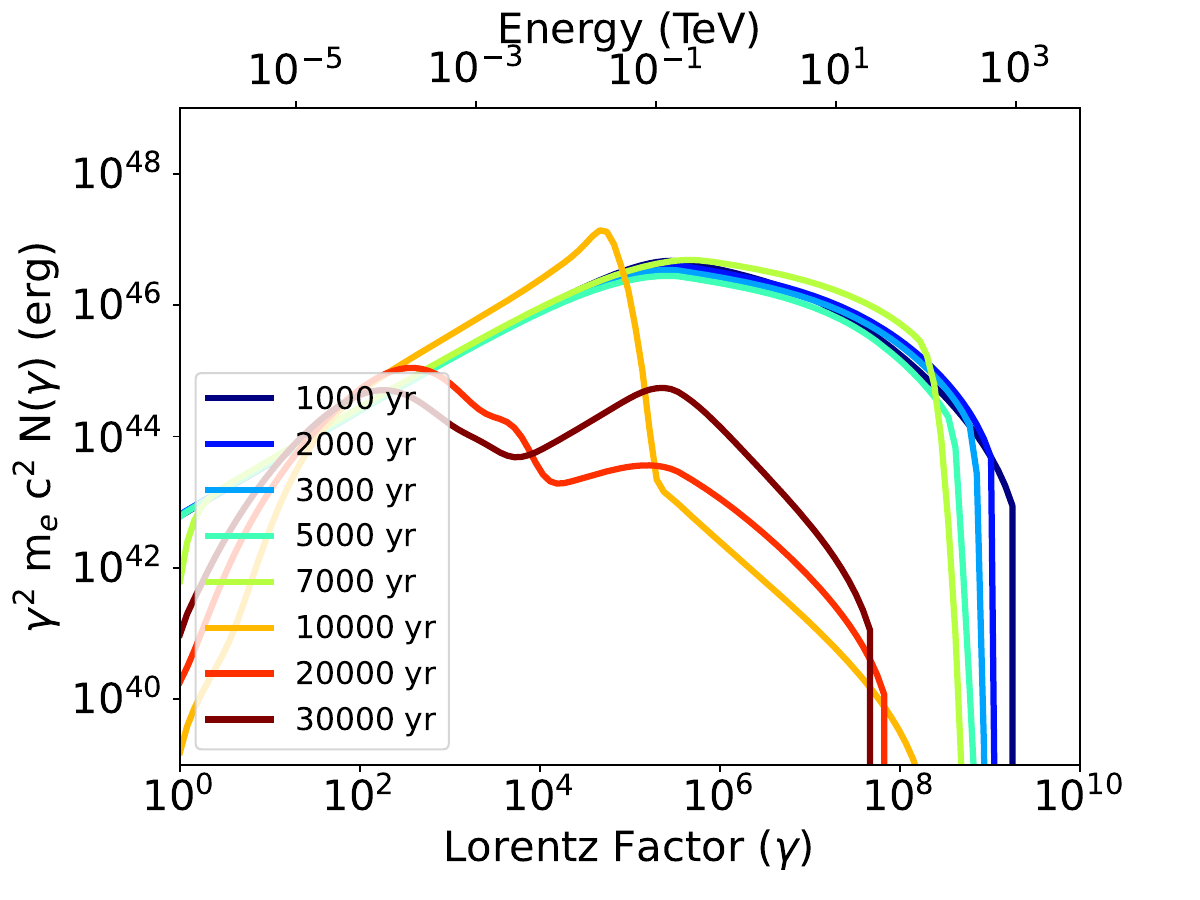}
\includegraphics[width=0.33\textwidth]{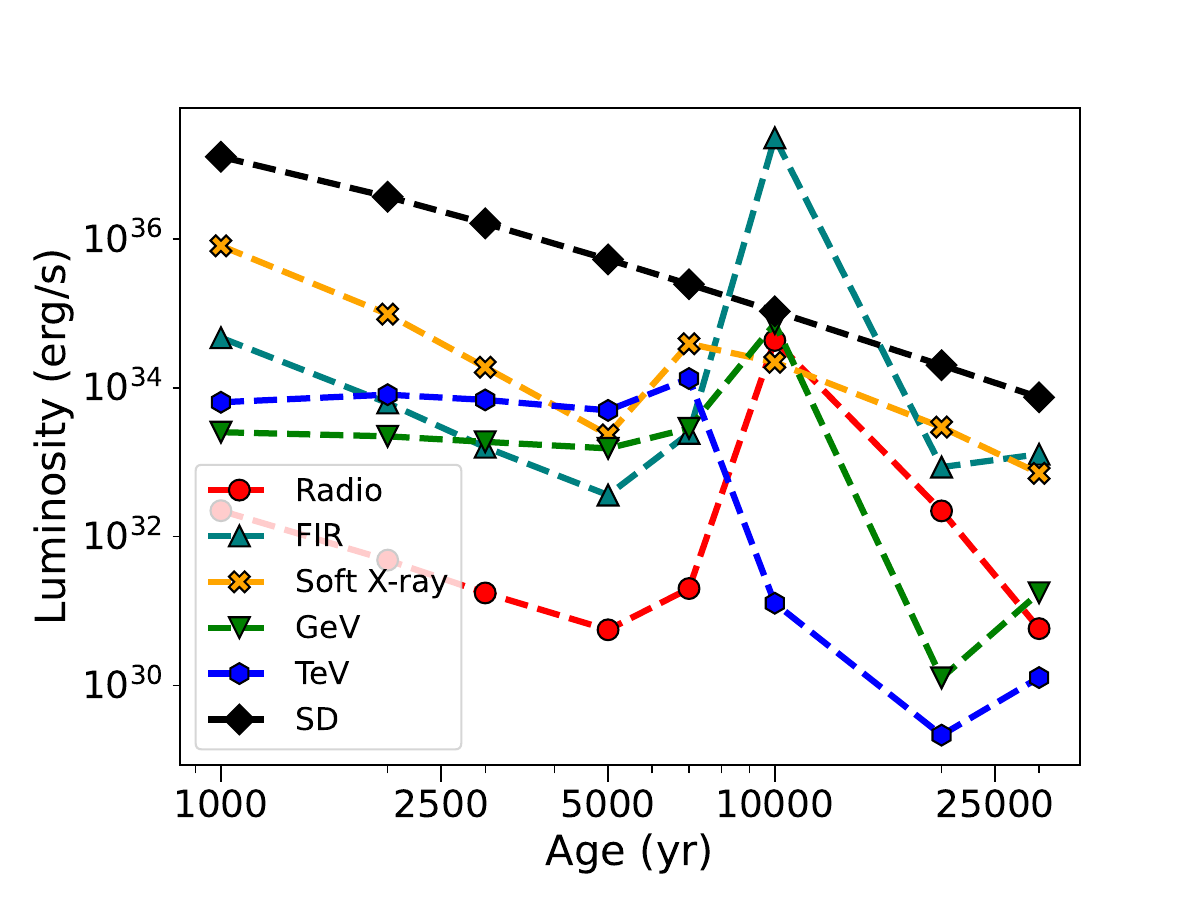}
\raisebox{2.5mm}{\includegraphics[width=0.33\textwidth]{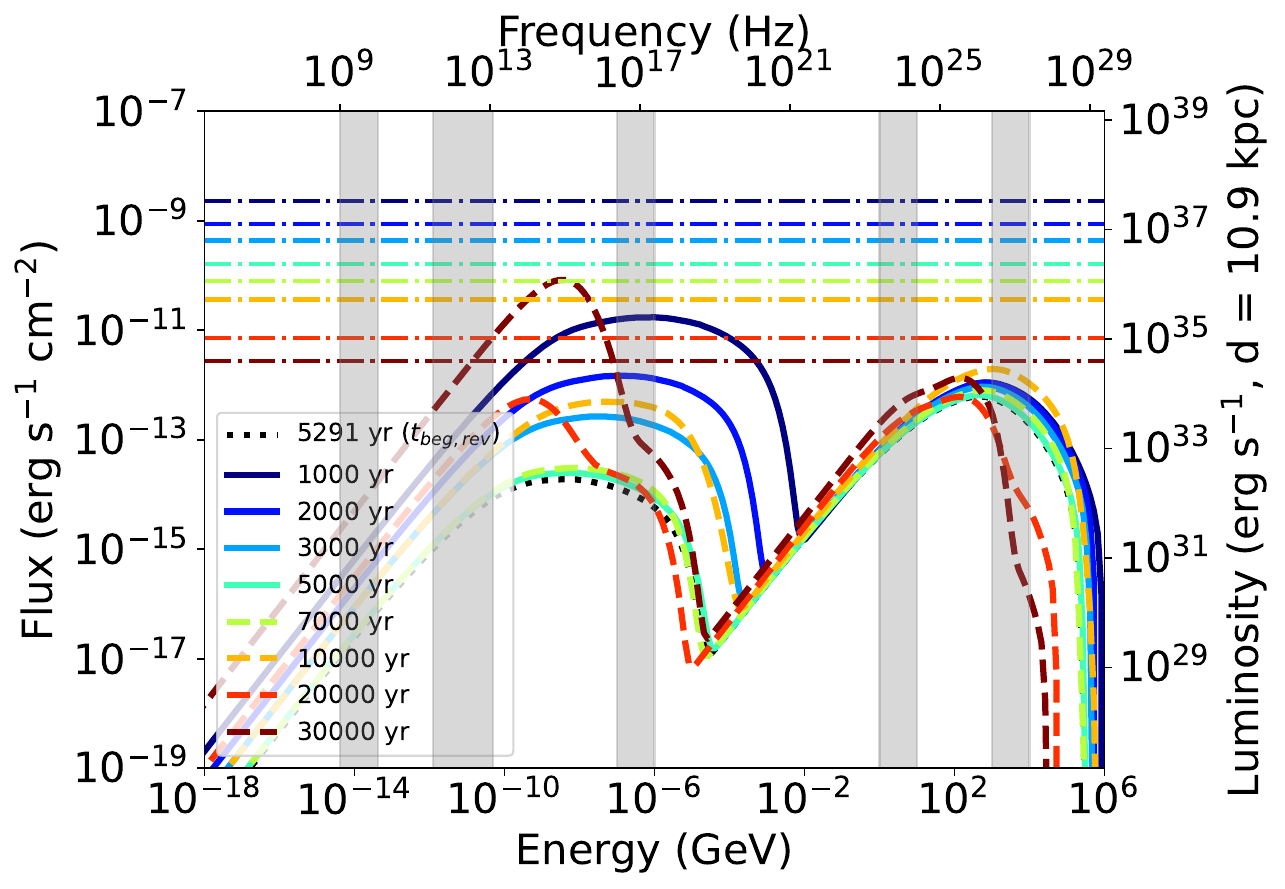}}
\includegraphics[width=0.33\textwidth]{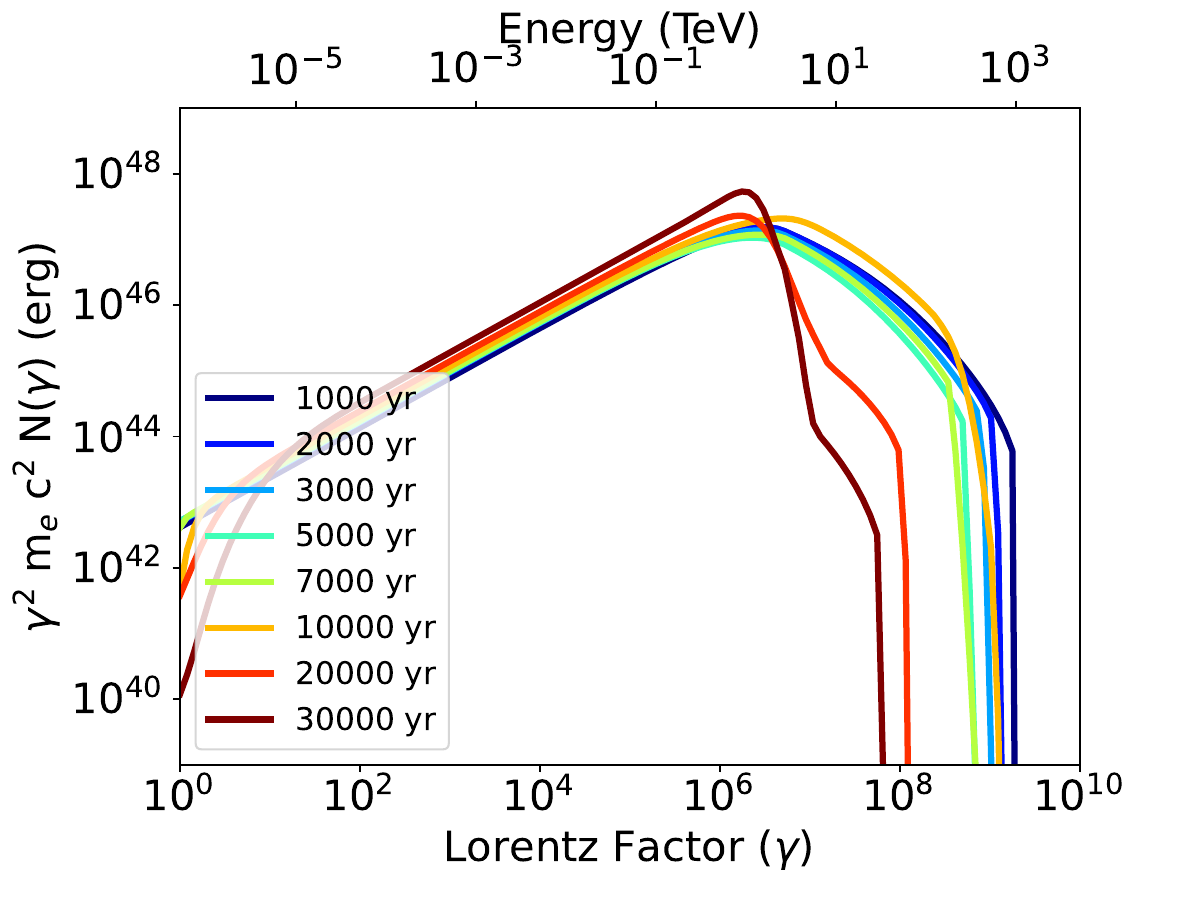}
\includegraphics[width=0.33\textwidth]{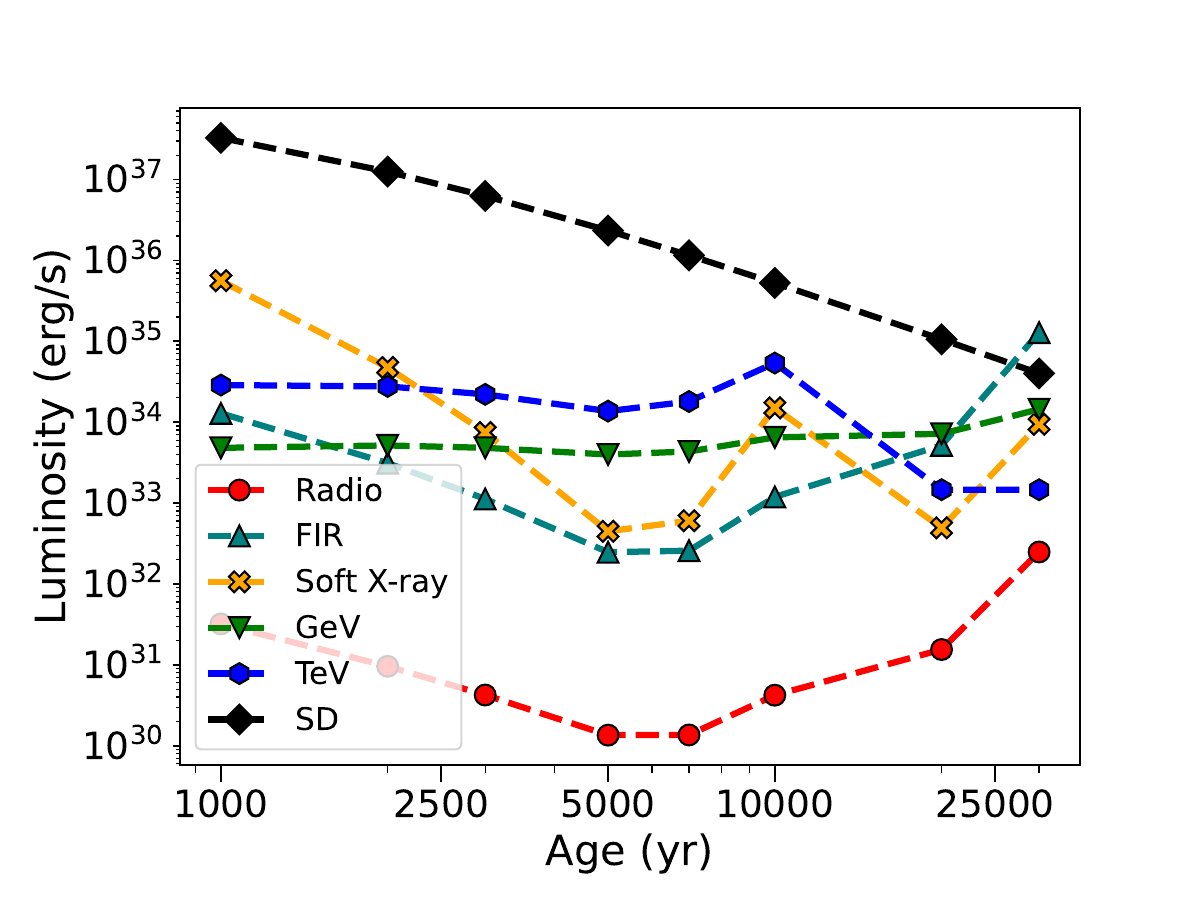}
\caption{MWL SEDs, electron spectra and integrated luminosity evolution at different frequency bands of two selected individual PWNe based on the CF are shown in this figure. 
The upper (lower) row corresponds to Example 1 with CF $\sim 50.3$ (Example 2 with CF $\sim 5.8$). 
The left panel in both rows shows the evolution of the SEDs with time. The solid (dashed) lines correspond to the SED before (after) the reverberation, whereas the dotted line corresponds to the SED at the reverberation age ($t_{\rm beg, rev}$). The grey-shaded regions in the MWL SED plot signify the frequency ranges for which the integrated luminosity has been extracted. The colored dot-dashed lines in the same plot indicate the spin-down luminosities at considered PWN age snapshots, and share a common line color as that for the MWL SEDs.
The middle panel in both rows shows the evolution of electron spectra at the considered snapshot ages, with the same color scheme as the MWL SEDs. 
The right panel of the figure contains the time evolution of integrated luminosities in the frequency ranges of radio (red), FIR (teal), soft X-ray (orange), GeV (green), and TeV (blue). 
The time evolution of spin-down luminosity (black) is also shown for comparison and to highlight the occurrence of superefficiency.
}
\label{fig: individual}
\end{figure*}

\begin{figure*}
    \centering
    \includegraphics[width=0.33\linewidth]{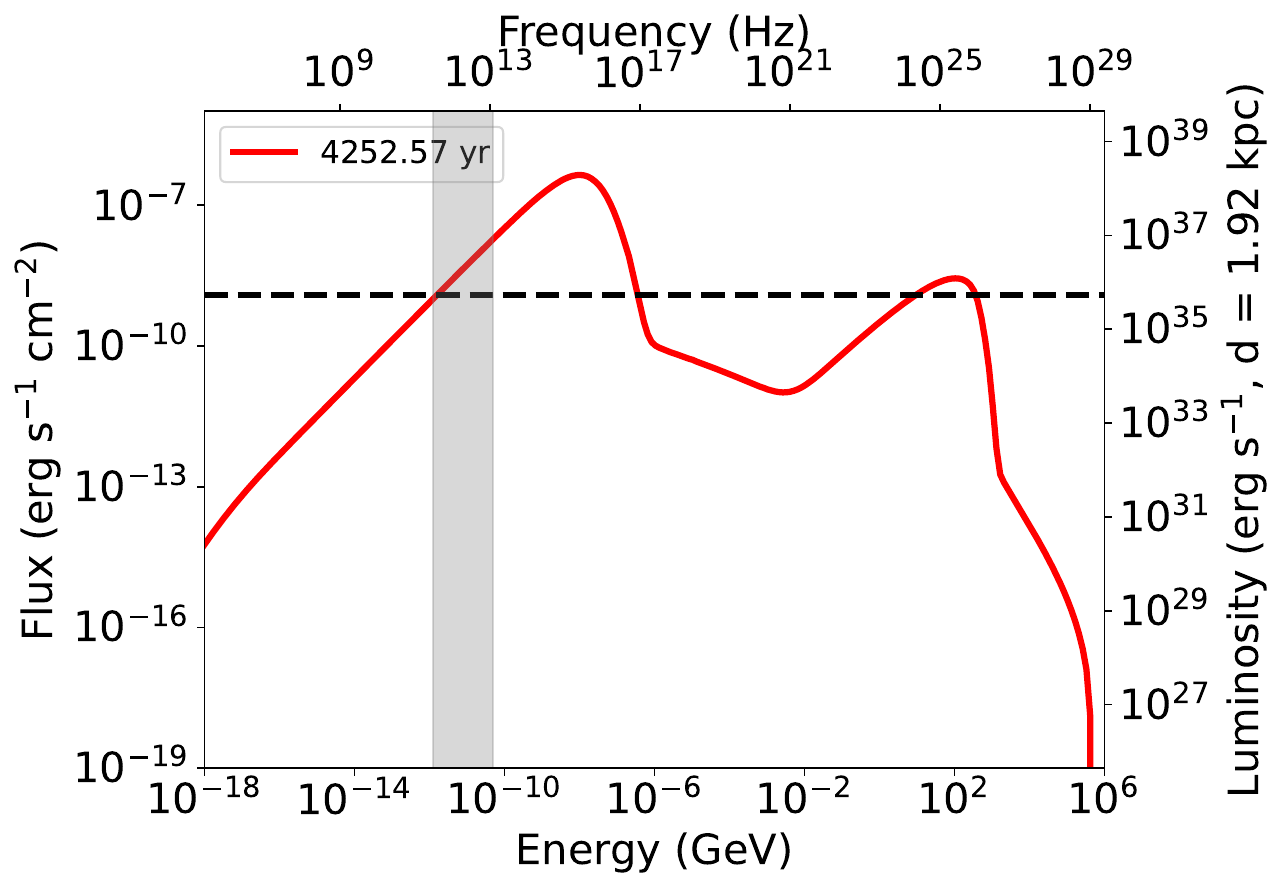}
    \includegraphics[width=0.31\linewidth]{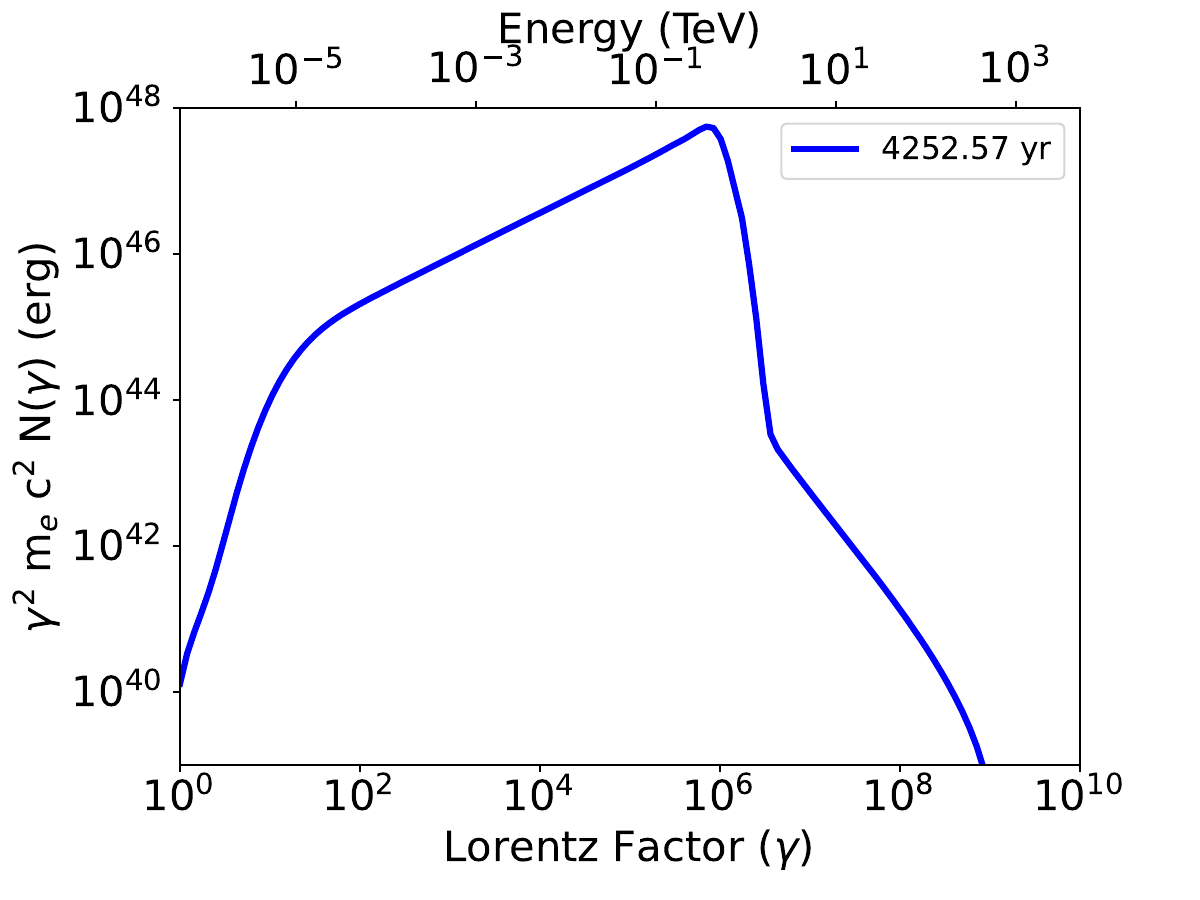}
    \includegraphics[width=0.31\linewidth]{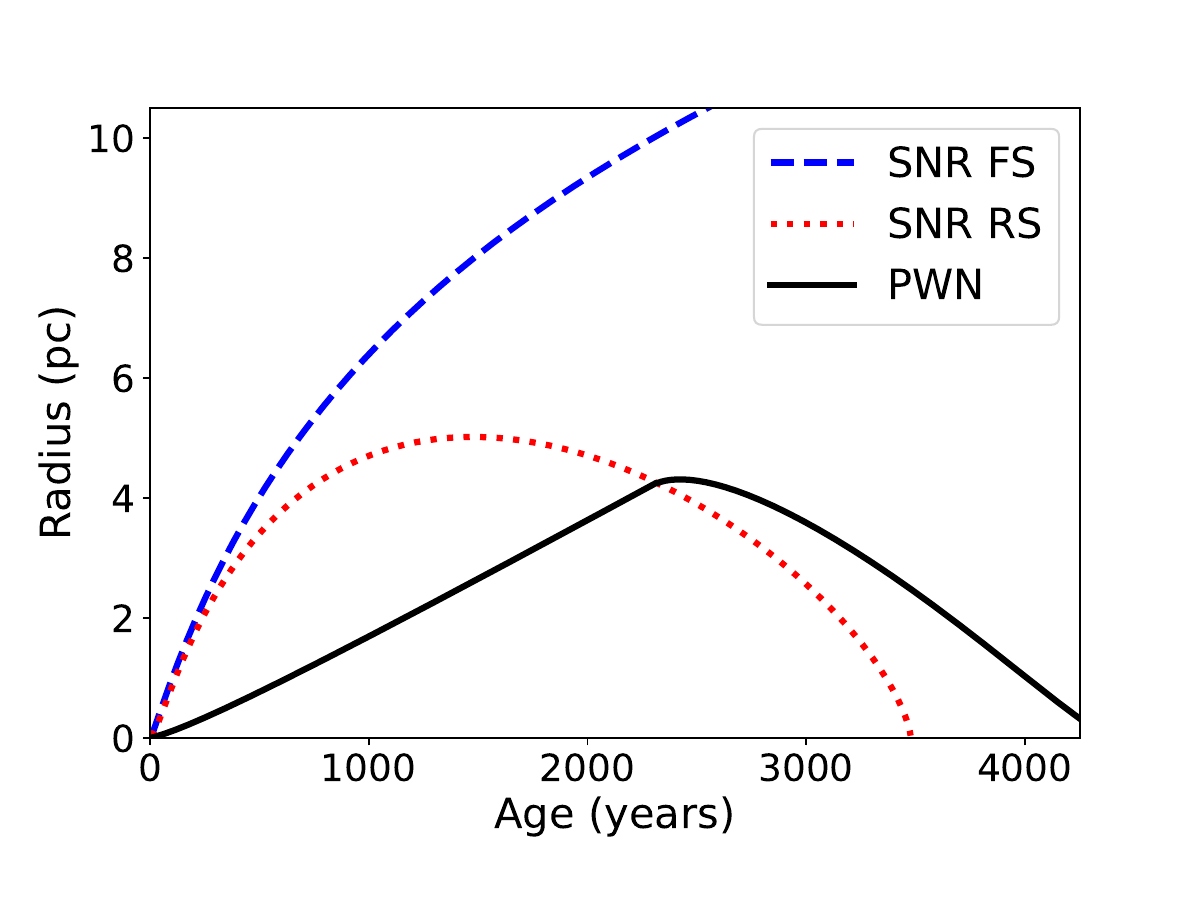}
    \includegraphics[width=0.33\linewidth]{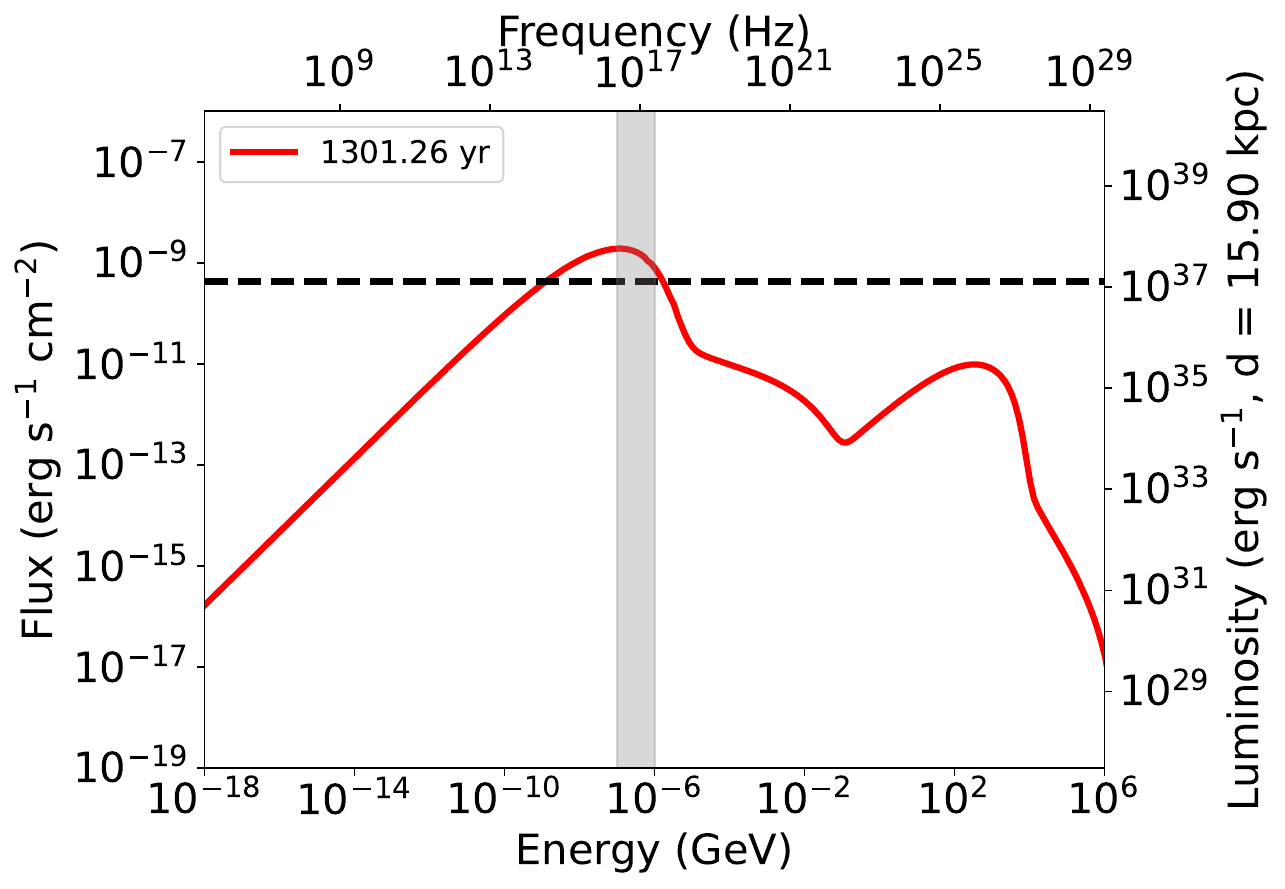}
    \includegraphics[width=0.31\linewidth]{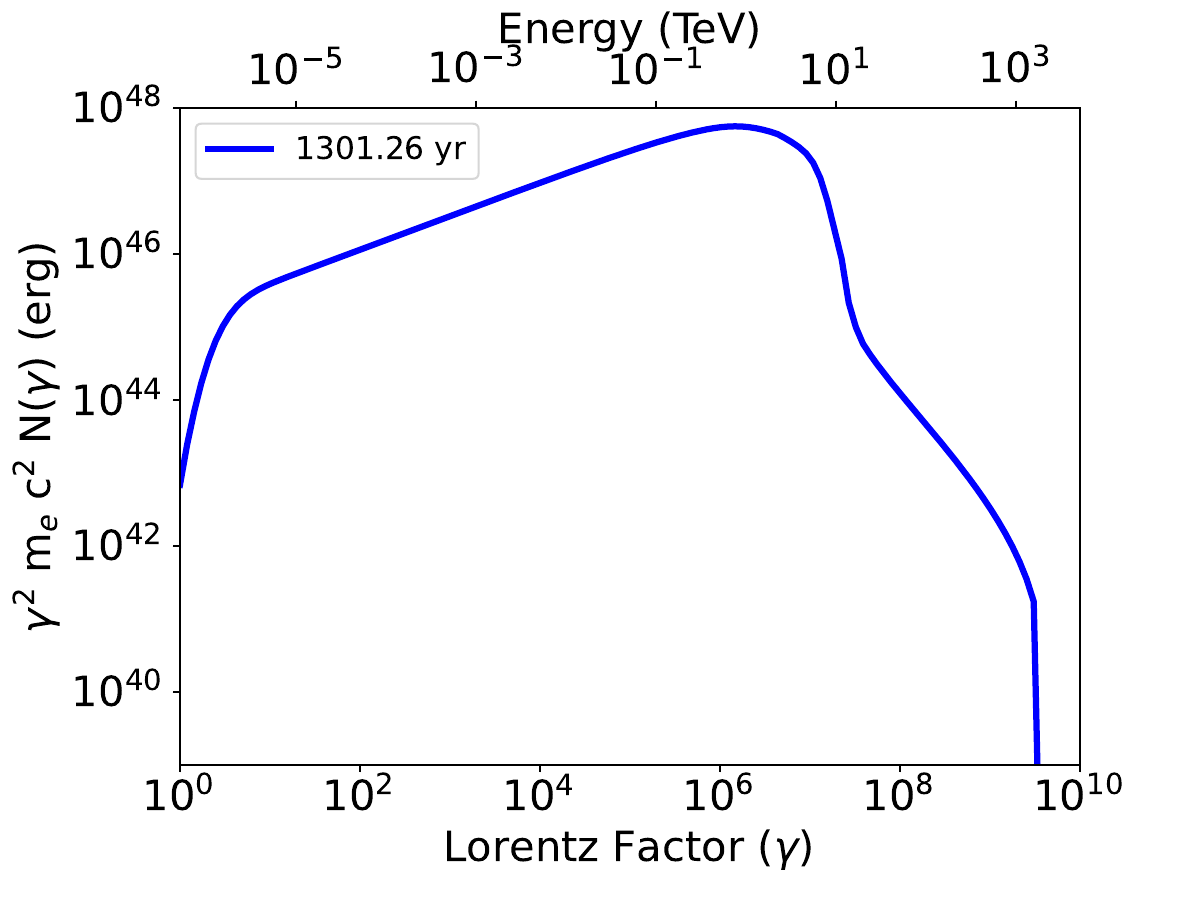}
    \includegraphics[width=0.31\linewidth]{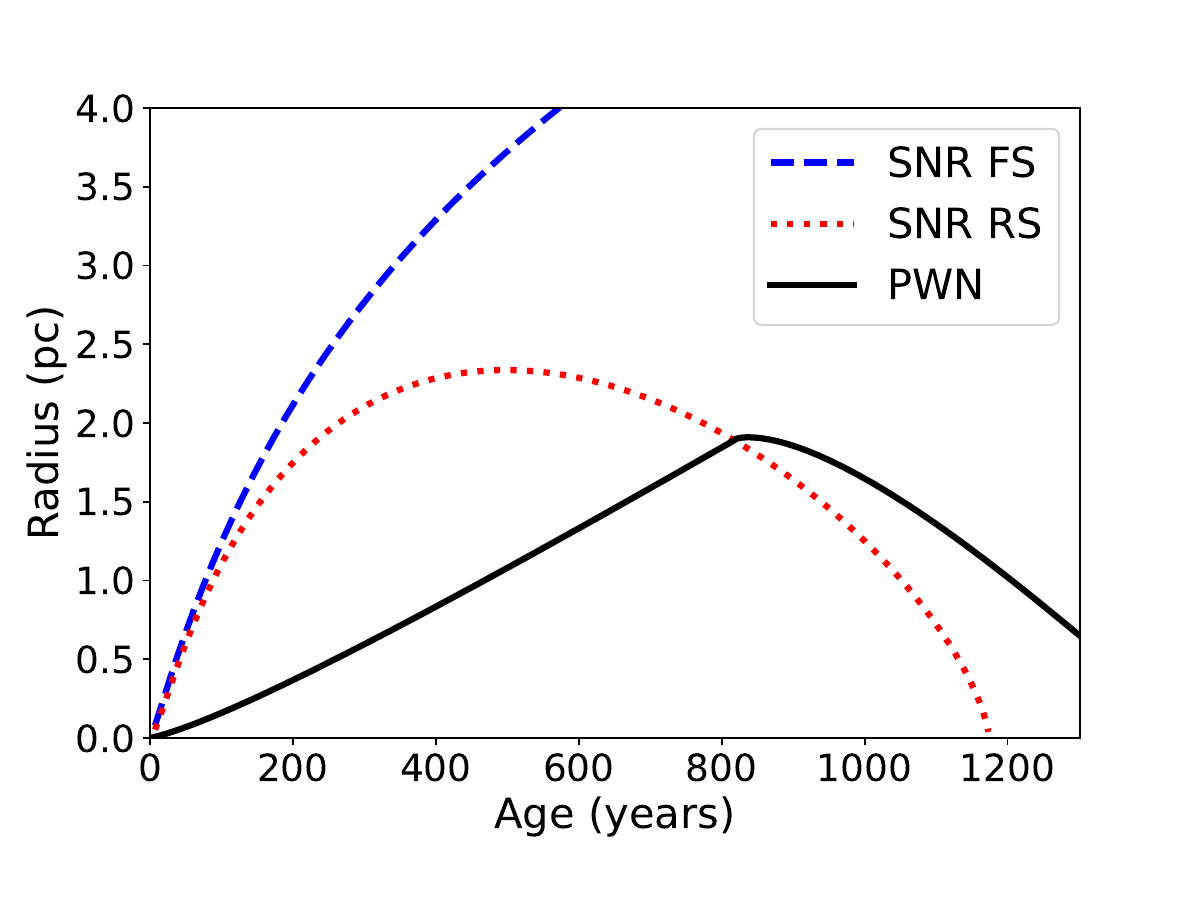}
    \includegraphics[width=0.33\linewidth]{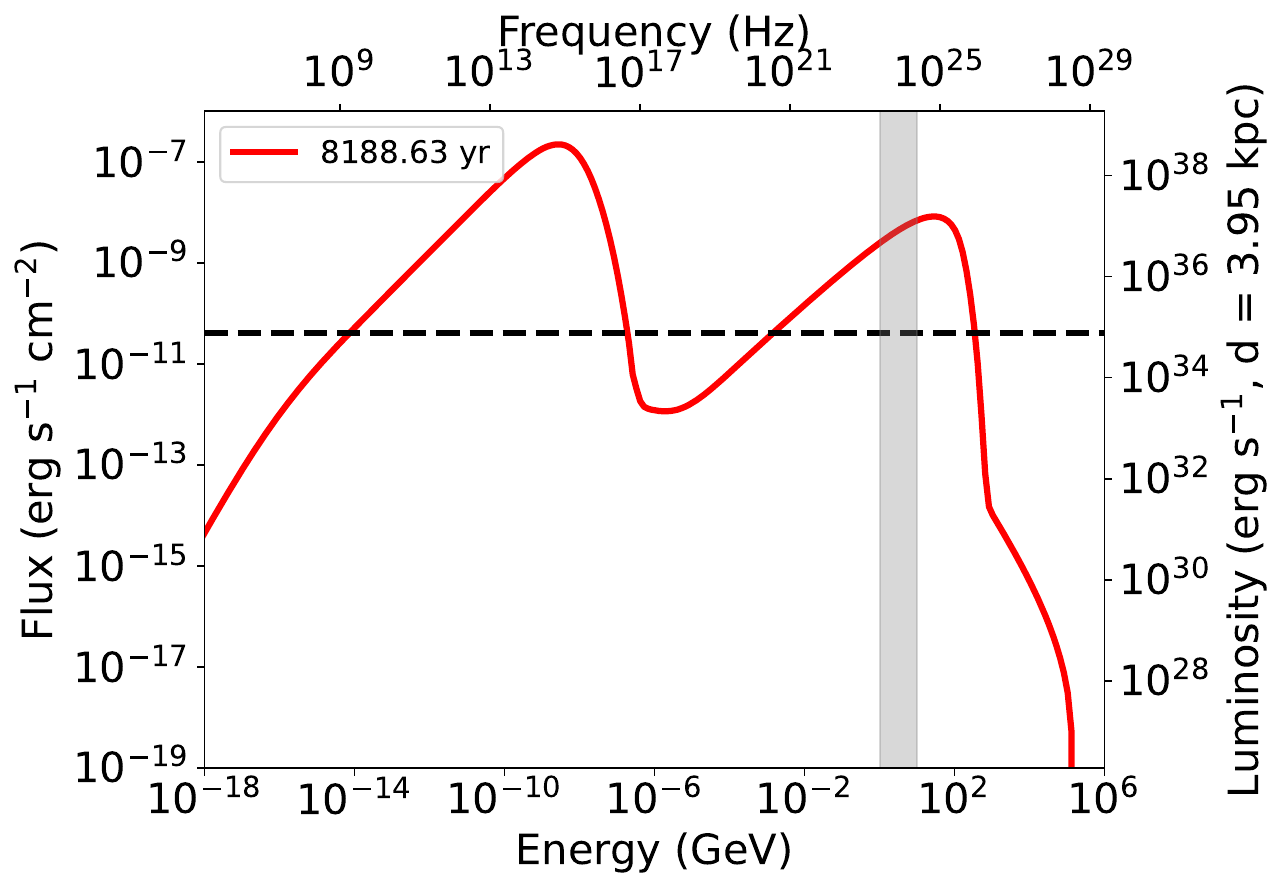}
    \includegraphics[width=0.31\linewidth]{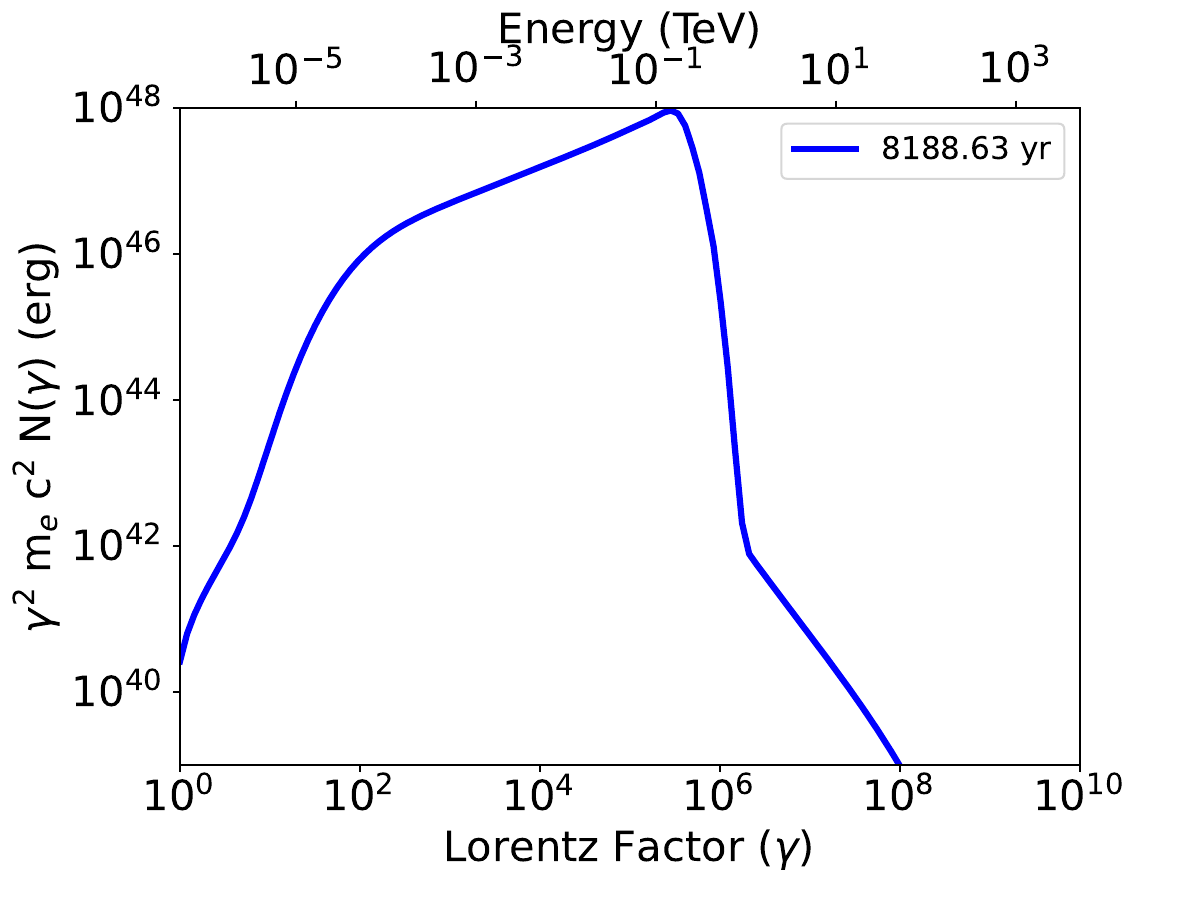}
    \includegraphics[width=0.31\linewidth]{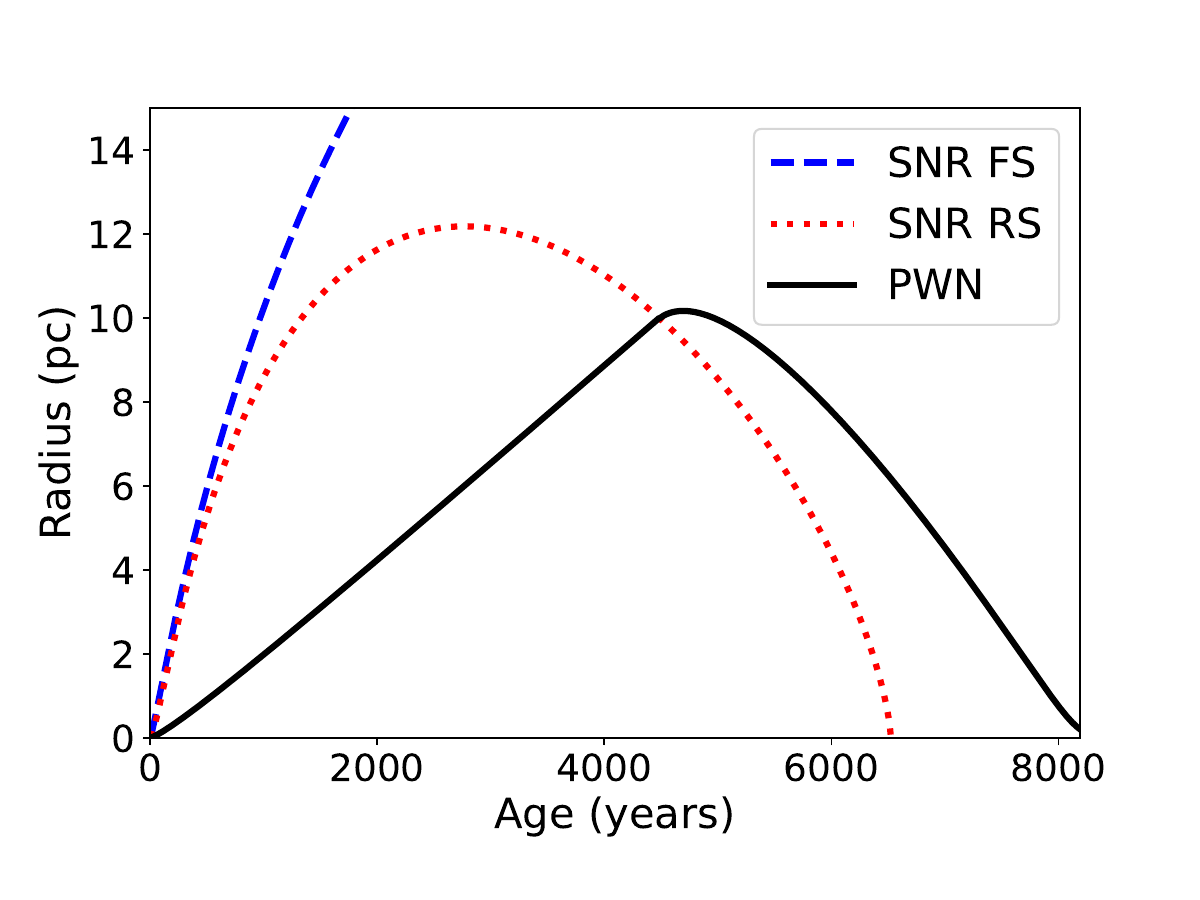}

    \caption{Representative examples of three individual sources that exhibit superefficiency at the current age in distinct energy bands: FIR (top panel), soft X-ray (middle panel), and GeV $\gamma$-rays (bottom panel). Each source is selected from the compressing phase of evolution to highlight the physical conditions driving enhanced radiative output. For each case, we show the SEDs (left panel), electron spectra (middle panel), and radii evolution (including the PWN and SNR) (right panel). The current age of the system is also provided in the plot. The grey shaded regions in the MWL SED plots signify the frequency bands where superefficiency was obtained, and the black dashed lines in the same plots represent the spin-down luminosity at the current age. 
}
    \label{fig: example_super}
\end{figure*}

\subsection{Estimation of superefficiency as a function of time}

To investigate the temporal evolution of superefficiency at the population level, we analyze the simulated PWN population at successive evolutionary snapshots. 
Considering a single population realization of 1600 sources, for each selected age in the time series, we generate the corresponding distributions of evolutionary phases and identify the subset of sources that satisfy the superefficiency criterion in each frequency band. 
These visualizations are constructed under the simplifying assumption that all sources are observed at the same evolutionary age, thereby isolating the intrinsic time dependence of superefficiency within the population.
Although this fixed-age assumption does not reflect the true age distribution of Galactic PWNe, which spans several orders of magnitude, it provides a controlled framework for assessing how the properties of the PWN population evolve with time and for quantifying how many sources would be superefficient at specific evolutionary stages. 
In particular, this approach allows us to disentangle evolutionary effects from population-mixing effects and to track the onset and eventual decline of superefficiency as a function of age. 
The resulting time-resolved pie charts and bar plots are shown in Fig. \ref{fig: pie_super_time_series}.

A clear trend emerging from these time series visualizations is that at sufficiently late times, nearly all superefficient events occur in the post-compression phase of PWN evolution. 
Even at relatively young ages, particularly in the first few snapshots shown in Fig. \ref{fig: pie_super_time_series}, a substantial number of superefficient sources are present across a broad range of frequencies once the first compression has ended. 
This highlights the importance of the initial compression episode in setting the conditions for superefficiency and demonstrates that enhanced radiative efficiency can persist well beyond the brief interval of active compression, especially at low frequencies.

We further find that the total number of superefficient sources at any given time is not constant over the course of the population's evolution. 
Instead, it varies systematically with age and energy band, exhibiting a pronounced increase when the population collectively enters the reverberation phase. 
This increase is most prominent during the first compression, when rapid magnetic field amplification and adiabatic particle reprocessing act in concert to boost the radiative output of a large fraction of systems. 
At later times and into the post-compression phase, the number of superefficient sources in intermediate-frequency bands (X-ray) gradually declines as magnetic field weakens and radiative losses deplete high-energy particles, while the low-frequency (FIR) superefficient population grows as higher-energy particles cool into this regime and accumulate.
To highlight the distribution of superefficient PWNe across different frequency bands as a function of time, Table \ref{tab:supereff_matrix_age} presents the matrix of superefficient source counts between frequency bands for four different ages. Each entry gives the number of sources that are simultaneously superefficient in the corresponding pair of bands for a given age.

To quantify these trends more explicitly, we present in Fig. \ref{fig: sup_percentage} the time evolution of the percentage of superefficient sources in each energy band. 
This representation illustrates how the relative prevalence of superefficiency depends on frequency and evolves as the population ages. 
Low-frequency bands show a continuous increase and a more extended period of elevated superefficiency fractions, reflecting the long lifetimes of the corresponding electron populations, whereas intermediate-frequency bands display comparatively sharper, more transient peaks tied closely to the timing of compression, following a gradual decline in number.
Additionally, higher frequency bands show a gradual increase in superefficiency deep into the post-compression phase.
Together, these results demonstrate that superefficiency is an intrinsically time-dependent phenomenon whose observational manifestation is strongly regulated by both evolutionary stage and photon energy.
Fig. \ref{fig: sup_percentage} also shows the time evolution of the fraction of sources 
that are exclusively superefficient at a given frequency band. 
By construction, sources contributing to a given curve in the plot are superefficient in that band while not being superefficient in any other frequency at the same epoch.

To study the superefficiency connection between three representative frequency bands, i.e., FIR, soft X-ray, and GeV, 
We 
analyzed the number of sources that are simultaneously superefficient in each pair of frequencies at different snapshot ages (see Table \ref{tab:supereff_matrix_age}).
%
We find a strong association between FIR and GeV superefficiency. Both bands are linked to comparatively long-lived lower-energy particles; the particles responsible for GeV IC emission can also radiate synchrotron emission in the infrared range for typical PWN magnetic fields, depending on the target photon field and magnetic field strength.
Therefore, systems that retain a large reservoir of such particles are more likely to appear superefficient in both FIR and GeV bands.
In contrast, the connection between soft-X-ray superefficiency and either FIR or GeV superefficiency is weaker. Soft-X-ray superefficiency is more closely tied to the transient compression phase, during which  magnetic field amplification and the availability of high-energy particles can briefly enhance the synchrotron luminosity. It is therefore more sensitive to the instantaneous dynamical state, the compression factor, and the survival of recently injected or less severely cooled high-energy particles, rather than to the longer-lived particle population that mainly controls the FIR and GeV trends.

Additionally, we show in Fig. \ref{fig: L0_tau0_age} how the locations of the superefficient sources evolve with the snapshot ages in the characteristic $\log_{10}(L_0/L_{\rm ch})$-$\log_{10}(\tau_0/t_{\rm ch})$ plane, for the same three representative frequency bands. 
A clear trend is observed for all three frequencies: at younger ages, the superefficient sources preferentially occupy the region of comparatively lower $L_0/L_{\rm ch}$ and higher $\tau_0/t_{\rm ch}$, whereas, with increasing age, an increasing number of superefficient systems also appears toward higher $L_0/L_{\rm ch}$ and lower $\tau_0/t_{\rm ch}$. 
This behavior reflects two complementary ways of achieving $L_{\rm band}/\dot{E}_{\rm SD}>1$. 
At early times, superefficiency is favored when the present spin-down luminosity is not too large and the injection timescale is sufficiently long, so that the nebula can accumulate a radiating particle population while being compared against a relatively moderate $\dot{E}_{\rm SD}$. 
At later times, however, systems born with larger $L_0/L_{\rm ch}$ and shorter $\tau_0/t_{\rm ch}$ have already injected a substantial amount of rotational energy early in their evolution, while their current spin-down power has declined significantly; the accumulated particle population can therefore remain luminous enough to exceed the contemporaneous $\dot{E}_{\rm SD}$. 
The original low-$L_0/L_{\rm ch}$, high-$\tau_0/t_{\rm ch}$ region does not totally disappear, because long-lasting injection can still maintain superefficiency in some systems, but the aging population increasingly admits sources from the high-$L_0/L_{\rm ch}$, low-$\tau_0/t_{\rm ch}$ region where the contrast between past injected energy and present spin-down power becomes larger.

This time-resolved analysis provides the evolutionary context for the current epoch results discussed in the previous section. 
Present-day distribution of superefficient PWNe reflects the cumulative outcome of reverberation-driven compression and post-compression evolution.
Low-frequency superefficiency is persisting long after the first compression and comparatively higher-frequency superefficiency primarily tracing the most recent or ongoing reverberation episodes. 
%

\subsection{Individual examples of PWNe based on compression factor}

To illustrate how the strength of reverse shock compression regulates both the onset and the spectral character of superefficiency, we compare two representative PWNe selected from the population, as was chosen in \cite{desarkar26}: Example 1, which undergoes strong compression with a compression factor ${\rm CF}=50.3$, and Example 2, which experiences only mild compression with ${\rm CF}=5.8$. 
Their SEDs, electron spectra, and time-dependent evolution of integrated luminosities are shown in Fig. \ref{fig: individual}. 
%

For the strongly compressed system (Example 1), the SEDs show an enhancement of emission as the nebula passes through its first compression. 
Above $\sim5\times10^{3}$ yr and during the first compression, the synchrotron component rises in the X-ray band, though it does not go above the spin-down luminosity at that epoch to achieve superefficiency.
The FIR component also shows an increase during the first compression, and continues to increase to achieve superefficiency at around $10^4$ yr. 
The IC component in the GeV range is also simultaneously amplified at the same time, and shows marginal superefficiency around the same epoch as FIR. 
This behavior is directly reflected in the luminosity evolution, where the X-ray luminosity increases rapidly but does not exceed the contemporaneous pulsar spin-down power. 
On the other hand, a very significant superefficiency is observed in the FIR band, while a marginal superefficiency is achieved in the GeV $\gamma$-ray band.
In this case, a larger compression factor produces a comparatively strong amplification of the magnetic field and substantial adiabatic heating of the particle population, so that electrons radiate efficiently. 
The combination of increasing particle energy density and enhanced magnetic field therefore drives the system into a state of superefficiency, primarily in the FIR band.
The electron spectra further reveal how this occurs. 
During active compression, the entire distribution is boosted in normalization for the X-ray emitting electrons. 
This leads to electrons producing elevated synchrotron emission in the X-ray band; however, it is not enough to achieve superefficiency, most likely because the magnetic field amplification (and the pre-reverberation magnetic field) is not strong enough, even after the first compression.
As the system evolves, the X-ray emitting electrons cool down to lower energies and accumulate, thus any possibility of superefficiency in the X-ray subsides.
Additionally, given that the higher energy particles lose energy and accumulate at lower energy levels, the superefficiency is observed at the FIR band at comparatively late times. 
Similarly, the accumulated relic particles radiate in GeV band via IC upscattering, therefore producing mild superefficiency in the GeV band during post-compression.

In contrast, the mildly compressed system Example 2 exhibits a much delayed radiative response to reverberation. 
Its SEDs display brightening well beyond the first compression, and a mild superefficiency is observed above $2 \times 10^4$ yrs, primarily in the FIR band. 
The luminosity evolution confirms that any episodes of superefficiency are confined to low photon energies (i.e., FIR) and at later times, with the radiative output in most bands remaining below the spin-down power. 
The weaker compression in this case is insufficient to boost synchrotron luminosity and to counteract radiative cooling at high energies.
This behavior is mirrored in the electron spectra as well. 
The overall normalization of the particle distribution is far less enhanced and delayed than in the strongly compressed case, while the high-energy electrons cool rapidly and accumulate in the lower energies, thus suppressing the potential for X-ray superefficiency, while the lower-energy accumulated electrons persist long enough to produce FIR superefficiency at a delayed time, well into the post-compression phase.
These two examples further demonstrate the evolutionary-stage-dependent population-level trends found in previous sections: FIR superefficiency can occur over a wide range of compression strengths, is not dependent on active compression and can manifest deep into the post-compression phase as it is dependent primarily on low-energy electron accumulation.
Instead, X-ray superefficiency requires strong and active compression (and also a higher level of pre-reverberation magnetic field) and is therefore comparatively rare. 

\subsection{Individual examples of PWNe based on superefficiency class}

To further illustrate the diversity of superefficient behaviour and its spectral imprint across the electromagnetic spectrum, we present in Fig. \ref{fig: example_super} three representative case studies selected from our population. 
Each source chosen is superefficient at the current epoch in distinct frequency bands-FIR, soft X-rays, and GeV $\gamma$-rays-thereby sampling different regions of the non-thermal SED and highlighting the variety of physical conditions under which superefficiency can arise. 
To emphasize the physical mechanisms responsible for superefficiency, we deliberately restrict this sample to sources that are currently undergoing their first and typically strongest compression episode, in order to isolate the direct impact of compression on the observed emission and avoid conflating it with longer-term post-compression effects.

For each of the three selected PWNe, we show a set of multi-panel plots that include the SEDs, the corresponding electron energy spectrum, and the temporal evolution of the relevant radii, including both the PWN boundary and the surrounding SNR structure. 
The SEDs illustrate how superefficiency manifests differently depending on the characteristic energies of the emitting particles, while the electron spectra provide insight into the balance between adiabatic energization and radiative cooling. 
The radius evolution panels, in turn, place these radiative signatures in their dynamical context by explicitly showing the ongoing contraction associated with reverberation.

A key distinction between the different classes of superefficient PWNe emerges in the shape and normalization of the underlying electron spectrum. 
FIR-superefficient systems are dominated by a strongly enhanced normalization at low electron energies, with little or no high-energy tail, reflecting the accumulation of long-lived, cooled particles that efficiently radiate at FIR frequencies. 
In contrast, X-ray superefficient PWNe require a pronounced high-energy tail and usually a normalization peak at comparatively higher energies, as synchrotron X-rays are produced only by freshly energized, rapidly cooling electrons during the peak of compression. 
GeV-superefficient sources occupy a more similar regime as the FIR case: since their emission is dominated by IC scattering, a strong enhancement in the population of GeV-emitting electrons is sufficient to produce superefficiency, even in the absence of an extended high-energy tail. 
These spectral differences provide a direct physical link between the observed superefficiency class and the stage and intensity of reverberation experienced by each system.

\subsection{Ensemble emission and particle spectra of superefficient PWNe across frequencies at a particular evolutionary epoch}

\begin{figure*}
    \centering
    \includegraphics[width=0.49\linewidth]{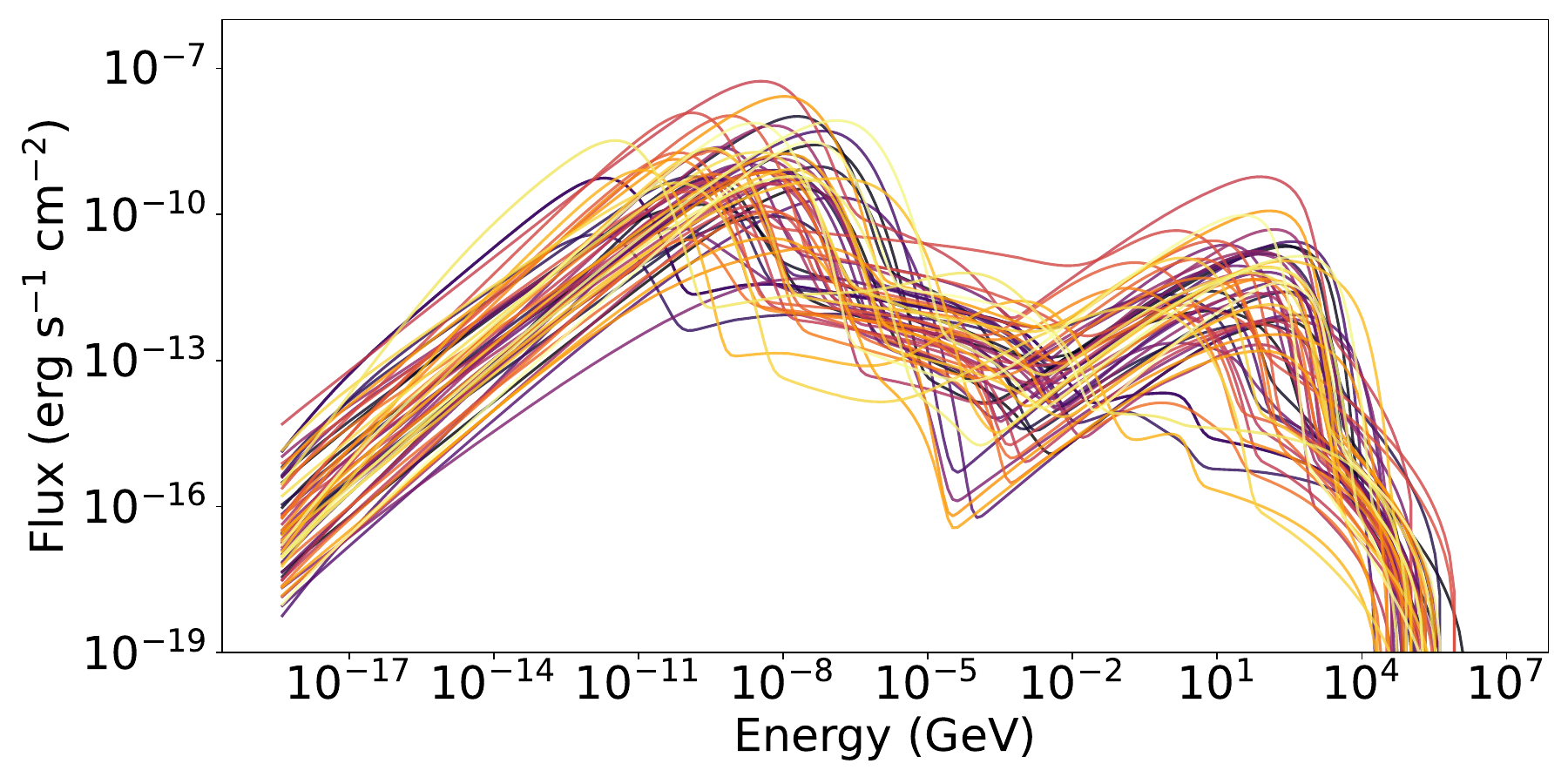}
    \includegraphics[width=0.49\linewidth]{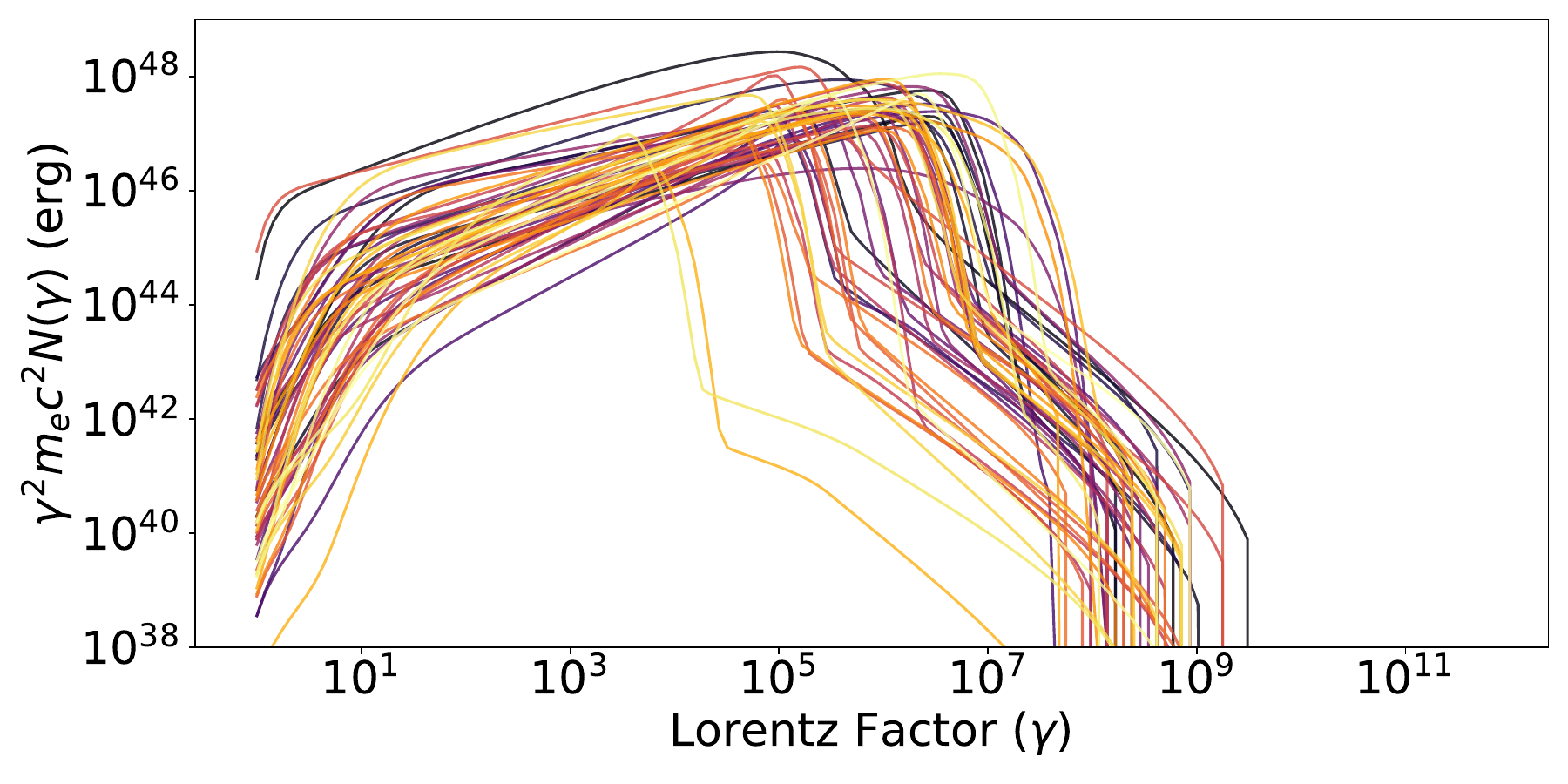}\\
    \vspace{0.15cm}
    \includegraphics[width=0.49\linewidth]{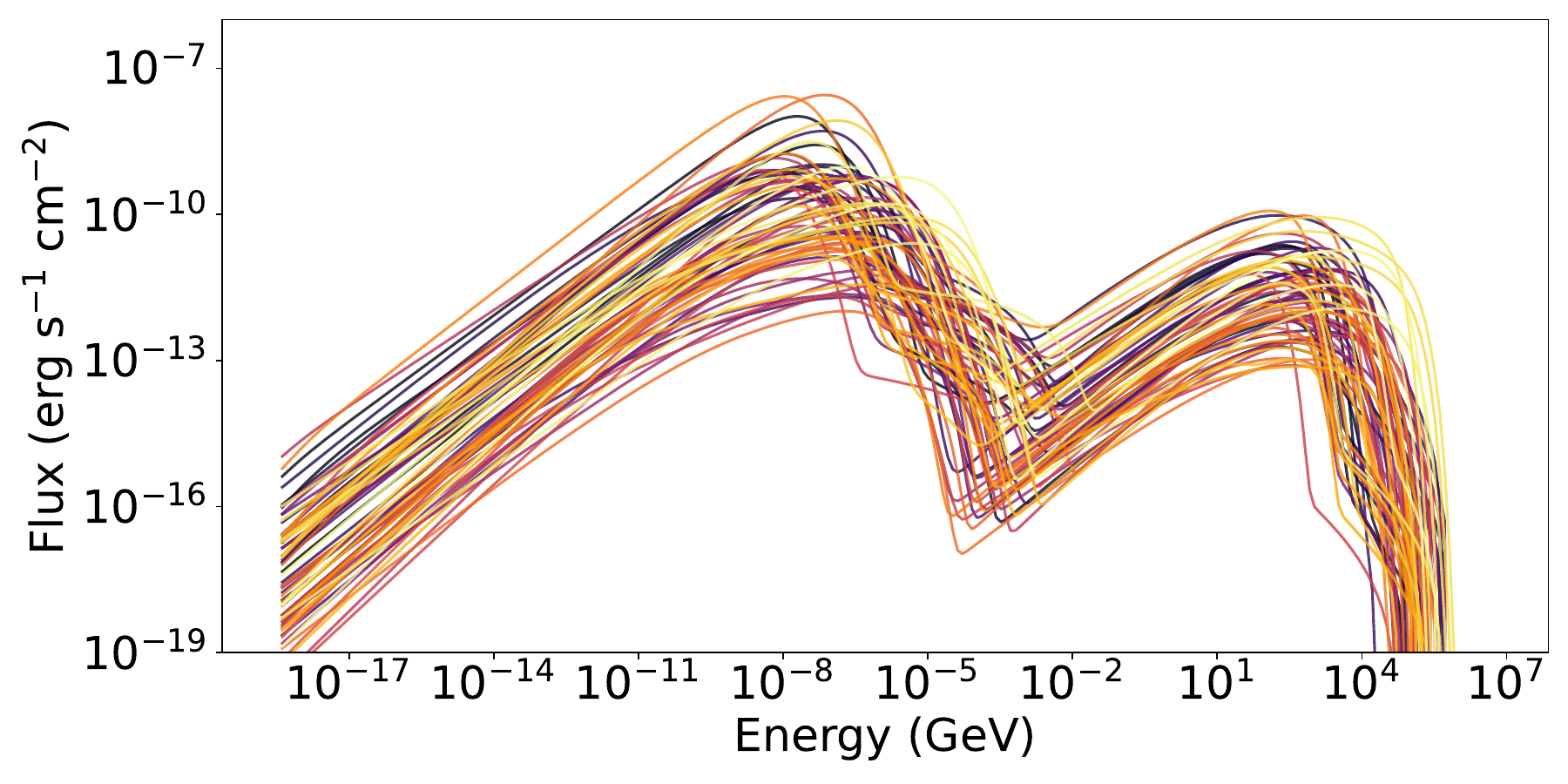}
    \includegraphics[width=0.49\linewidth]{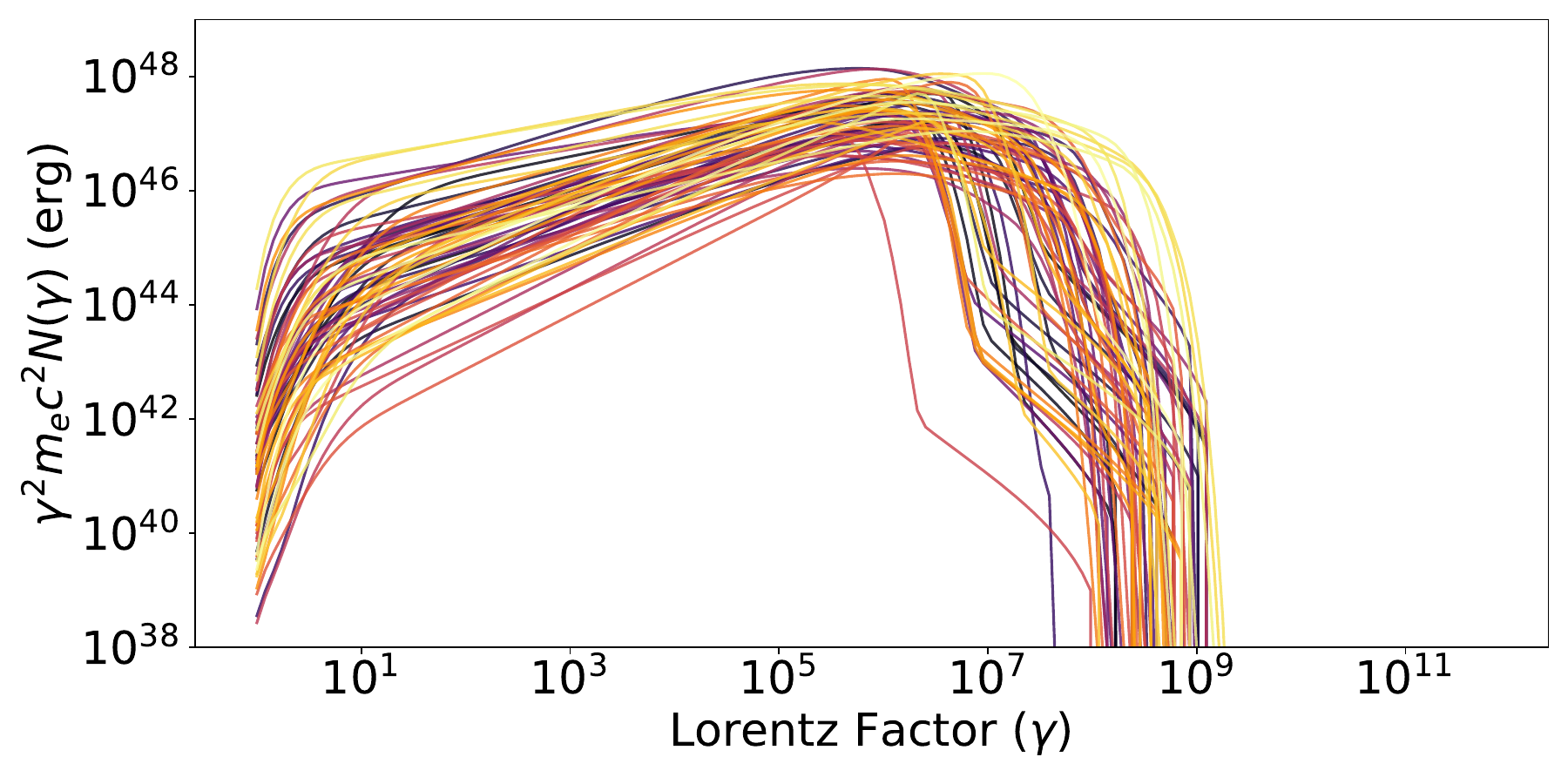}\\
    \vspace{0.15cm}
    \includegraphics[width=0.49\linewidth]{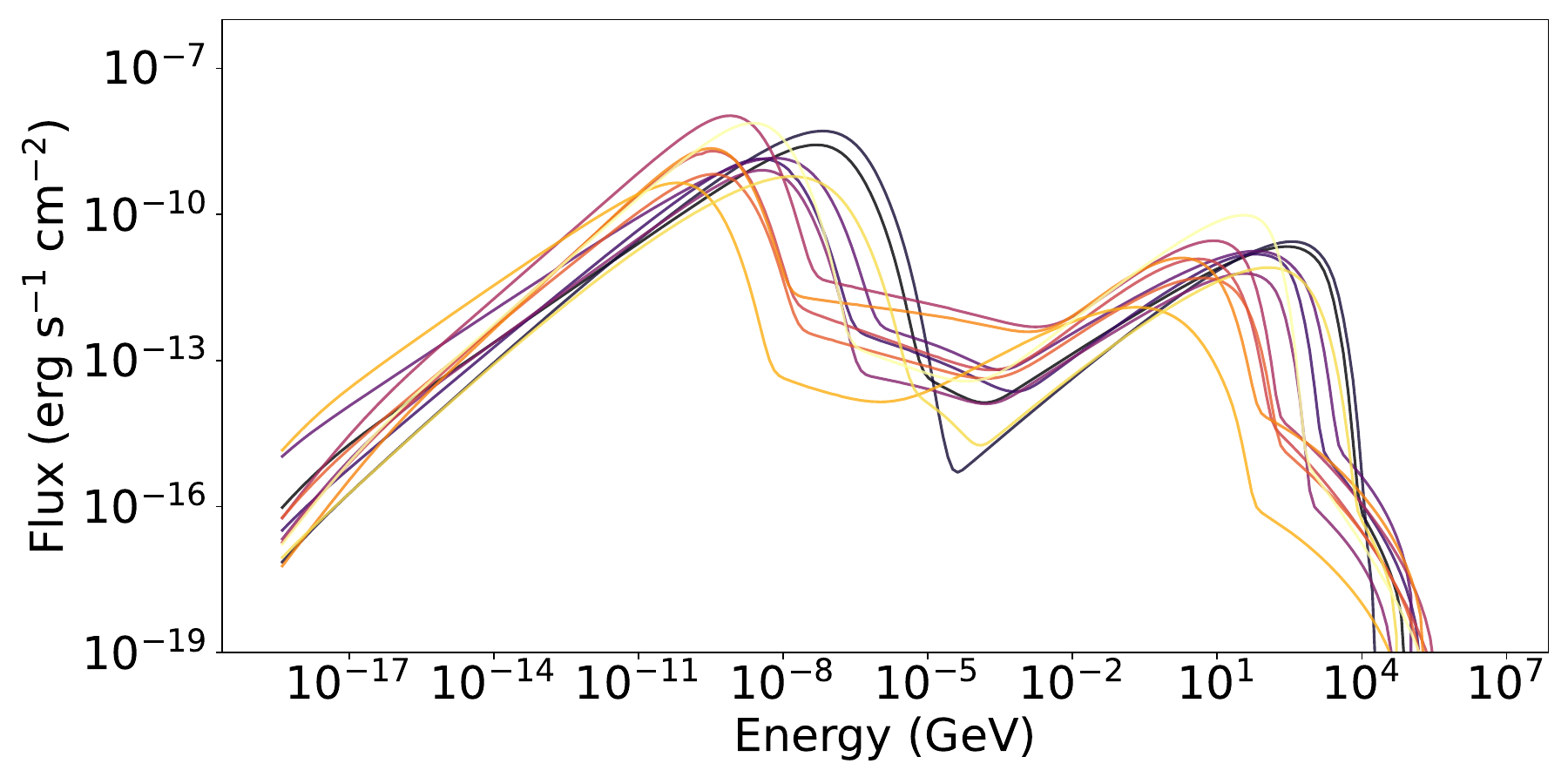}
    \includegraphics[width=0.49\linewidth]{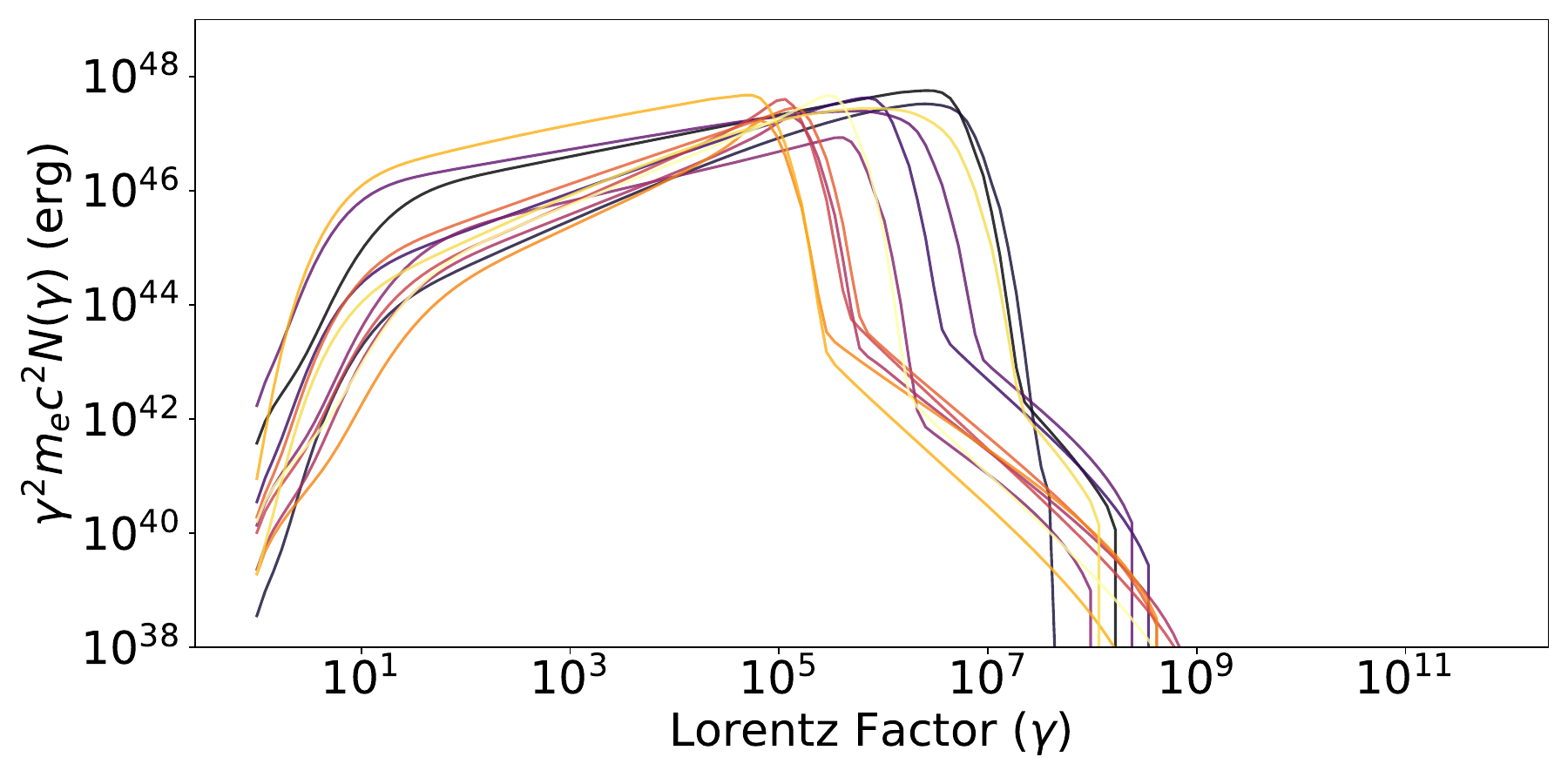}
    \caption{Grouped spectral energy distributions (left column) and corresponding electron spectra (right column) for PWNe exhibiting superefficiency during the compression phase in a single random realization of 1600 sources. The panels show sources selected for superefficiency in the FIR (top row), soft X-ray (middle row), and GeV $\gamma$-ray (bottom row) bands at 7000 years. Note that a continuous colormap is used only to distinguish the different curves: each color corresponds to an individual PWN and has no additional physical meaning. }
    \label{fig: grouped_super_5000}
\end{figure*}

In Fig. \ref{fig: grouped_super_5000}, we examine the collective radiative and particle properties of PWNe that exhibit superefficiency at a fixed evolutionary age of 7000 years. 
This epoch is chosen because it corresponds to the phase where the number of PWNe undergoing active compression is maximal in the population. 
We further restrict the sample to PWNe that are simultaneously superefficient in a given band and in the compressing phase. 
This selection is physically motivated, as during active compression, the adiabatic heating and magnetic field amplification are at their peak, making the reverberation-driven superefficiency both strong and physically meaningful.
We aim to ensure that the ensemble spectral properties capture the intrinsic imprint of reverberation rather than evolutionary fading.

We consider three representative energy bands-FIR, X-rays, and GeV $\gamma$-rays. 
For each band, we construct grouped SEDs and electron spectra by combining all compressing PWNe that satisfy the superefficiency criterion in that band at 7000 yr. 
A robust pattern emerges in this ensemble-based approach. 
Sources that are superefficient in a given energy band exhibit highly similar SED shapes and electron spectra, demonstrating that superefficiency in each band is associated with a characteristic electron population and cooling regime. 
FIR-superefficient PWNe are dominated by large populations of low-energy electrons with long radiative lifetimes, which accumulate efficiently and produce enhanced synchrotron emission once the magnetic field is amplified. 
Their grouped electron spectra peak at relatively low Lorentz factors with comparatively modest extension to higher energies, and their SEDs show strong FIR bumps.

X-ray superefficient sources, in contrast, show much harder and more sharply peaked electron spectra extending to higher Lorentz factors. 
Their grouped SEDs are characterized by intense synchrotron emission in the X-ray band, reflecting the fact that X-ray superefficiency requires a transient population of freshly injected, ultra-relativistic electrons radiating in strongly amplified magnetic fields. 
Because these particles cool rapidly in sufficiently enhanced magnetic fields, the X-ray superefficiency should be confined to a short time window during active compression, explaining why it is generally rarer in post-compression stage.

GeV-superefficient PWNe occupy an intermediate regime. Their grouped electron spectra peak at moderately high energies, but are less extreme than those of X-ray superefficient sources. 
The grouped SEDs show strong IC components in the GeV band, produced by electrons that are energetic enough to upscatter ambient photon fields but not so energetic that synchrotron losses immediately quench their emission. 
This allows GeV superefficiency to persist for longer than X-ray superefficiency and to extend beyond the peak of compression and into the post-compression phase.

\subsection{Ensemble emission and particle spectra of FIR superefficient PWNe across CF ranges}

\begin{figure*}
    \centering
    \includegraphics[width=0.49\linewidth]{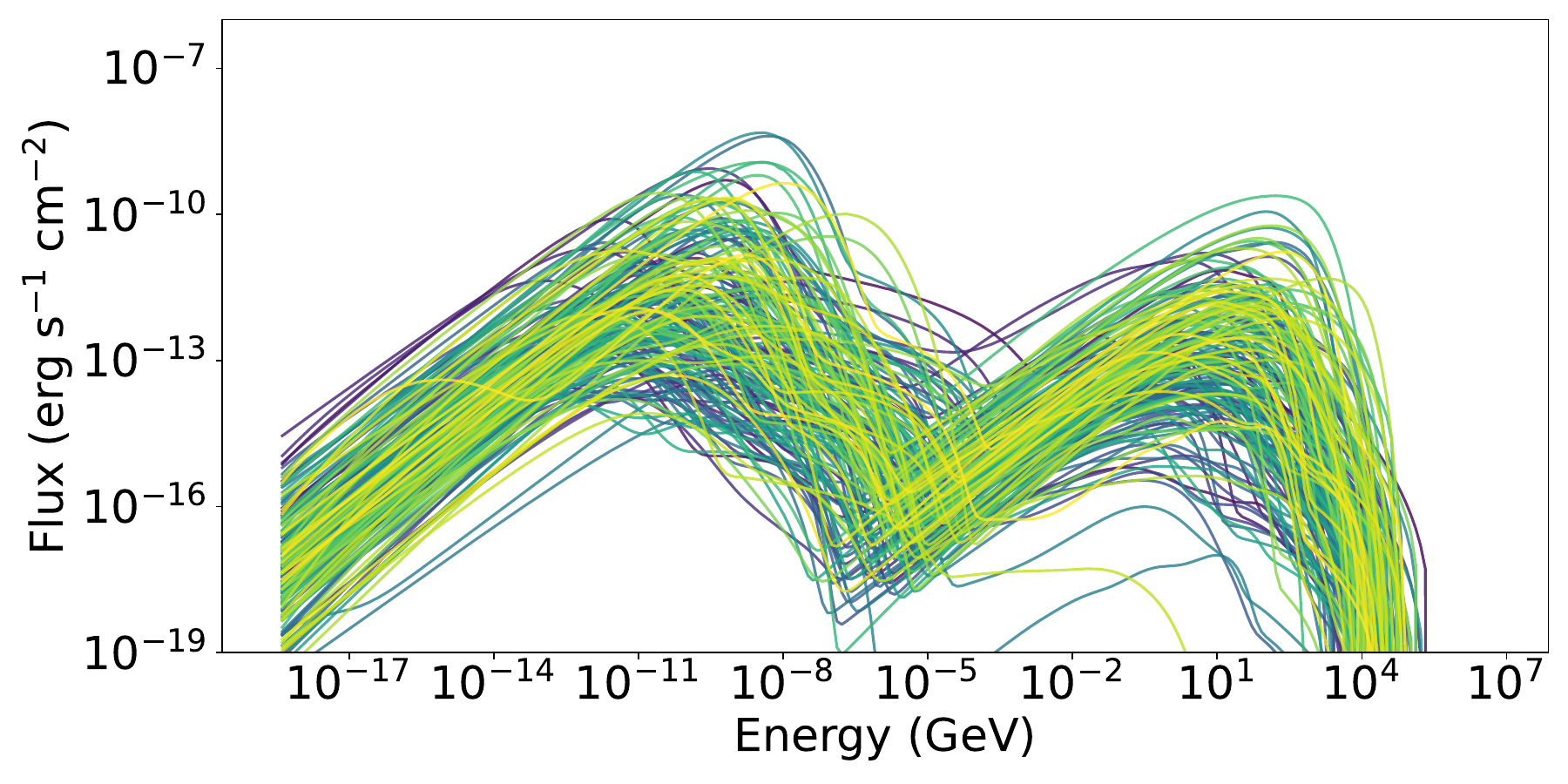}
    \includegraphics[width=0.49\linewidth]{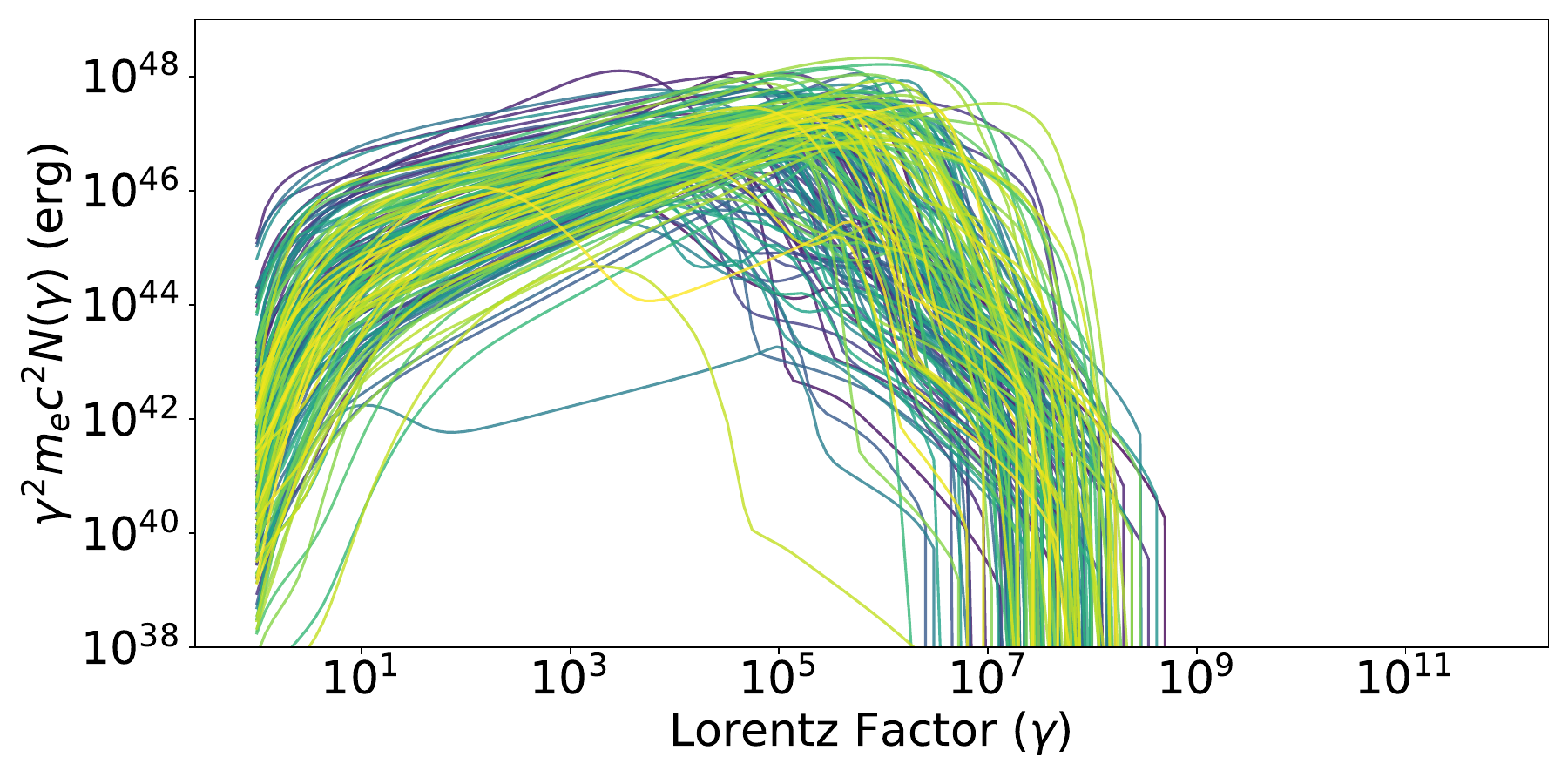} \\
    \includegraphics[width=0.49\linewidth]{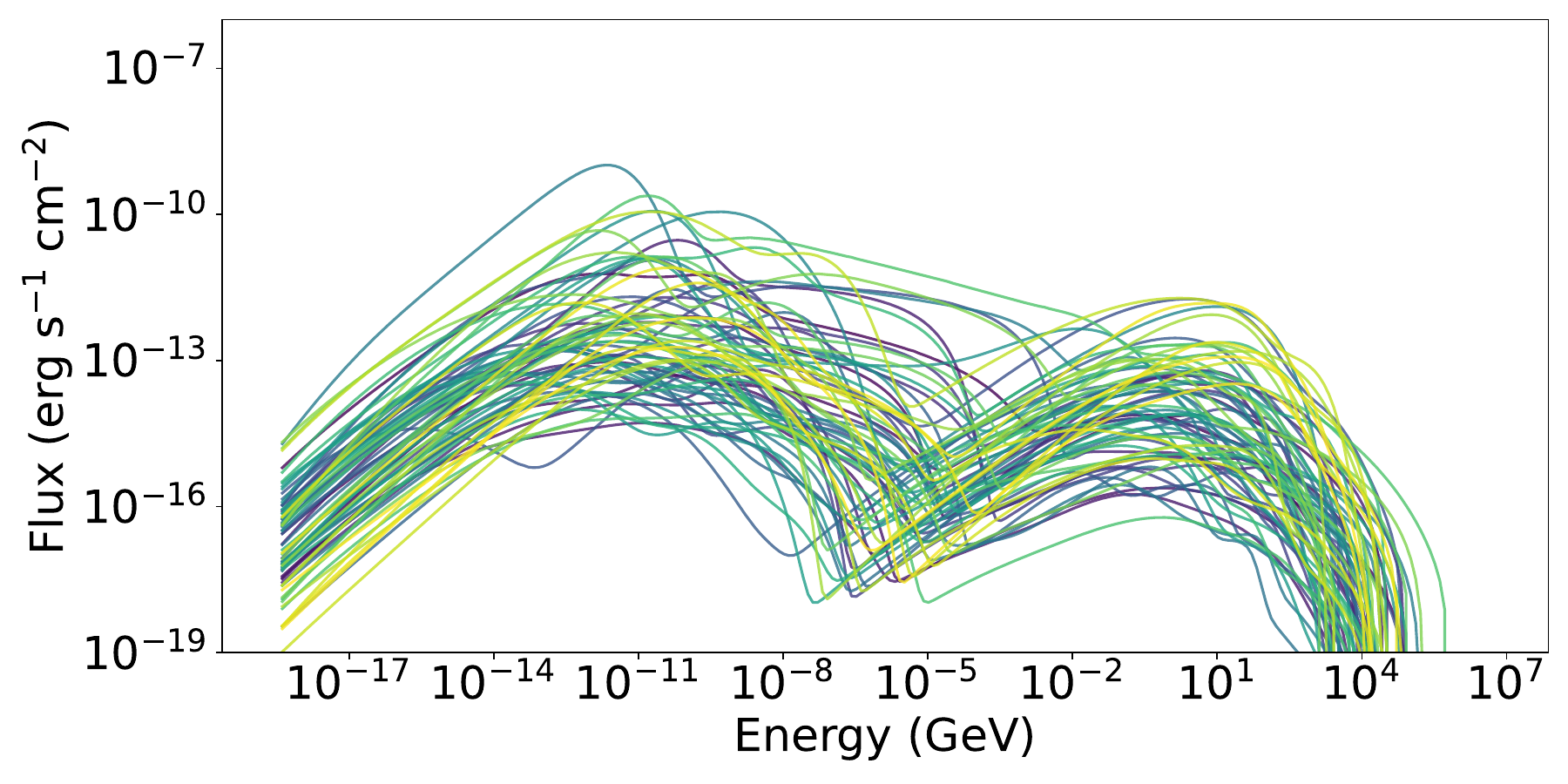}
    \includegraphics[width=0.49\linewidth]{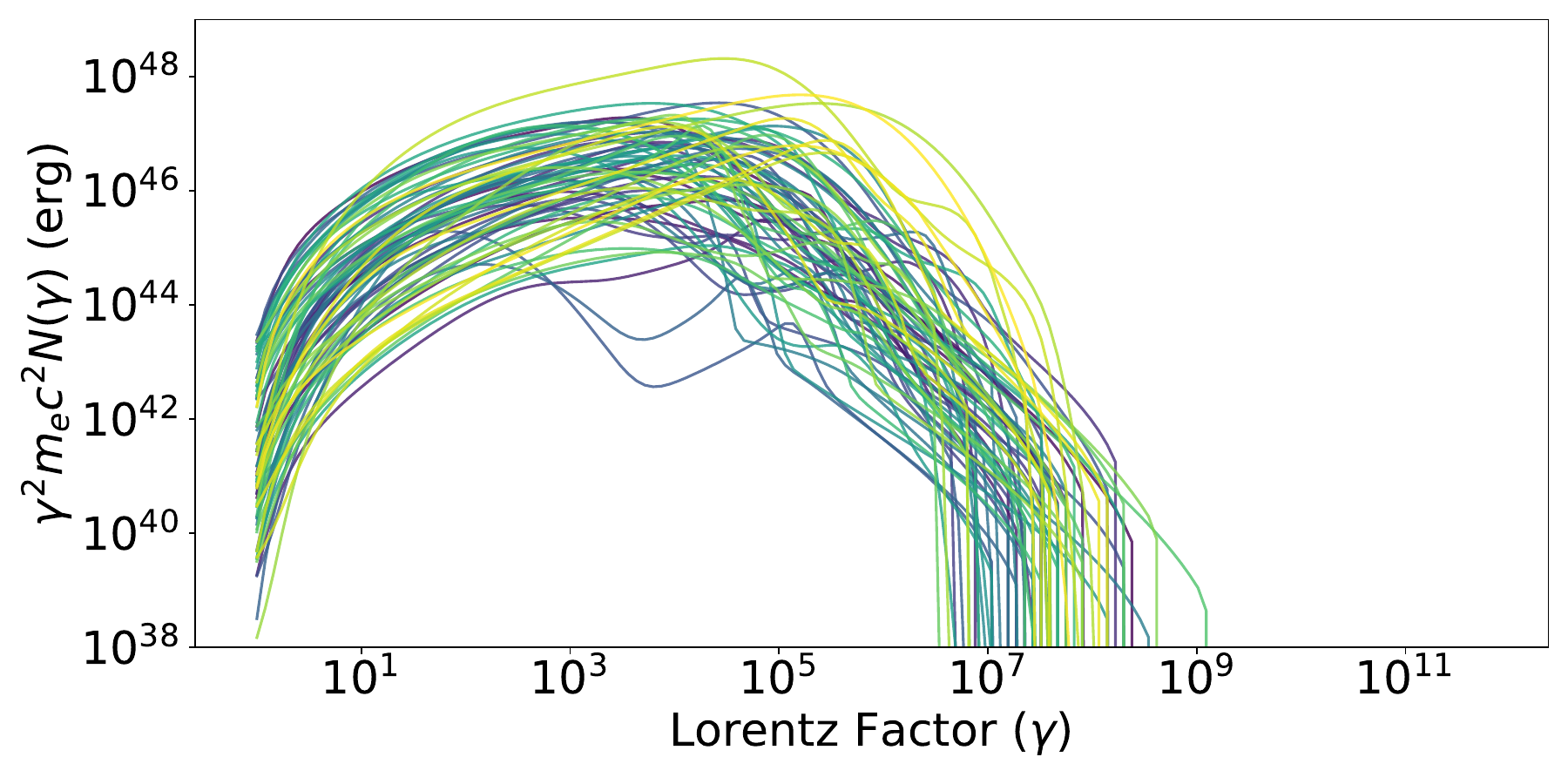} \\
    \includegraphics[width=0.49\linewidth]{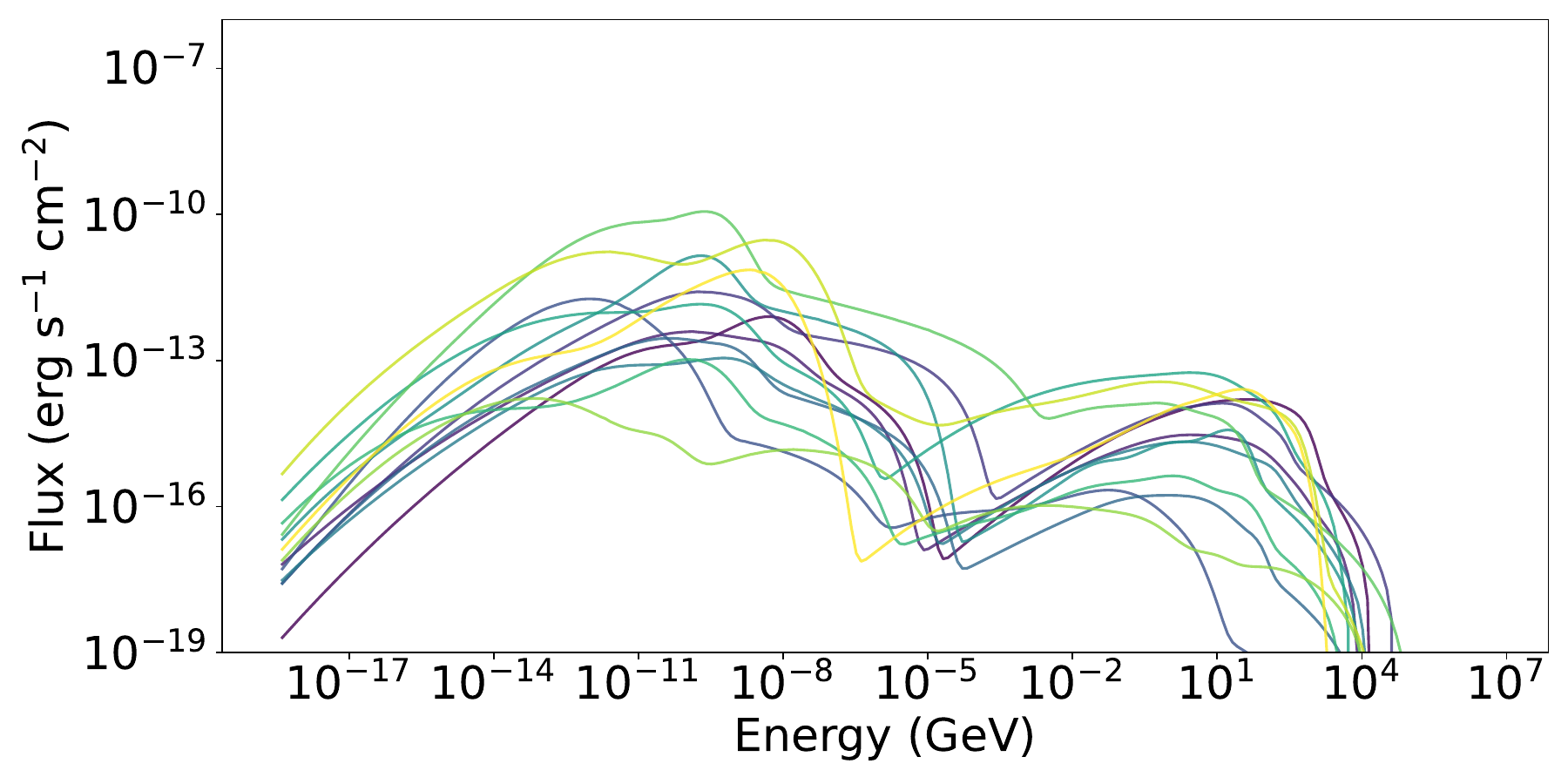}
    \includegraphics[width=0.49\linewidth]{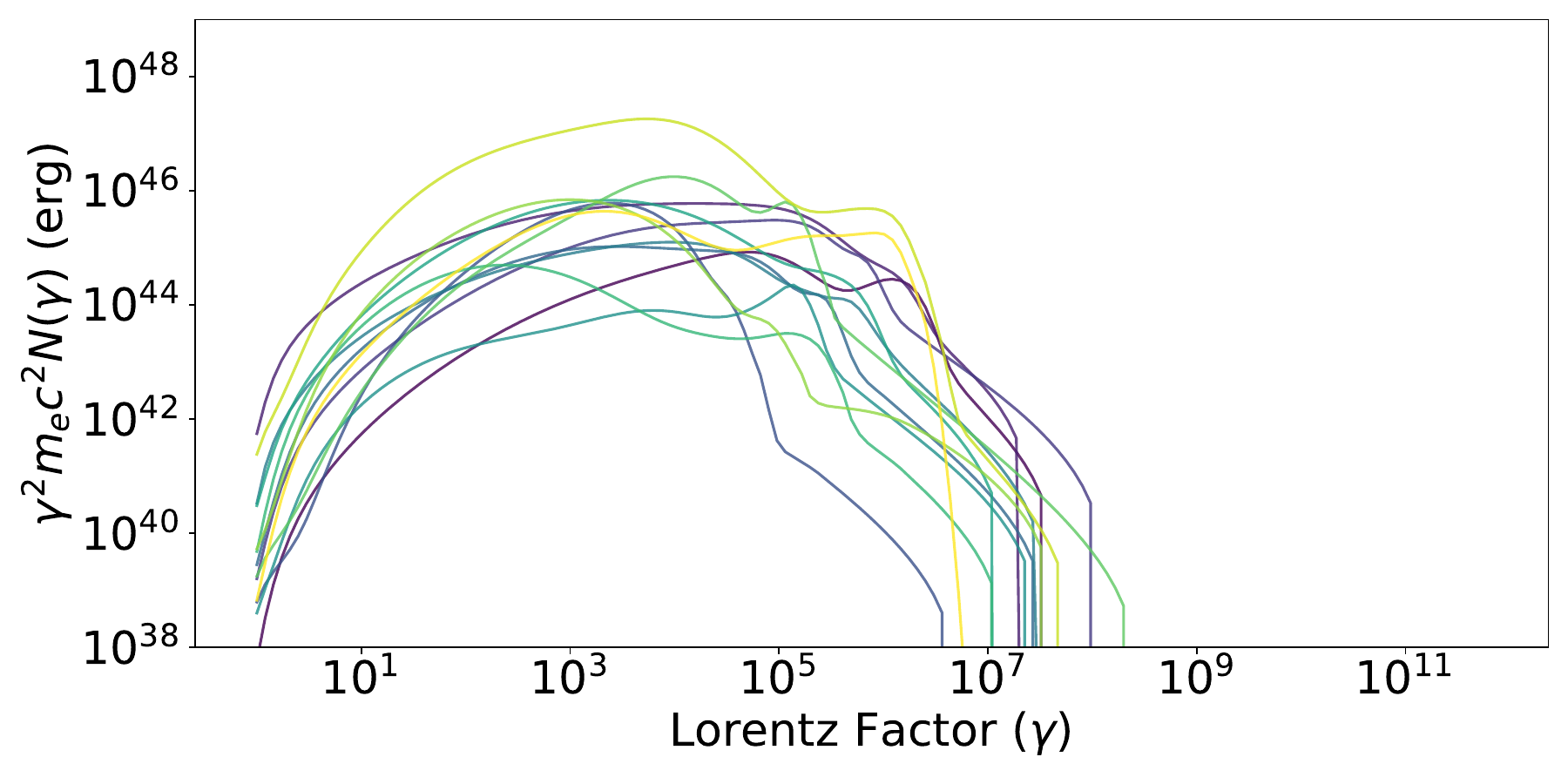} \\
    \includegraphics[width=0.49\linewidth]{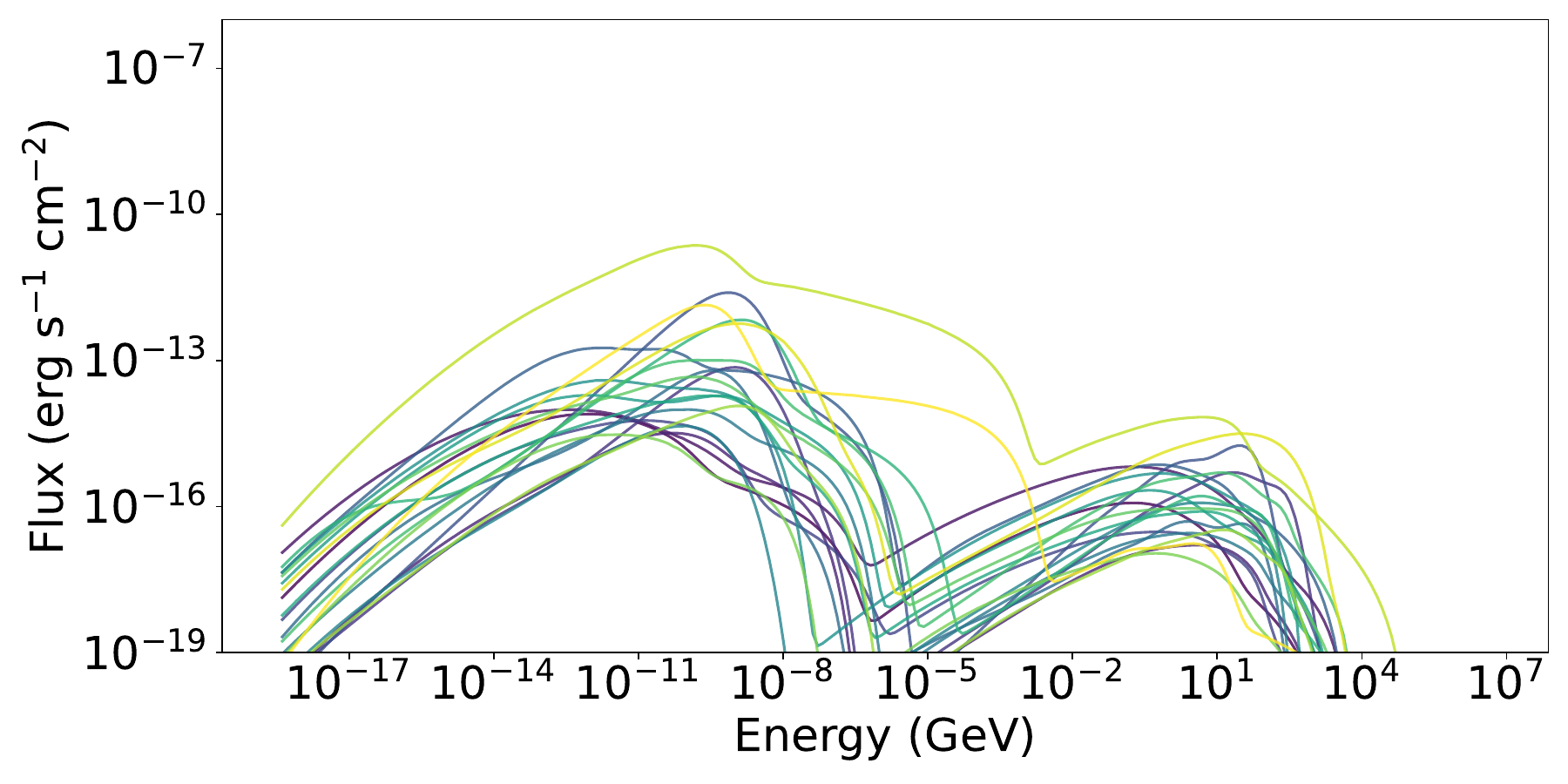}
    \includegraphics[width=0.49\linewidth]{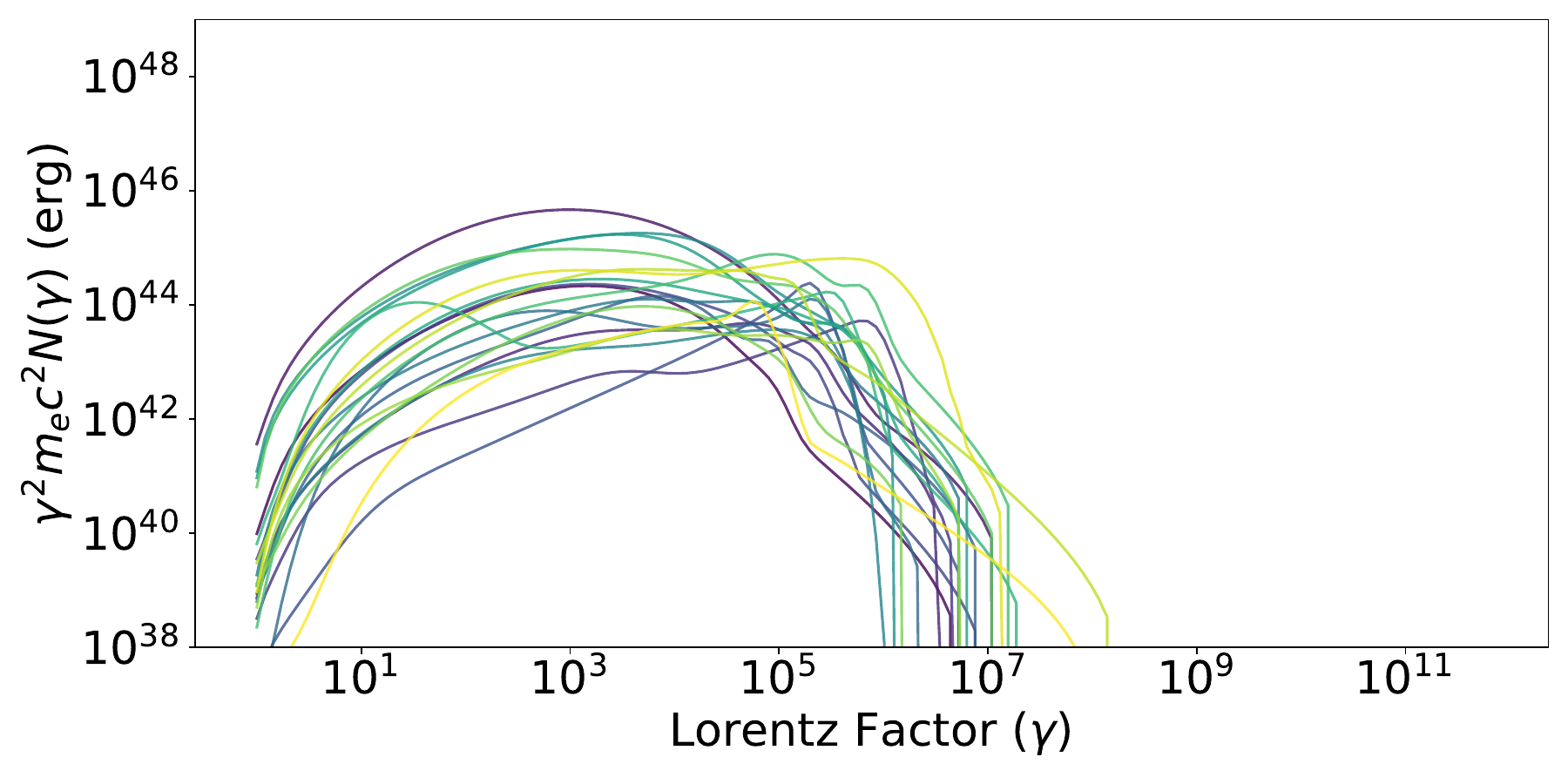}
    \caption{Grouped SEDs (left panel) and electron spectra (right panel) for sources exhibiting superefficiency in FIR frequency range at the current epoch considered from a single random realization of 1600 sources, but for four different CF ranges: CF $< 10$ (first row), $10 < \mathrm{CF} < 50$ (second row), $50 < \mathrm{CF} < 100$ (third row) and CF $> 100$ (fourth row). Note that a continuous colormap is used only to distinguish the different curves: each color corresponds to an individual PWN and has no additional physical meaning.}
    \label{fig: grouped_super_CF}
\end{figure*}

To investigate how the strength of reverberation controls the spectral imprint of FIR superefficiency, we group all PWNe that are superefficient in the FIR band at the current epoch into four ranges of compression factor: CF $<10$, $10<{\rm CF}<50$, $50<{\rm CF}<100$, and CF $>100$. 
We focus on the FIR band for this analysis because it contains the largest number of superefficient sources at the present epoch. Moreover, since the CF is defined only after a system has entered the post-compression phase, there are comparatively fewer X-ray– or GeV-superefficient PWNe in that stage to support a statistically meaningful ensemble study.
For each range, we construct ensemble SEDs and electron spectra by combining all sources that satisfy the FIR-superefficiency criterion within that CF range. 
The resulting grouped SEDs and electron spectra are shown in Fig. \ref{fig: grouped_super_CF}.

As the CF increases, both the number of FIR-superefficient sources and their FIR flux levels decrease, and the characteristic peak of the FIR-emitting electron spectra shift to lower Lorentz factors. 
In the CF $<10$ case, FIR-superefficient PWNe are numerous and display the highest FIR fluxes. 
Their grouped SEDs show strong FIR peaks, and the corresponding electron spectra are peaked at and extended until moderate electron Lorentz factors. 
These systems experience only mild compression, so the magnetic field is moderately enhanced without inducing severe synchrotron cooling. 
As a result, long-lived low-energy electrons accumulate efficiently and radiate strongly in the FIR band, making this regime optimal for sustained FIR superefficiency.

In the intermediate range $10<{\rm CF}<50$, the number of FIR-superefficient sources begins to decline, and the grouped FIR SEDs become systematically fainter. 
The normalization peaks of the electron spectra shift toward lower Lorentz factors, indicating that stronger compression and the associated magnetic field amplification enhance synchrotron losses, progressively depleting the FIR-emitting electron population.
FIR superefficiency therefore persists, but in fewer systems.

For $50<{\rm CF}<100$, this trend strengthens further. 
The FIR-superefficient sources are even lower in number, and the electron spectra show substantial depletion of low-energy particles, characterized by a peak shift towards even lower Lorentz factors. 
Stronger compression in this regime produces larger magnetic field amplification, which efficiently removes FIR-emitting electrons through synchrotron cooling. 
As a result, FIR superefficiency becomes increasingly difficult to sustain.

The CF $>100$ case contains the most extremely compressed systems and the smallest number of FIR-superefficient sources. 
Here, magnetic field amplification is so strong that synchrotron cooling becomes catastrophic even for low-energy, FIR-emitting electrons, rapidly draining the particle reservoir that would otherwise power FIR emission. 
Consequently, both the FIR flux and the number of FIR-superefficient PWNe are minimized in this regime.

Across all CF ranges, FIR superefficiency is therefore governed by a competition between particle accumulation and synchrotron destruction, modulated by the compression strength. 
Mildly compressed systems maximize the survival of FIR-emitting electrons, while extreme compression depletes these electrons via severe radiative cooling and shifts them into even lower energies.

\subsection{Equipartition is not a stable state across the evolution}

\begin{figure}
    \centering
    \includegraphics[width=0.99\columnwidth]{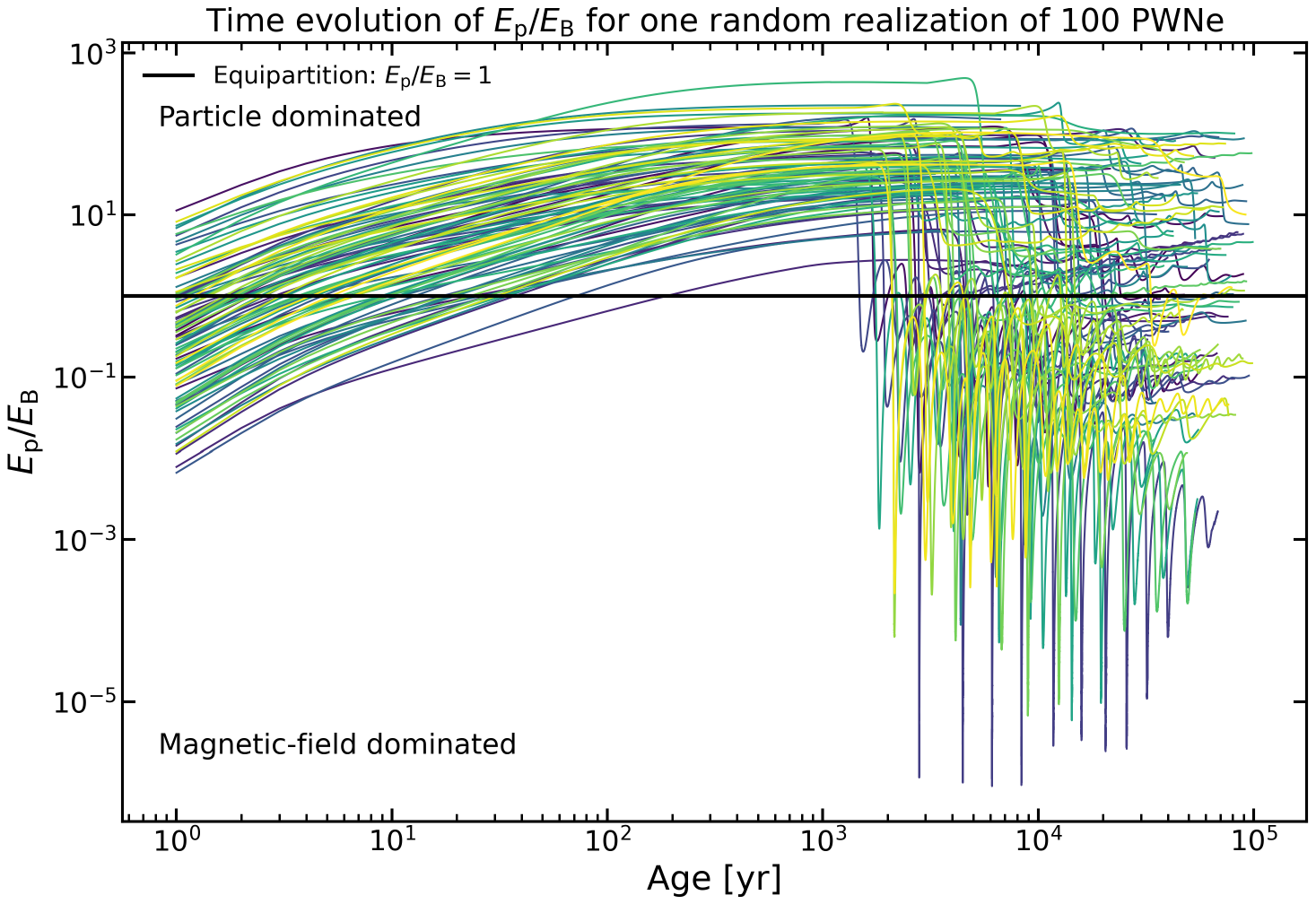}
    \caption{Example of particle-to-magnetic energy ratio for 100 randomly selected sources used in the synthesis. Note that a continuous colormap is used only to distinguish the different curves: each color corresponds to an individual PWN and has no additional physical meaning.}
    \label{fig: eqipart}
\end{figure}

Fig. \ref{fig: eqipart} shows the time evolution of the ratio between particle and magnetic energy, $E_{\rm p}/E_{\rm B}$, for one random realization of 100 PWNe chosen as an example.
Here, $E_{\rm B}(t) = (B(t)^2/8\pi)  (4\pi R(t)^3/3) $ and $E_{\rm p}(t)$ is obtained from integrating the particle spectrum in energy. 
The horizontal line marks equipartition, $E_{\rm p}/E_{\rm B}=1$. 
The systems begin with a broad range of energy partitions and rapidly evolve away from equipartition, with most sources becoming particle dominated during the early free-expansion stage. 
Once reverberation starts, compression amplifies the magnetic field and can drive sharp, sometimes repeated excursions toward magnetically dominated states. 
However, these crossings of equipartition are transient: no source remains close to $E_{\rm p}/E_{\rm B}=1$ for an extended period. 
The population therefore shows that equipartition is not a stable evolutionary attractor, but rather a temporary condition that can be crossed during dynamical reprocessing, especially during and after reverse-shock interaction.

\section{Concluding remarks}\label{conclusion}

Superefficiency represents a natural, albeit extreme, consequence of reverberation driven by the interaction between the PWN and the reverse shock. 
Through magnetic field amplification, adiabatic particle reprocessing, and the redistribution of the particle population, reverberation can temporarily elevate the radiative output of a PWN above the instantaneous spin-down power of its central pulsar, profoundly reshaping its observational appearance across the electromagnetic spectrum.

By addressing superefficiency within a population synthesis framework, this work moves beyond isolated case studies and theoretical expectations to provide a quantitative, statistically robust picture of how frequently superefficiency should arise in the Galaxy and how its manifestation depends on frequency and evolutionary stage.
We have shown that superefficient PWNe occur across the phase-space, and is not 
associate to a particular region of parameters.
We have also shown that superefficiency is intrinsically multi-band in nature: it is most prevalent at low frequencies-particularly in the FIR-while becoming increasingly transient toward the X-ray regime, where superefficiency is generally confined to brief intervals of active compression. 
The systematic separation between compressing and post-compression phases demonstrates that superefficiency is a sequence of physically distinct radiative regimes tied to the dynamical evolution of the nebula.

Our comparison between different population assumptions and evolutionary models highlights the critical role of realistic reverberation physics. 
In particular, the different predictions obtained with \texttt{TIDE+L} and \texttt{TIDE} models underscore the importance of accurately tracking particle evolution and magnetic field amplification during compression in order to capture the prevalence of superefficiency. 
Without such a treatment, the extent of low-frequency (FIR) superefficiency or the phase selectivity of intermediate-frequency (X-ray) superefficiency cannot be accurately captured.

Beyond quantifying the prevalence of superefficiency and its dependence on frequency and evolutionary stage,
this work establishes a unified physical framework for the temporal evolution of superefficient fractions, time-dependent individual source behavior, and epoch- and CF-dependent ensemble spectral properties. %
These results show that reverberation-driven compression and particle reprocessing govern both the number of superefficient PWNe and their MWL manifestation.

Together with \cite{desarkar26}, which focused on the detectability and average visibility of the Galactic PWN population, this study links typical PWNe to their most extreme radiative episodes. 
By unifying population statistics, evolutionary modeling, compression physics, and MWL phenomenology, this work intends to establish a more comprehensive framework for understanding superefficiency and for placing these extreme but relatively common events within their broader Galactic context.

\begin{acknowledgements}
      This work has been supported by the grant PID2024-155316NB-I00 funded by MICIU /AEI /10.13039/501100011033 / FEDER, UE, and CSIC PIE 202350E189. This work is also partly supported by the Spanish program Unidad de Excelencia María de Maeztu CEX2020-001058-M, financed by MCIN/AEI/10.13039/501100011033, and by the MaX-CSIC Excellence Award MaX4-SOMMA-ICE, and also supported by MCIN with funding from European Union NextGeneration EU (PRTR-C17.I1).												
      ADS is supported by the grant Juan de la Cierva JDC2023-052168-I, funded by MCIU/AEI/10.13039/501100011033 and by the ESF+. 		
      DFT acknowledges the T. D. Lee Institute, where part of this research was done, for hospitality.					
      BO and NB acknowledge partial support by the INAF Mini-grant HYPNOTIC87A: Hidden Young Pulsar Nebula Occupying The Inner Core of 87A and by the European Union – NextGenerationEU RRF M4C2 1.1 grant PRIN-MUR 2022TJW4EJ.
\end{acknowledgements}

   \bibliographystyle{aa} 
   \bibliography{aa.bib} 

\end{document}